\documentclass[preprint]{elsarticle}
\usepackage{amssymb,amsmath,graphicx,amscd,epsfig}
\usepackage[utf8]{inputenc}
\usepackage[T1]{fontenc}
\usepackage[colorlinks=true, pdfstartview=FitV, linkcolor=blue, citecolor=red, urlcolor=magenta, breaklinks=true]{hyperref}
\usepackage[all]{xy}
\usepackage{color}
\usepackage{orcidlink}
\usepackage{subeqnarray}
\usepackage{booktabs}   
\usepackage{array}      
\usepackage{float}     
\usepackage{graphicx}  

\newcommand{\bb}{\bibitem}
\newcommand{\bes}{\begin{subequations}}
\newcommand{\ees}{\end{subequations}}
\def\ben{\begin{eqnarray}}
\def\een{\end{eqnarray}}
\newcommand{\bens}{\begin{subeqnarray}}
\newcommand{\eens}{\end{subeqnarray}}

\newcommand{\PC}[1]{\ensuremath{\left(#1\right)}}
\def\be{\begin{equation}}
\def\ee{\end{equation}}

\def\tanh{\text{tanh}}
\def\sech{\text{sech}}
\def\exp{\text{exp}}
\def\cos{\text{cos}}

\def\arctanh{\text{arctanh}}

\begin{document}

\begin{frontmatter}

\title{Compact, long-range and vacuumless regimes controlled by scalar fields in two-dimensional flat and five-dimensional warped geometries}

\author{Elisama E. M. Lima \,\orcidlink{0000-0001-7166-9156}}
\address{Federal Institute of Education, Science and Technology of Bahia (IFBA), 
47808-006, Barreiras, BA, Brazil}

\author{Dionisio Bazeia \,\orcidlink{0000-0003-1335-3705}}
\address{Department of Physics, Federal University of Para\'iba (UFPB), Jo\~ao Pessoa, PB 58051-970, Brazil}

\begin{abstract} 
We investigate the presence of compact, long-range, and vacuumless regimes in models described by real scalar fields in 2D flat and 5D  warped geometries. In the two-dimensional case, we study in detail several models that support localized solutions. When embedded into a warped $\text{AdS}_5$ background, these fields generate distinct kinds of brane profiles.
\end{abstract}

\end{frontmatter}

\section{Introduction} \label{intro}

Topological defects such as kinks, domain walls, vortices, and monopoles play an essential role in many areas of physics, from condensed matter systems to high-energy field theory and cosmology~\cite{B1,B2,B3,B4,B5,BB1,BB2,BB3,BB4}. In $(1+1)$ dimensions, kinklike configurations arising from real scalar fields are examples of such localized structures.
Among the wide variety of scalar models, special attention has been devoted to those supporting \emph{compactons}, which are field solutions completely localized in a finite spatial region, originally introduced in Ref.~\cite{rosenau}. Compactons have since been studied in different contexts~\cite{C1,C2,compactkinks,marques2014,lima2017}, including their appearance in braneworld geometries where they may lead to configurations connecting thick and thin brane behaviours~\cite{Br1,Br2,Br3,bra,Br4,Br5,Br6,Br7,Br8,Br9}, called \emph{hybrid branes}~\cite{marques2014,lima2017}. Furthermore, some models support \emph{long-range} defects, for which the field approaches the vacuum with polynomial behaviour, while others develop \emph{vacuumless} configurations, where the minima of the potential are removed to infinity \cite{ChoVacuumless1999,ChoV,dbazeia1999,Gomes2012,BMM2018,Blinov2022}. These distinct regimes control how energy and gravity are localized, motivating the systematic comparison of those field theories.

In this work, we propose classes of scalar field models characterized by a positive integer parameter $n\in \mathbb{Z}^+$ and, in a more general formulation, by an additional deformation $\alpha$. By varying these parameters, the models can interpolate between smooth kinks, compactons, long-range, or vacuumless structures. It allows us to understand how the properties of the localized solutions evolve. Through a comparative analysis, their implications are investigated in both flat and warped $AdS_5$ geometries.

We start the investigation in Sec. \ref{sec-2}, with a general description of models driven by a single real scalar field, including classical stability of static solutions and the mass aspects. In Sec.~\ref{sec-3} we discuss three distinct deformations of the standard $\phi^4$ model in $(1+1)$ dimensions. Starting from the $\phi^4$ kink at $n=1$, each kind of potential evolves in a different way as $n$ increases: the first smoothly transforms the kink into a compacton; the second eliminates the classical mass at the vacua to generate long-range kinks with highly interactive polynomial tails; and the third, inspired by a generalization of the Starobinsky inflationary potential, drives the system towards a vacuumless configuration with runaway minima. We then introduce the $\alpha$-deformation~\cite{compactkinks,marques2014} in Sec.~\ref{sec-4}, which alters these scenarios through a square root modification of the potential. In the large-$\alpha$ limit, two families of compact solutions are obtained in a standard field theory, where the compactification originates from the deformation of the potential rather than from nonlinear kinetic terms, as proposed in Ref.~\cite{lima2024}. This regime further demonstrates that, while this mechanism is capable of transforming standard kinks into compactons, long-range systems resist compactification due to the flatness of their massless vacua.

The investigation continues in Sec. \ref{sec-5}, where we embed these scalar field constructions into a five-dimensional warped geometry of the Randall–Sundrum type~\cite{Br1,Br2,Br3,bra}. By implementing the first-order formalism, we obtain analytical expressions for the warp factor and energy density for each case. Although the first model was previously analyzed in the braneworld setting~\cite{marques2014}, the remaining models and generalizations have not yet been explored in the warped geometry context. Here, we demonstrate how the warped geometry regularizes the divergence of the vacuumless limit giving rise to a Gaussian brane profile, and how the $\alpha$-generalized scenarios generate novel hybrid branes, characterized by thick-to-thin transitions along the extra dimension at finite values of $n$. We then finish the work in Sec. \ref{sec-6}, summarizing the main results and adding comments concerning other investigations of current interest.

\section{Generalities} \label{sec-2}

In this section, we introduce general aspects of models describing a single real scalar field in $(1+1)$-dimensional spacetime, governed by the standard Lagrangian density
\be
\label{lagran1}
{\cal L} = \frac{1}{2} \partial_\mu \phi \, \partial^\mu \phi - V(\phi),
\ee
where $V(\phi)$ represents the scalar field potential. For simplicity, we adopt natural units $\hbar=c=1$, and work with dimensionless fields and spacetime coordinates. The metric tensor is $\eta_{\mu\nu} = \text{diag}(1, -1)$. The invariance under translations in spacetime results in the energy-momentum tensor taking the form
\be
T^{\mu\nu} = \partial^\mu \phi \, \partial^\nu \phi - \eta^{\mu\nu} \mathcal{L}.
\ee
The equation of motion given by the Euler–Lagrange equation reads
\be
\partial_\mu \partial^\mu \phi + \frac{dV}{d\phi} = 0,
\ee
and for static field configurations this reduces to
\be
\frac{d^2 \phi}{dx^2} = \frac{dV}{d\phi}. 
\label{soe}
\ee
We are interested in localized solutions interpolating between different minima of $V(\phi)$; then the solutions must satisfy the boundary conditions $\phi(x \to \pm\infty) = v_\pm$, where $v_\pm$ are neighboring minima of the potential. 

A simpler way to handle the equation of motion is by introducing an auxiliary function $ W(\phi)$, often referred to as \textit{superpotential}, such that
\be
V(\phi) = \frac{1}{2} \left( \frac{dW}{d\phi} \right)^2
\ee
This formulation allows the second-order differential equation \eqref{soe} to be reduced to first-order, following the Bogomol'nyi-Prasad-Sommerfield (BPS) procedure \cite{bogomol}: 
\be
\frac{d\phi}{dx} = \frac{dW}{d\phi}
\label{foebogo}
\ee
The energy density associated with the field is given by the $T_{00}$ component of the energy-momentum tensor,  i.e., $\rho = T_{00}$. For static configurations, we find
\be
\rho(x) =\frac12 \left(\frac{d\phi}{dx}-\frac{dW}{d\phi}\right)^2+ \frac{dW}{dx}.
\label{rho}
\ee
When the static field obeys the first-order equation \eqref{foebogo}, the total energy becomes
\be
E = \left| W(\phi(x \to \infty)) - W(\phi(x \to -\infty)) \right|.
\ee
The solutions obtained from the first-order formalism not only satisfy the original second-order equation, but also correspond to field configurations that minimize the total energy, which is determined entirely by the asymptotic values of the field and does not depend on how the solution behaves in between the vacua \cite{bogomol,Bazeia}.

The stability of the solutions is examined by considering small fluctuations around the static solution, $\phi(x,t) = \phi(x) + \sum_k\eta_k(x) \cos(\omega_k t)$, with $k\in \mathbb{Z}$, which leads to a Schrödinger-like equation 
\be
\left(- \frac{d^2}{dx^2} + U(x) \right) \eta_k(x) = \omega_k^2 \, \eta_k(x),
\ee
where the stability potential is defined as
\be
\label{spx}
U(x) = \frac{d^2 V}{d\phi^2} =  \left( \frac{d^2W}{d\phi^2} \right)^2 + \frac{dW}{d\phi} \cdot \frac{d^3W}{d\phi^3},
\ee
evaluated at the static solution $\phi = \phi(x)$. This determines the spectrum of the perturbations, and the system is considered linearly stable if all eigenvalues $\omega_n^2$ are non-negative. In the BPS framework, the Schrödinger-like equation can be factorized, ensuring the existence of linearly stable solutions, as discussed in Ref.~\cite{Bazeia}.

The mass of a fluctuation around a vacuum is determined by the second derivative of the potential at the minima
\be
m^2_{\pm} = \left. \frac{d^2 V}{d\phi^2} \right|_{\phi = v_\pm}
\label{mass}
\ee
For symmetric models, such as the $\phi^4$ theory \cite{Kevrekidis2019}, 
$m^2_- = m^2_+ = m^2$. When this quantity is finite, it controls the asymptotic exponential behaviour of the kink towards the  vacua, i.e., the right tail of the solution behaves like $v_{+}-\phi(x) \sim e^{-m x}$ \cite{Gomes2012}. If the mass increases further and further, the solution reaches the minima faster, which could lead to compacton structures \cite{marques2014}. The stability potential $U(x)$ that asymptotically approaches $m^2$ tends to behave as an infinite well, localizing the fluctuations within a finite region.

Otherwise, when the derivatives of the potential vanish at the minima up to higher orders, it makes the classical mass vanish, giving $m^2 = 0$. In this case, the field does not reach the asymptotic region exponentially but rather with a power-law behaviour that falls off slower than the exponential one \cite{Gomes2012,BMM2018,Blinov2022}. Thus the tail of the solution is affected, which follows a polynomial form. These kinks support long-range interactions and are often referred to as long-range or highly interactive structures. More recently, Ref.~\cite{superlongrangeKinks} investigated \textit{super long-range} kinks, in which derivatives of all orders vanish at the minima, giving rise to logarithmic tails. The distinction among the different types of asymptotic behaviour plays a central role in determining the physical aspects of the solutions, including localization, stability, and interaction properties.

\section{Localized Solutions in Scalar Field Models} \label{sec-3}

Let us now analyse models in the context of standard scalar field theory that supports localized solutions characterized by a positive integer parameter $n$. Depending on how this parameter appears in the potential, the corresponding solutions may transform the usual $\phi^4$-kink into compact, long-range, or vacuumless structures. Below, we will study three different models. The first one allows a smooth transition to a compacton, resulting in a strictly localized configuration \cite{marques2014}.  In contrast, the second model, originally introduced in the context of modified dynamics~\cite{twin2012}, is here reinterpreted under standard kinematics. In this setting, the resulting solutions remain smooth for all values of $n$,  but exhibit long-range behaviour when $n>1$. The third model has a potential with runaway minima, which becomes vacuumless for very large $n$ and gives rise to an unlimited solution~\cite{ChoVacuumless1999,ChoV,dbazeia1999}. These distinct features will later be explored within a more general framework and then implemented in braneworld scenarios.

\subsection{Model A: connecting kink and compacton}

\begin{figure}%
\centering
\includegraphics[scale=0.25]{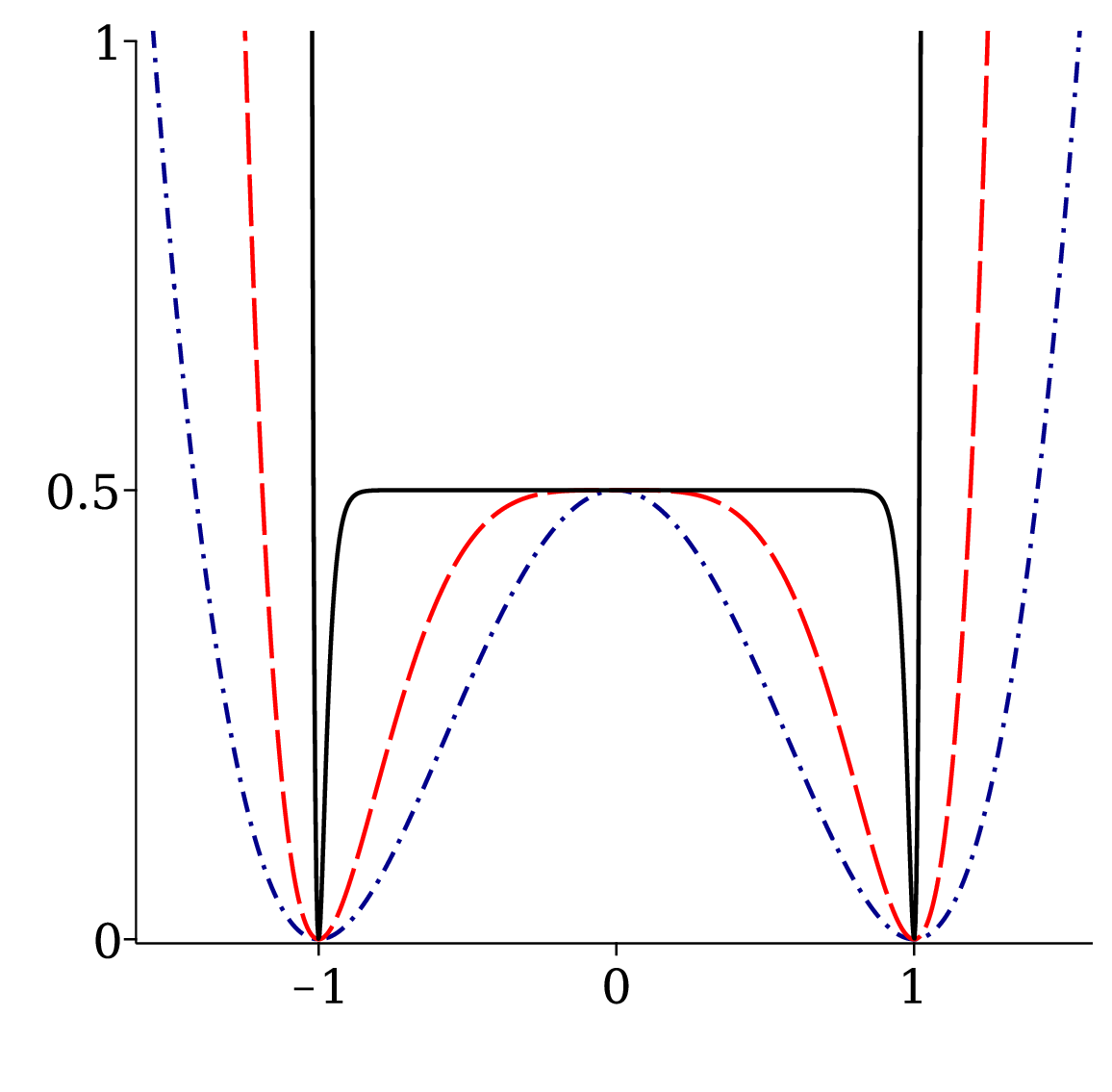}
\includegraphics[scale=0.25]{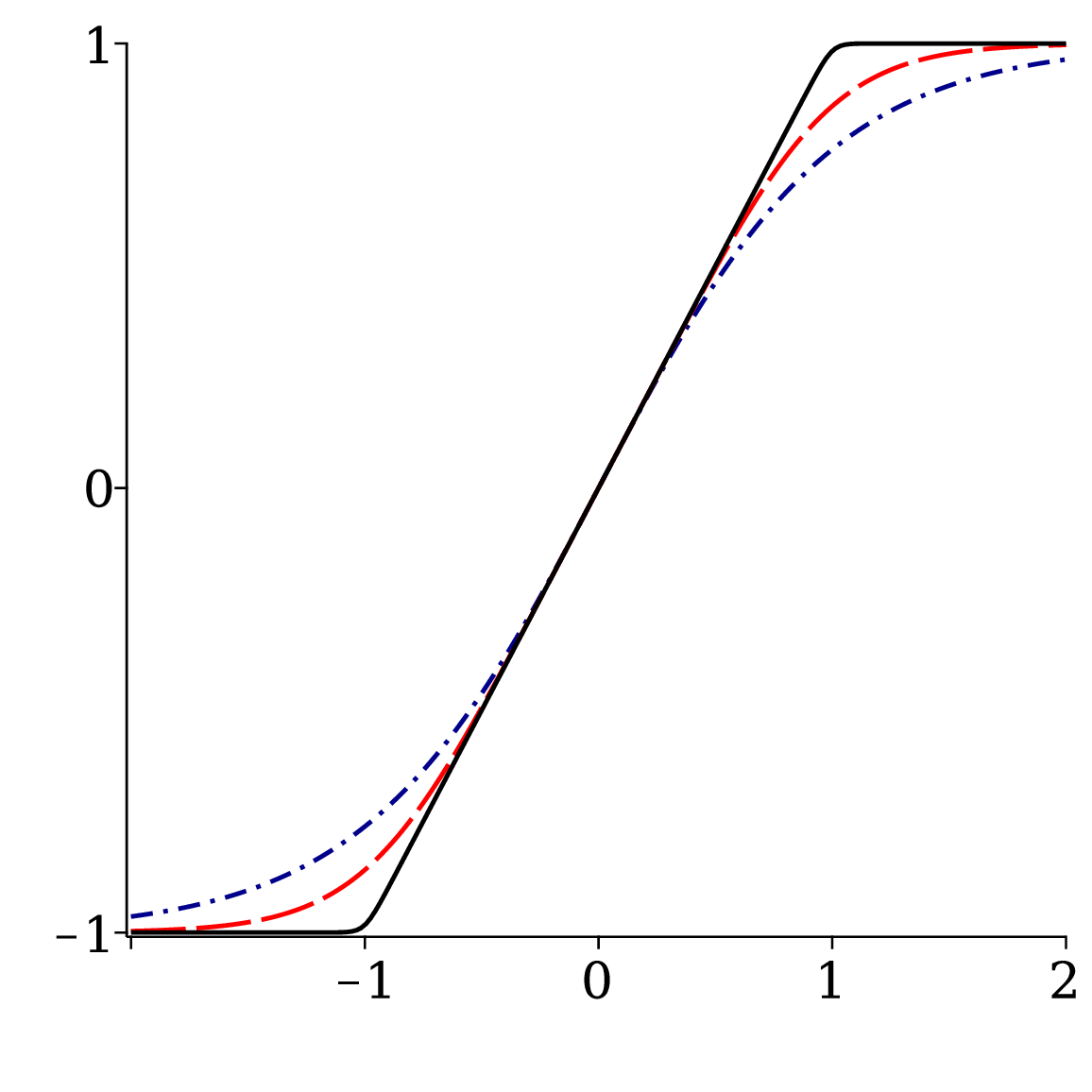}
\caption{The potential $V_{A}(\phi)$ (left) and the kink solution $\phi(x)$ (right) for $n=1,2,20$, represented by dash-dotted (blue), dashed (red), and solid (black) lines, respectively.}
\label{fig1a}
\end{figure}

\begin{figure}%
\centering
\includegraphics[scale=0.25]{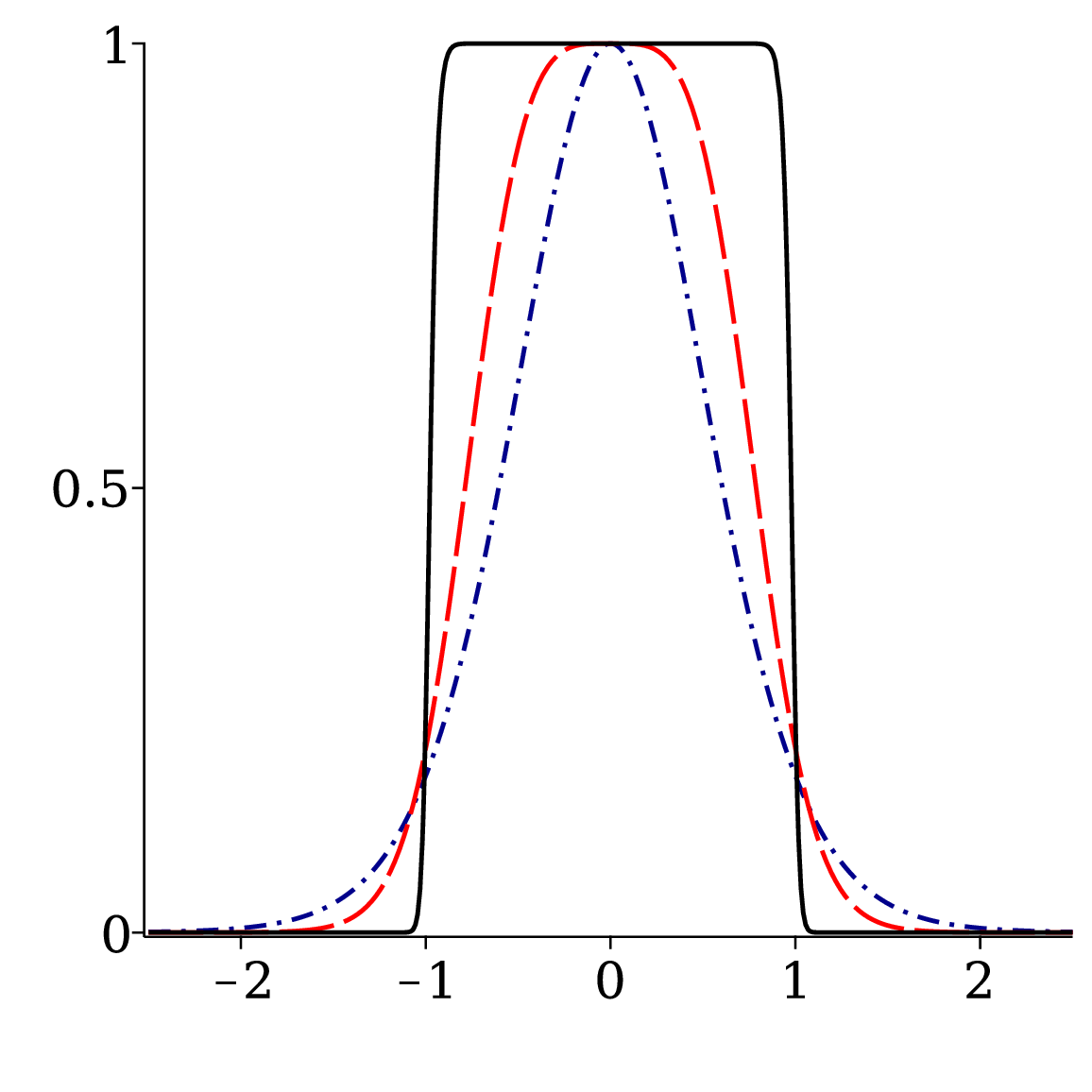}
\includegraphics[scale=0.25]{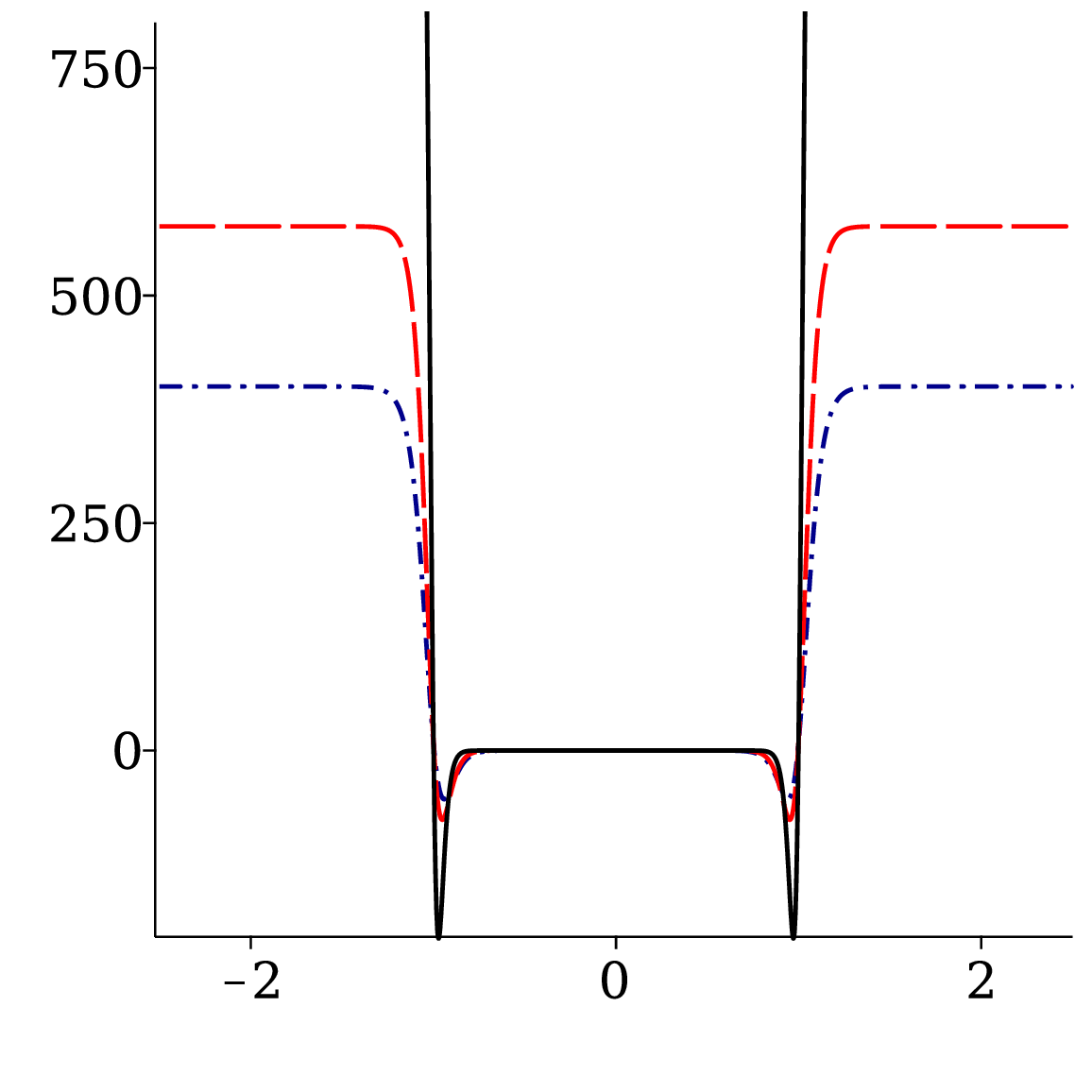}
\caption{The energy density $\rho(x)$ (left) for the same values of $n$ used in Fig.~\ref{fig1a}, and the stability potential $U(x)$ (right) for $n=10,12,20$ represented by the same line styles.}
\label{fig1b}
\end{figure}

The first potential suggested  is
\be
V_{A}(\phi)=\frac{1}{2}\left(1-\phi^{2n}\right)^2,
\label{Vn}
\ee
where $n \in \mathbb{Z}^+$ controls the shape of the potential and other characteristics associated with the resulting solution profiles (see Figs.~\ref{fig1a}--\ref{fig1b}) \cite{marques2014}. For $n=1$, this model reduces to the standard $\phi^4-$theory: $V(\phi)=\frac{1}{2}\left(1-\phi^{2}\right)^2$, supporting the kink solution $\phi=\tanh(x)$, the energy density $\rho(x)=\sech^4(x)$ and the modified P\"oschl-Teller stability potential, $U(x)= 4-6\,\sech^2(x)$, with two bound states $\omega_0=0$ and $\omega_1^2=3$ \cite{teller,flugge}. Regardless the value of $n$, the  expression \eqref{Vn} exhibits minima at the positions $\phi_{min}=v_\pm=\pm 1$, but the mass varies according to $m^2=4n^2$.

\begin{figure}%
\centering
\includegraphics[scale=0.36]{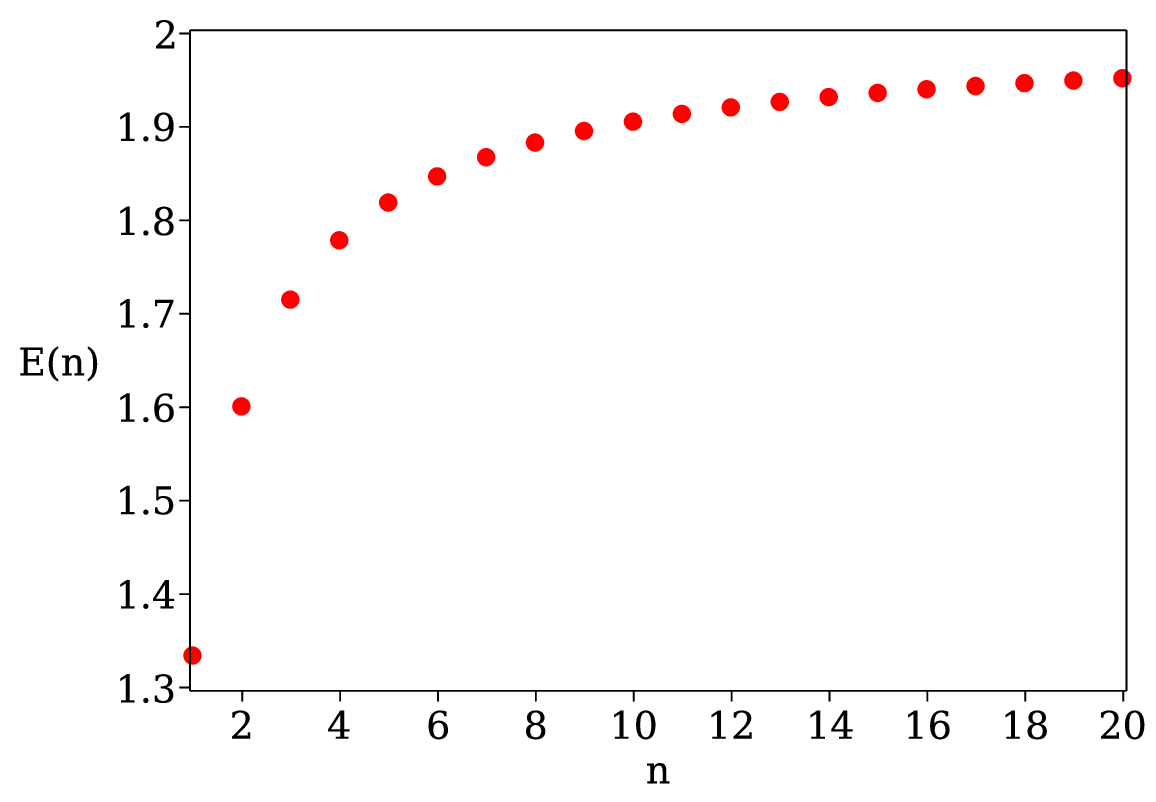}
\caption{Total energy $E_A$ as a function of the parameter $n$.}
\label{fig1c}
\end{figure}

The first-order equation associated with this model is given by
\be
\frac{d\phi}{dx} = 1 - \phi^{2n}.
\label{foe1}
\ee
The corresponding function $W$ is 
\be
W(\phi) = \phi - \frac{\phi^{2n+1}}{2n+1}.
\label{W1}
\ee
The total energy is obtained from the difference
\be
E_{A} =W(1)-W(-1)= \frac{4n}{2n+1},
\label{E1}
\ee
which for $n=1$ is $E = {4}/{3}$, gradually increasing  until reaching the value $E=2$ in the limit $n \to \infty$, as depicted in Fig.~\ref{fig1c}. Equation \eqref{foe1} leads to an implicit expression for the solutions in terms of the Lerch transcendent function ${\Phi}$ 
\be
\phi(x) \cdot {\Phi}\left( \phi(x)^{2n},1,\frac1{2n}\right)=2nx,
\ee
or equivalently in terms of the hypergeometric function ${ _2F_1}$
\be
\phi(x) \cdot {_2F_1}\left(1,\frac1{2n};1+\frac1{2n};\phi(x)^{2n}\right)=x.
\label{solVnstandard}
\ee
The energy density can be expressed at the solution as $\rho(x) =\phi'^2 =  \left(1 - \phi^{2n}\right)^2$. However, since $\phi(x)$ is obtained only by an implicit form, $\rho(x)$ cannot be written explicitly as a direct function of $x$.

The hypergeometric function for $n=1$ can be rewritten in terms of the inverse hyperbolic tangent as
\be
{ _2F_1}\left(1,\frac{1}{2};\frac{3}{2};\phi^{2}\right) = \frac{\arctanh(\phi)}{\phi},
\label{f1arctanh}
\ee
whose equivalence can be confirmed either by comparing the series expansions of both sides or directly using known mathematical properties among special functions \cite{mathrefs}. The relation \eqref{f1arctanh} allows us to explicitly get the analytic solution  $\phi(x)=\tanh(x)$,  corresponding to the $\phi^4$ profile. 

On the other hand, in the limit $n\to\infty$, the hypergeometric function tends to unity for any $|\phi|<1$,  
\begin{equation}
\lim_{n\to\infty} { }_2F_1\!\left(1,\frac{1}{2n};1+\frac{1}{2n};\phi^{2n}\right)=1,
\end{equation}
so that Eq.~\eqref{solVnstandard} reduces to the simple relation $\phi(x)=x$ within the interval $x\in(-1,1)$.  In this regime, the field configuration becomes linear in the core and exactly saturates to the vacuum values at the boundaries, producing a compact kink. The corresponding energy density is constant, $\rho(x)=1$, inside the compact support and vanishes outside. This behaviour contrasts with the case $n=1$, where the solution is a smooth kink.

The implicit solution expressed in Eq.~\eqref{solVnstandard} determines how the parameter $n$ affects the profile of the defect. For $n=1$, the hypergeometric function reduces to the standard $\phi^4$ kink, where the field smoothly connects the minima. However, for larger values of $n$, the solution changes to a localized compacton with a finite spatial support.
To better understand how this transition occurs, let us examine the asymptotic behaviour of the solution, which can be directly obtained from the first-order equation~\eqref{foe1}. As we move away from the origin at $x = 0$, the field approaches the vacua exponentially, following $1 - \phi(x) \sim e^{-2n x}$ for $x \to \infty$, while the energy density goes as $\rho(x) \propto e^{-4n x}$. This shows that, as $n$ increases, the solution that initially approached the minima only asymptotically becomes progressively more localized near the origin, until it reaches the minima at finite values of $x$ as $n \to \infty$. In this regime, a compact structure is observed. The associated linear stability potential tends towards an infinite well, increasing the number of bound states (see right panel of Fig.~\ref{fig1b}). Thus, the parameter $n$ smoothly transforms an infinitely extended kink into a strictly localized compacton.

\subsection{Model B: kinks with polynomial tails}

\begin{figure}%
\centering
\includegraphics[scale=0.25]{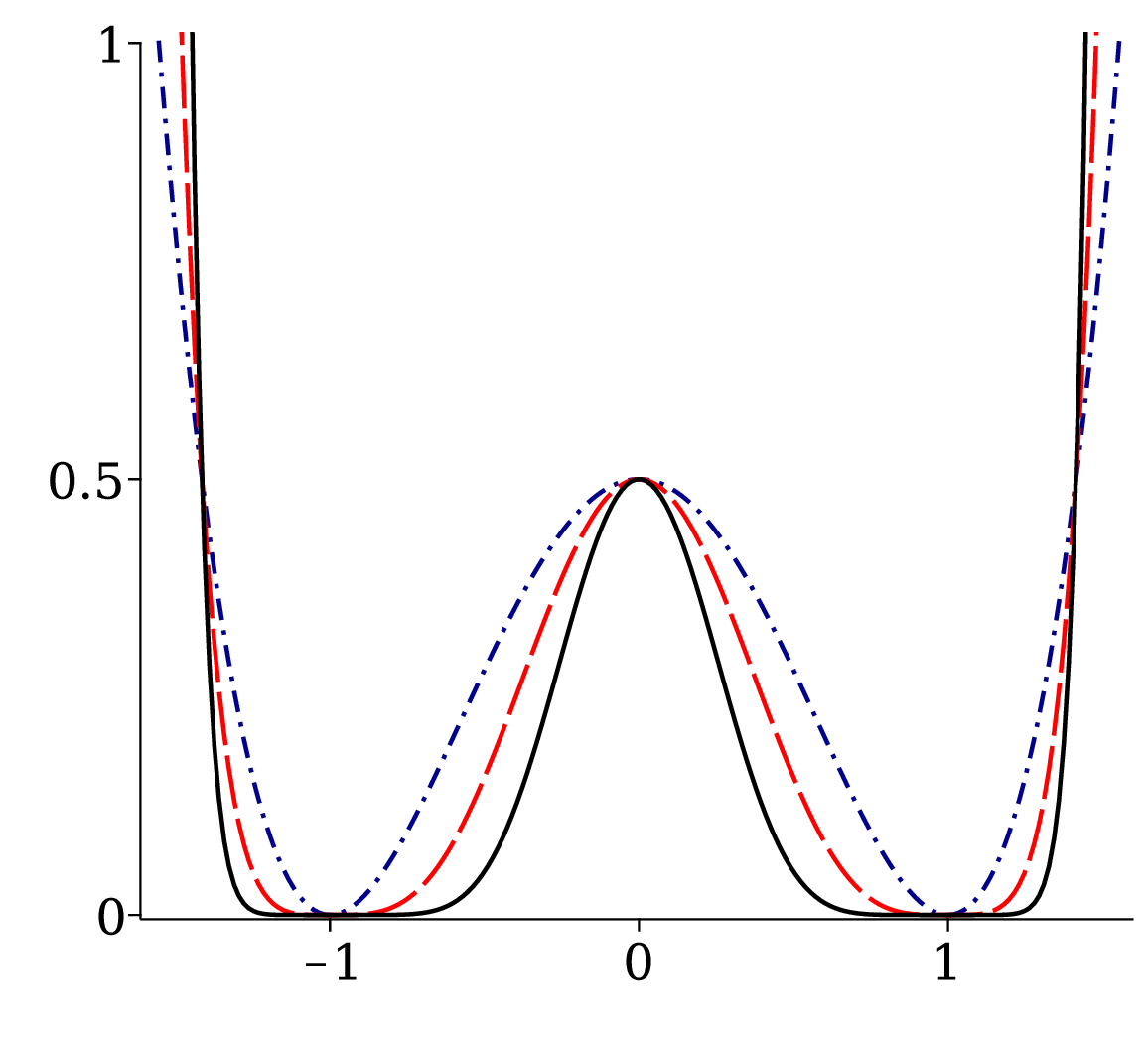}
\includegraphics[scale=0.24]{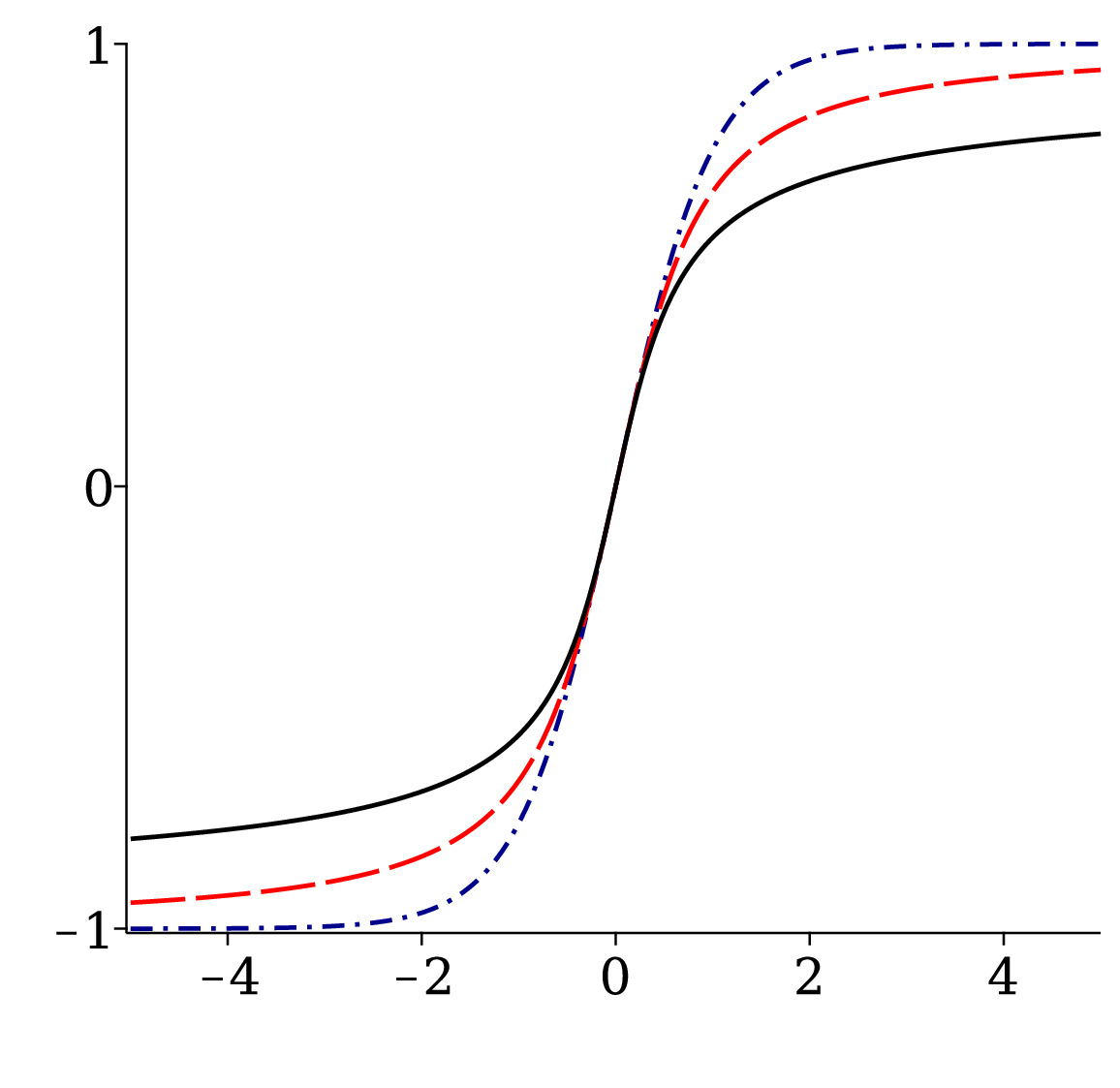}
\caption{The potential $V_{B}(\phi)$ (left) and the kink solutions $\phi(x)$ (right) for $n=1,2,4$, represented by dash-dotted (blue), dashed (red), and solid (black) lines, respectively.}
\label{fig2a}
\end{figure}

The second model is defined by
\be
V_{B}(\phi) = \frac{1}{2} \left(1 - \phi^2 \right)^{2n}.
\label{V2}
\ee
Different from the construction proposed in Ref.~\cite{twin2012}, which investigates twinlike models in a noncanonical perspective, we work here with fields described within the canonical scenario. In comparison with the potential given in Eq.~\eqref{Vn}, where the exponent $n$ acts directly on the powers of $\phi$, it is now applied to the entire structure $(1 - \phi^2)$. Although both models have degenerate minima at the same points,  $v_\pm= \pm 1$, and reduce to the $\phi^4$ form when $n = 1$, they exhibit markedly distinct behaviour as $n$ increases. An important difference lies in the local curvature at the vacua, see Fig.~\ref{fig2a}. For $n = 1$, the second derivative of $V_{B}(\phi)$ at the minima yields the usual mass $m^2 = 4$. However, for $n > 1$, the classical mass is zero and the potential becomes flatter around the vacua. The first nonzero derivative ${d^{k} V_B}/{d\phi^{k}}$ at the minima is of order \(k=2n\) for $n \geq 1$,  and is given by
\be
\left.\frac{d^{2n}V_B}{d\phi^{2n}}\right|_{\phi=\pm 1}
= \frac{1}{2}\sum_{j=n}^{2n} \binom{2n}{j} (-1)^j \frac{(2j)!}{(2j-2n)!}
\label{nonvan}
\ee
The associated first-order equation is
\begin{equation}
\frac{d\phi}{dx} = \left(1 - \phi^2 \right)^n,
\label{foe2}
\end{equation}
which admits the implicit solution 
\begin{equation}
\phi(x) \cdot {_2F_1} \left( \frac{1}{2}, n; \frac{3}{2}; \phi(x)^2 \right) = x.
\label{sol_phi_model2}
\end{equation}
The identity in~\eqref{f1arctanh} combined with the symmetry property of hypergeometric functions under permutation of the first two arguments, shows that the usual $\phi^4$ solution is recovered for $n = 1$. For large $x$, the field $\phi(x)$ approaches the vacuum according to a power-law for $n > 1$. Specifically,  one finds 
\be
1 - \phi(x) \sim \frac{1}{(2^n(n-1)x)^{1/(n - 1)}}  \quad  (x \to \infty)
\ee
which have polynomial tails ~\cite{Gomes2012,BMM2018,Blinov2022}. 
The parameter $n$ controls the order of the first nonvanishing derivative of the potential at its minima and, consequently, determines how far the solution extends. These results differ from those reported in~\cite{twin2012}, which adopts modified dynamics while keeping the solution unchanged. In the present case, as $n$ increases, the vacua are reached more slowly, and the solutions develop increasingly long-range tails. In Fig.~\ref{fig2a}, the potential $V_{B}(\phi)$ and the corresponding solutions~\eqref{sol_phi_model2} are shown for some values of $n$.

\begin{figure}%
\centering
\includegraphics[scale=0.36]{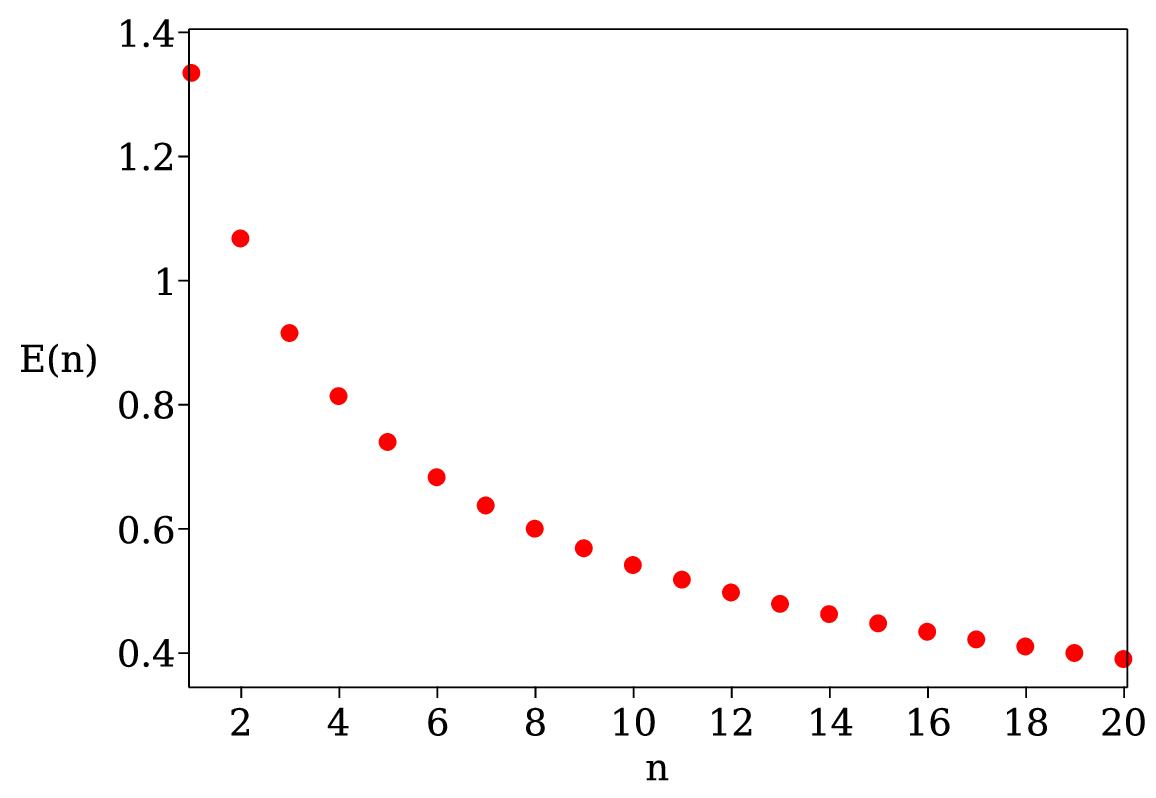}
\caption{Total energy $E_B$ for some values of $n$; note that $E_B \to 0$ in the limit $n \to \infty$.}
\label{fig2b}
\end{figure}

\begin{figure}%
\centering
\includegraphics[scale=0.25]{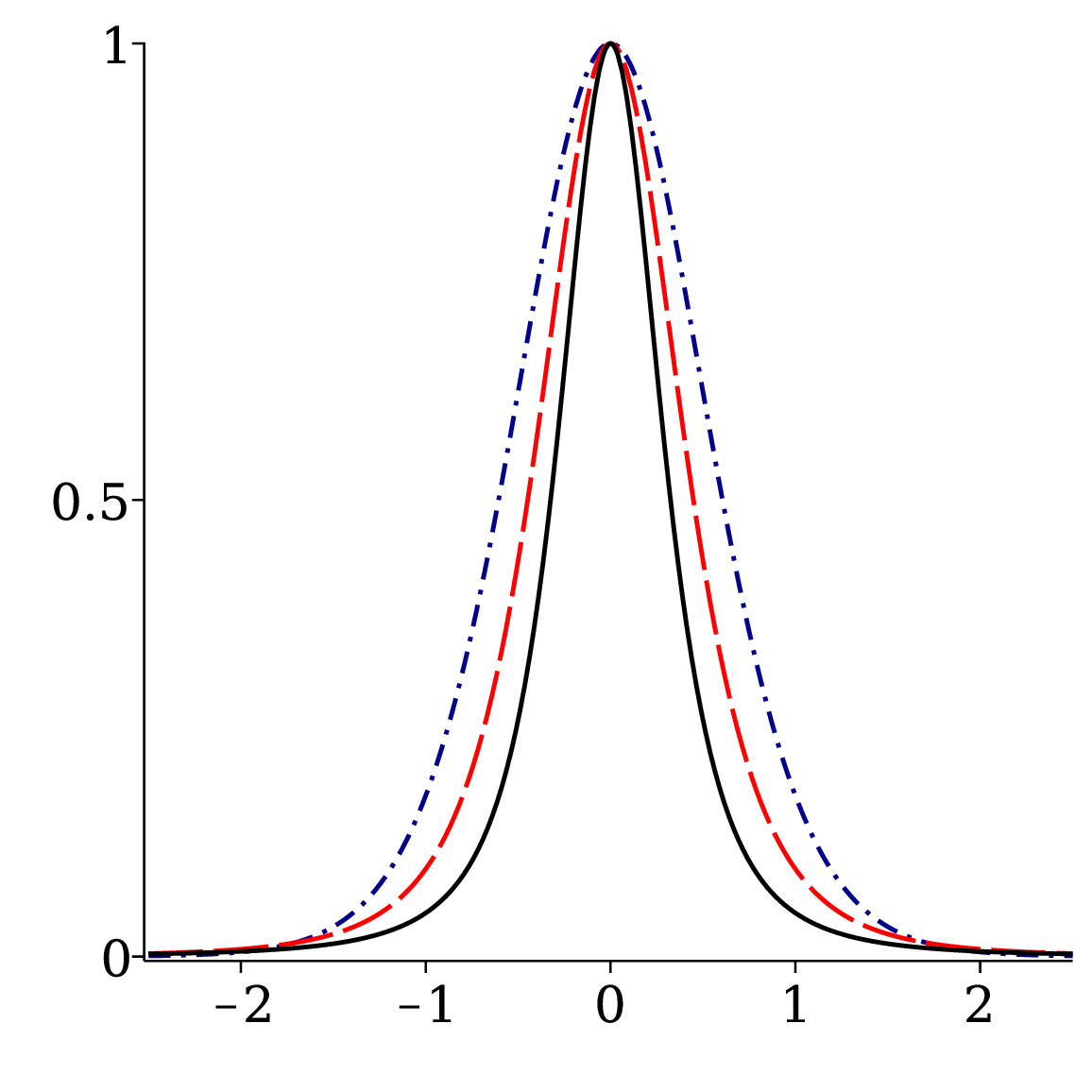}
\includegraphics[scale=0.24]{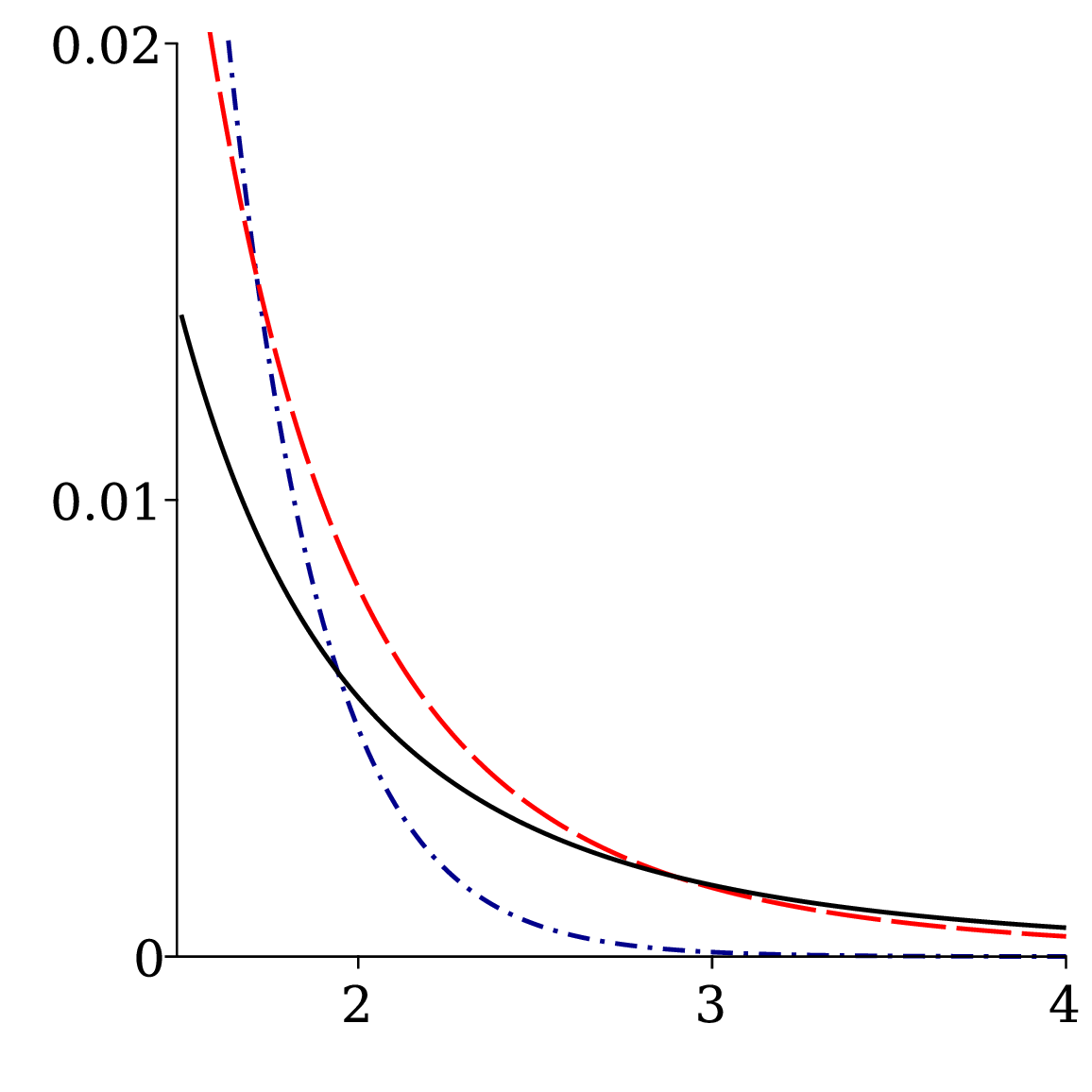}
\caption{The energy density $\rho(x)$ for the same values and line styles used in Fig.~\ref{fig2a}. The left panel shows its general behaviour, and the right panel depicts its behaviour far from the core.}
\label{fig2c}
\end{figure}

\begin{figure}%
\centering
\includegraphics[scale=0.36]{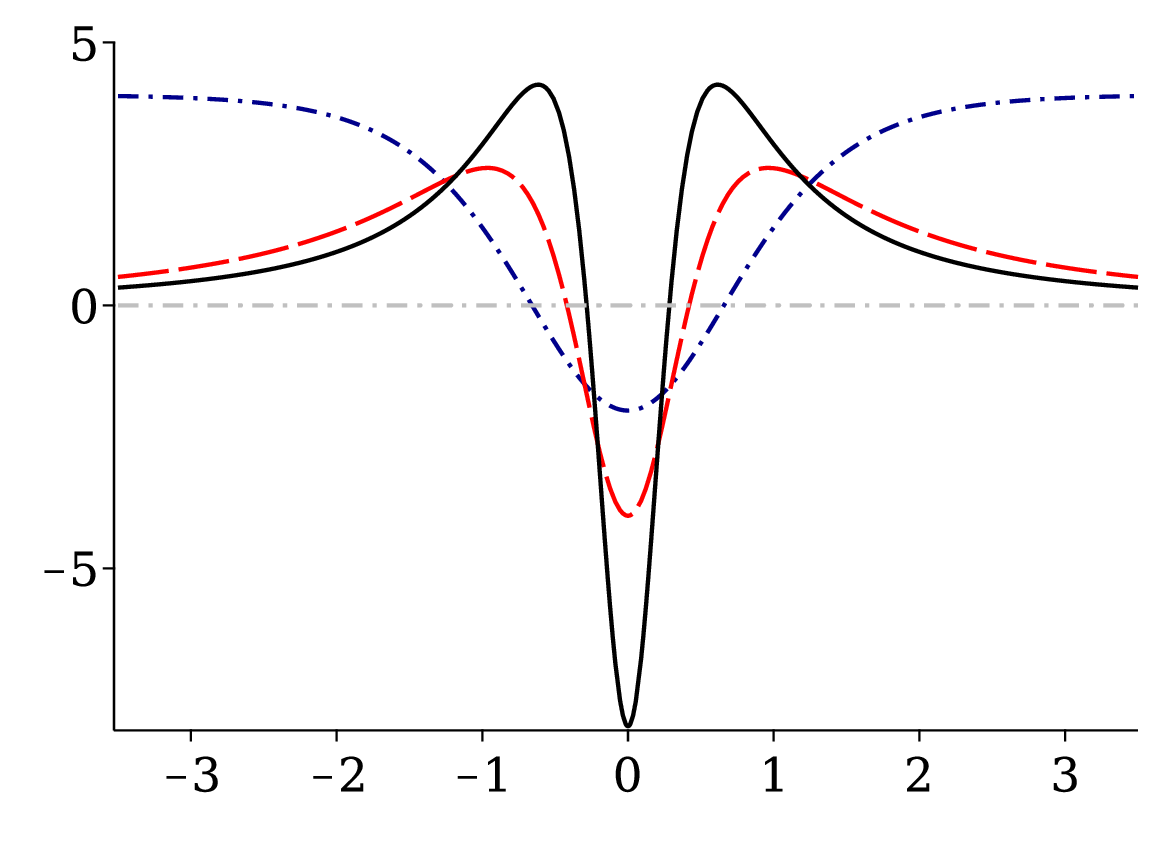}
\caption{The stability potential $U(x)$ of the model $V_B(\phi)$ for the same values of parameter and line styles used in Fig.~\ref{fig2a}.}
\label{fig2d}
\end{figure}

The corresponding function $W(\phi)$ is
\begin{equation}
W(\phi) = \phi \cdot {_2F_1} \left( \frac{1}{2}, -n; \frac{3}{2}; \phi^2 \right).
\label{Wtwin}
\end{equation}
For the specific values $n=1,2$; the function $W(\phi)$ reduces to polynomial forms:
\begin{align}
W_1(\phi) &= \phi - \frac{\phi^3}{3} \label{Wn1} \\
W_2(\phi) &= \phi - \frac{2}{3} \phi^3 + \frac{1}{5} \phi^5 \label{Wn2}
\end{align}

From this, we obtain the total energy for any $n$, which is
\be
E_{B}=2\,W(1)=2\cdot {_2F_1} \left( \frac{1}{2}, -n; \frac{3}{2}; 1 \right)
\ee
The hypergeometric function at unity simplifies to Gamma functions, giving
\begin{equation}
E_B = \frac{\sqrt{\pi} \, \Gamma(n + 1)}{\Gamma\left(n + \frac{3}{2} \right)}.
\end{equation}
Figure~\ref{fig2b} depicts the decreasing behaviour of the total energy as $n$ gets larger, showing that configurations with longer tails engender lower energies. For large $n$, we can use Stirling’s approximation to obtain $E_B \sim {\sqrt{\pi}} \,{{n^{-1/2}}}$. 

For the asymptotic behaviour of the energy density $\rho(x)$ and the stability potential $U(x)$, we have
\[
\rho(x) \propto
\begin{cases}
e^{-4|x|}, & n = 1 \quad \text{(short-range tail)},\\[6pt]
|x|^{-\frac{2n}{\,n-1\,}}, & n > 1 \quad \text{(long-range tail)}.
\end{cases}
\]
and
\[
\begin{aligned}
n = 1:& \quad U(x) \to 4 \quad \text{(exponential approach)}, \\[4pt]
n > 1:& \quad U(x) \sim \dfrac{n(2n - 1)}{(n - 1)^{2}} \, \dfrac{1}{x^{2}} 
\quad \text{(power-law)}.
\end{aligned}
\]

 Figure~\ref{fig2c} shows that, close to the origin, a larger $n$ results in a faster falloff of the energy density, but farther from the centre, the falloff becomes slower and the energy is distributed over a wider spatial region, in agreement with the asymptotic behaviour of the solutions. The stability potential, which for $n=1$ results in a modified Pöschl–Teller type, acquires a volcano shape for $n>1$ vanishing asymptotically, as displayed in Fig.~\ref{fig2d}. In this case, it only supports a single bound state, which is the zero mode $\eta_{0}\propto \phi'(x)$.

\subsection{Model C: Inspired by the Starobinsky inflaton potential}

\begin{figure}%
\centering
\includegraphics[scale=0.25]{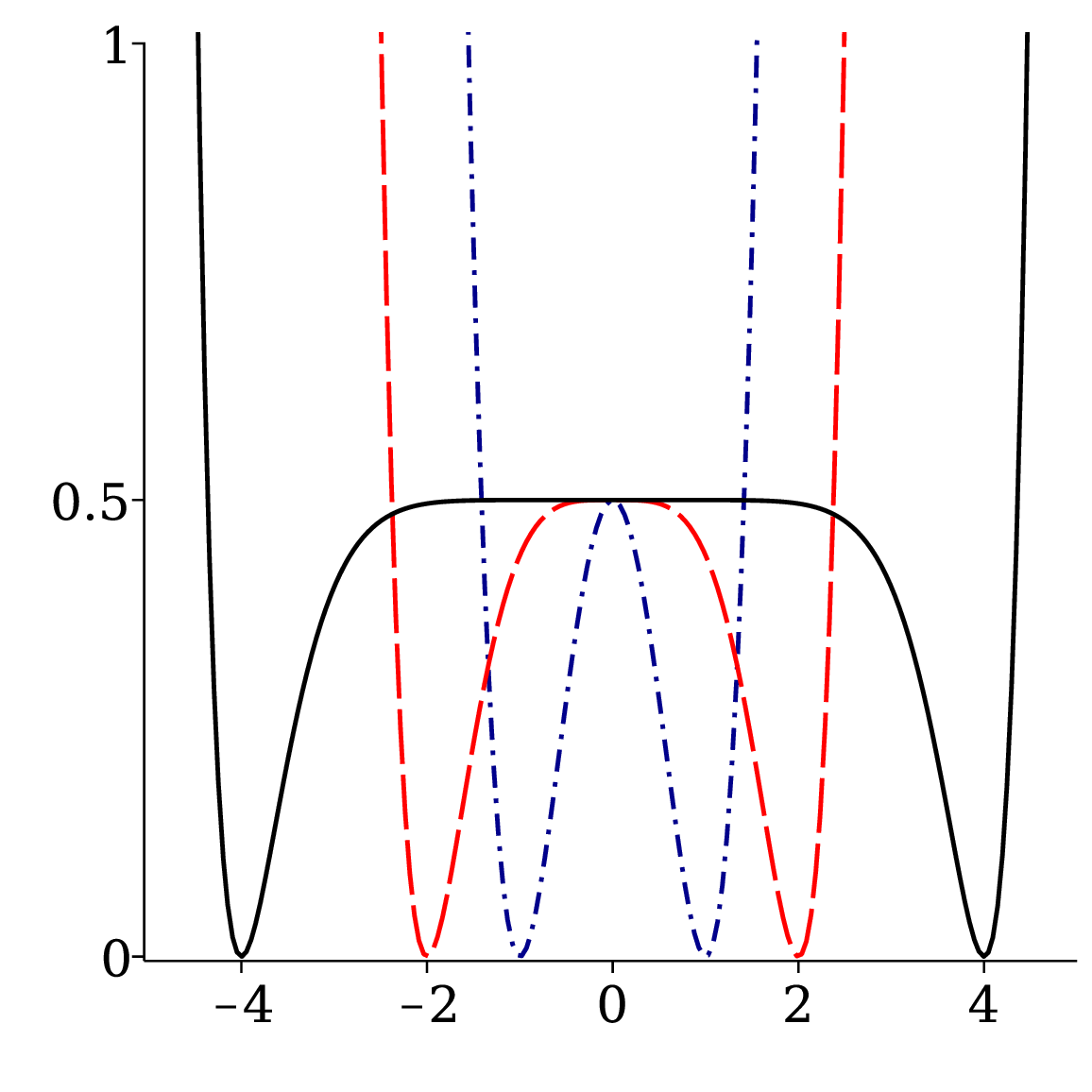}
\includegraphics[scale=0.24]{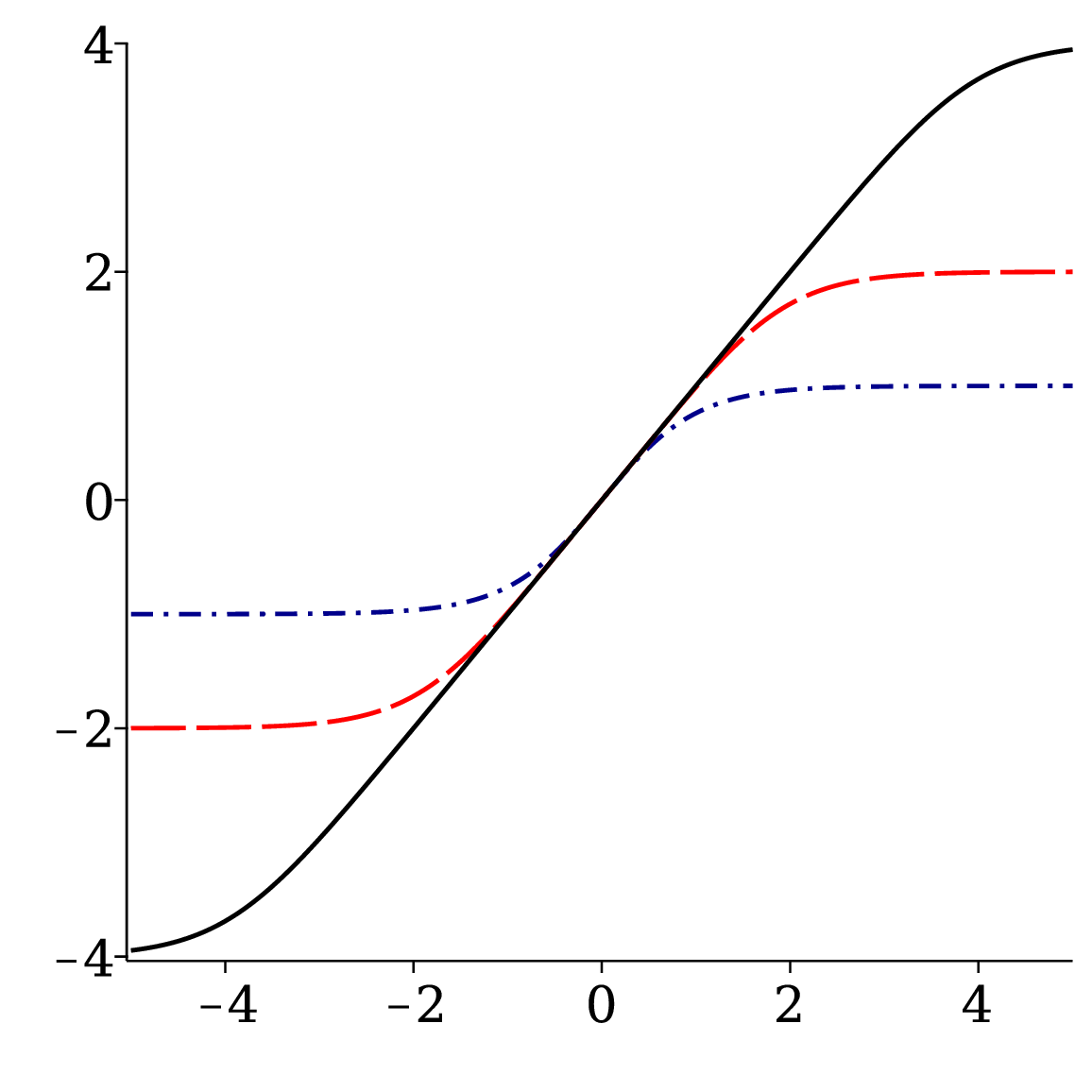}
\caption{The potential $V_C(\phi)$ (left) and the kink solutions $\phi(x)$ (right), for $n=1,2,4$, represented by dash-dotted (blue), dashed (red) and solid (black) lines, respectively.}
\label{fig3a}
\end{figure}

The next potential investigated in this work is inspired by a generalization of the Starobinsky inflationary model \cite{epjp-2021}, originally constructed to describe cosmological inflation in the early universe \cite{starobinsky}. We consider a form in terms of the parameter integer $n$, which supports kinklike solutions~\cite{lima2022,lima2024}. While the deformations used in \cite{lima2022,lima2024} were asymmetric, in this work we rewrite the potential in a symmetric manner to better suit our analysis of braneworld scenarios, to be studied in Sec. \ref{sec-5}, where the scalar field $\phi$ plays a central role in modeling the brane structure. In the case of a two-dimensional flat geometry, the potential is given by  
\be
V_{C}(\phi)=\frac12\Big(1-\Big(\frac{\phi}n\Big)^{2n}\Big)^2.
\label{Starobinskyg}
\ee 
Some structural similarity with Eq.~\eqref{Vn} might be seen,  particularly the dependence on the exponent $2n$. However, a distinct rescaling on the field argument $\phi/n$ is introduced, which shifts the position of the minima and consequently alters how the solutions approach their asymptotic limits \cite{lima2024}. Unlike the first model, in which the mass depends on $n$, here the mass remains constant at $m^2=4$. Additionally, the potential $V_{C}(\phi)$ has minima at $\phi_{min}=\pm n$, keeping the maximum at $\phi_{max}=0$. This leads to solutions where the field spreads more widely as $n$ increases, while the minima move farther apart, as illustrated in Fig.~\ref{fig3a}. The model~\eqref{Starobinskyg} gradually transforms the usual $\phi^4$ into a \textit{vacuumless potential}~\cite{ChoVacuumless1999,ChoV,dbazeia1999}, in which the two minima go to infinity as $n\to \infty$.  

\begin{figure}%
\centering
\includegraphics[scale=0.25]{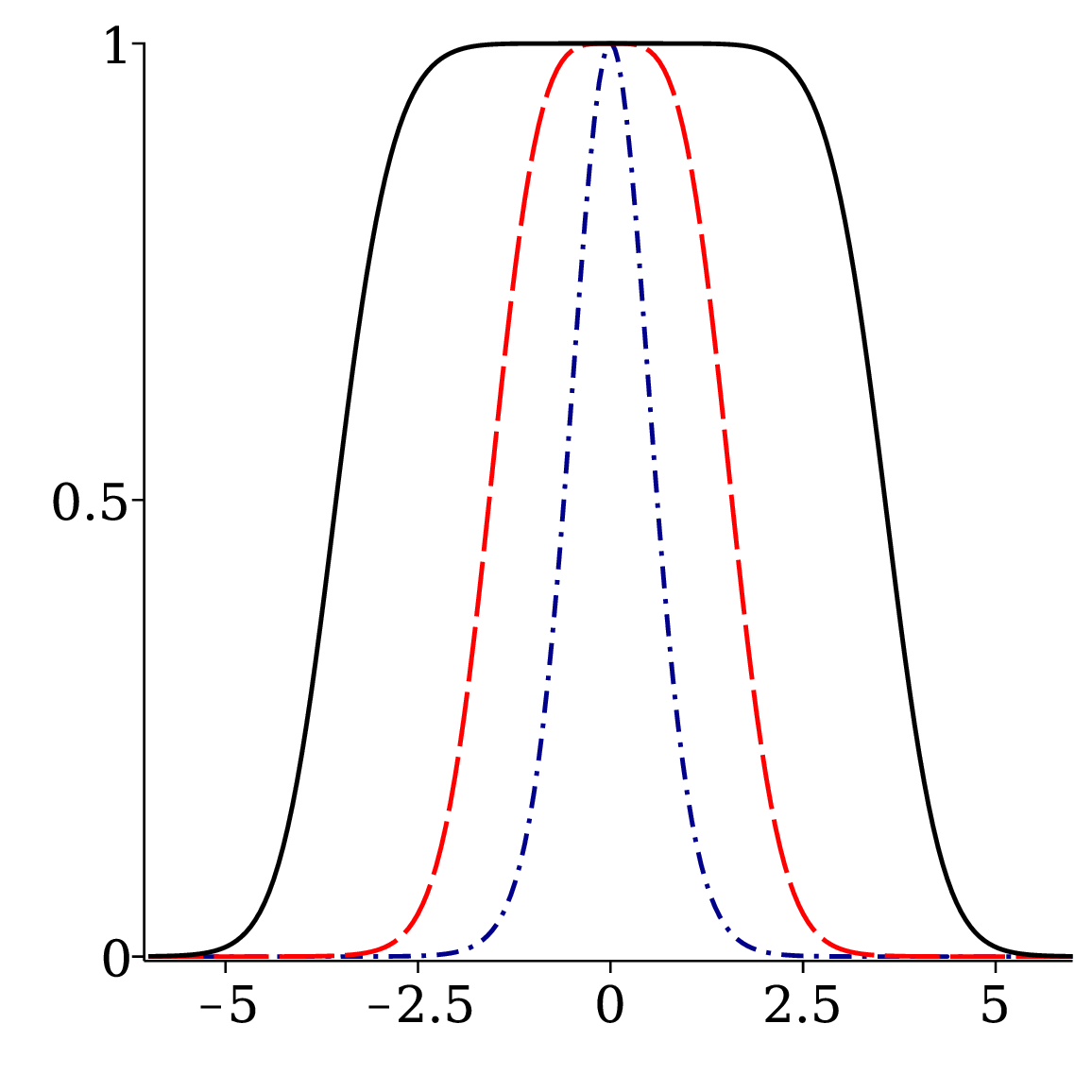}
\includegraphics[scale=0.24]{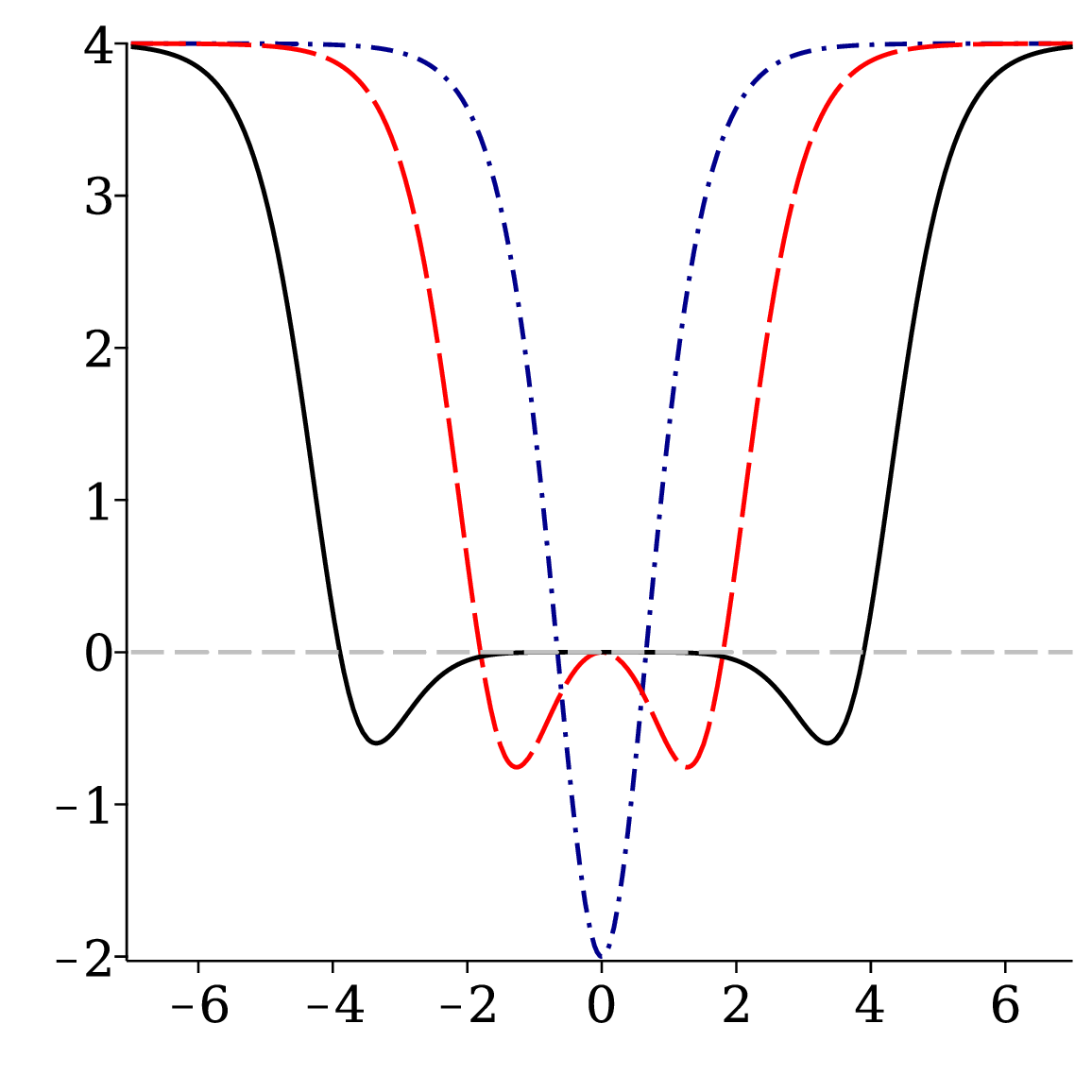}
\caption{The energy density $\rho(x)$ (left) and the stability potential $U(x)$ (right) for the model $V_C(\phi)$, with the same line styles and values of parameter used in Fig.~\ref{fig3a}.}
\label{fig3b}
\end{figure}

Furthermore, there is a correspondence between $V_{C}(\phi)$ and the potential introduced in Ref.~\cite{epjp-2021}, known as the $\beta$-Starobinsky, which emerges from the brane inflationary scenario also discussed in Ref.~\cite{JCAP2018}. 
In that approach, the inflation is driven by a field associated with an extra dimension in braneworld cosmology, naturally leading to a class of generalized potentials \cite{epjp-2021}. To better understand how $V_{C}(\phi)$  reproduces the form of the $\beta$-Starobinsky potential described in~\cite{epjp-2021}, we can apply the field parameterization:
\begin{equation}
\phi \to n - \frac{\phi}{\sqrt{6}M_{\rm Pl}}, \quad n \to \frac{1}{2\beta}, \quad V_{C}(\phi) \to \frac{V(\phi)}{2V_0},
\end{equation}
to find
\be
V(\phi)=V_0\left(1-\left(1-\sqrt{\frac23}\beta\frac{\phi}{M_{Pl}}\right)^{\frac1\beta}\right)^2
\label{Vbeta}
\ee
where $\beta$  is a real parameter, $M_{Pl}$ is the Planck mass, and $V_0$ is the potential amplitude. In the cosmological context, the Planck mass sets the fundamental scale for gravitational interactions, while  $\beta$ controls the shape of the potential, modulating the inflation of the early universe~\cite{epjp-2021}. The expression~\eqref{Vbeta} generalizes the Starobinsky inflationary potential~\cite{ketov2020, ketov2025}, reducing to the canonical form in the limit \( \beta \to 0 \)
\be
V(\phi) \to V_0 \left(1 - e^{- \sqrt{{2}/{3}}\,{\phi}/{M_{\rm Pl}}} \right)^2
\ee
Such modification provides a broader framework for understanding how the conventional model can be adapted to fit various cosmological possibilities, potentially enhancing its agreement with observational data \cite{epjp-2021}. 

In the present work, however, we focus on discussing topological solutions in scalar field theory, where the Lagrangian density has the conventional form given in Eq.~\eqref{lagran1}. Then, we employ the first-order formalism to express the function \( W(\phi) \) as
\be
W(\phi) = \phi - \frac{n}{2n+1} \left( \frac{\phi}{n} \right)^{2n+1}
\label{W2}
\ee
The energy associated with the topological solutions of the potential \( V_{C}(\phi) \) is then given by
\be
E_{C} = W(n) - W(-n) = \frac{4n^2}{2n+1}.
\label{E2}
\ee
There is a direct relation between the energies in Eqs.~\eqref{E1} and \eqref{E2}, namely $E_{C} = n E_{A}$. As a consequence, the energy $E_{C}$ diverges in the vacuumless limit ($n \to \infty$), in contrast to model \eqref{Vn}, which reaches a finite value.

The first-order equation
\be
\frac{d\phi}{dx} = 1 - \left( \frac{\phi}{n} \right)^{2n}
\ee
leads to kinklike solutions, which can be expressed implicitly as
\be
\label{solstarstandard}
\phi(x) \cdot {\Phi} \left( \left( \frac{\phi(x)}{n} \right)^{2n}, 1, \frac{1}{2n} \right) = 2n x,
\ee
or yet,
\be
\phi(x) \cdot {_2F_1} \left( 1, \frac{1}{2n}; 1 + \frac{1}{2n}; \left( \frac{\phi(x)}{n} \right)^{2n} \right) = x.
\label{solstarolike}
\ee
This expression is somewhat similar to the one emerging from model $V_A(\phi)$, which is obtained in \eqref{solVnstandard}. However, as $n$ increases, solution \eqref{solstarolike} connects two minima that are moving to $\pm n$, assuming a linear profile $\phi(x) = x$ in the regime $n\to\infty$. This is shown in Fig.~\ref{fig3a}. The hypergeometric solution in Eq.~\eqref{solstarolike} reveals how the rescaling of the field broadens the defect rather than compactifying it, where its argument is modulated by $\phi/n$. The classical mass remains constant ($m^2=4$) for all values of $n$, ensuring that the field tails maintain the exponential behaviour $e^{-2|x|}$.

Figure~\ref{fig3b} shows energy density and the stability potential. As $n$ varies, the width of the central region in $\rho(x)$ changes consistently with the kink profile, becoming unlimited in the vacuumless regime. For $n>1$, the falloff of the energy density becomes slower around the centre, in contrast with the polynomial one observed in Fig.~\ref{fig2c}. Regarding the stability potential, variations in $n$ affect both the depth and width of the central well through the core profile,  but asymptotically $U(x)$ always approaches the constant value $U(x)\to 4$. As $n$ approaches infinity, the vacua move to $\pm n\to\pm\infty$,  and the exponential tails reside very far from the centre, giving rise to a solution that looks vacuumless. In this case, one has $\phi(x)/n\ll1$ for any fixed $x$, then
\(
\rho(x)=\bigl(1-(\phi/n)^{2n}\bigr)^{2}\to 1
\qquad \text{and} \qquad
U(x)=V_{\phi\phi}\bigl(\phi(x)\bigr)\to 0,
\)
i.e., a broad \emph{core plateau} with $\rho\simeq 1$ and $U\simeq 0$ arises.

The distinct behaviors observed across the three kinds of models (A, B, and C) reveal how the placement of $n$ leads the system to three distinct regimes. In Model A, $n$ scales the classical mass as $m^2 = 4n^2$, forcing the field to reach its vacua increasingly fast until it compactifies. The Model B, when implemented under standard kinetics, presents a physical behaviour completely different from the nonstandard scenario found in Ref~\cite{twin2012}; here, any value of $n > 1$ eliminates the classical mass ($m^2 = 0$) while opening up higher-order non-vanishing derivatives, leading to solutions with polynomial tails. In Model C, the parameter rescales the field argument to drive the potential toward a vacuumless profile with runaway minima. This establishes a formulation where one can select whether a topological defect experiences a compact support, highly interactive polynomial tails or vacuumless behaviour.

\section{Deformation via the $\alpha$-Parameter} \label{sec-4}

The models investigated in the previous section can be written in a more general way through the inclusion of a real parameter $\alpha$,
\be
V_{\alpha}(\phi)=\frac{1}{2\alpha}\left(\sqrt{1+4\alpha \left(1+\frac12\alpha\right)V(\phi)}-1\right),
\label{Vmod}
\ee
where $V(\phi)$ corresponds to the original potential \cite{compactkinks,marques2014}. The minima of $V_{\alpha}(\phi)$ remain unchanged for any $\alpha$, while the deformation introduced modifies the potential profile between these points.

For $\alpha$ very small, we get
\be
V_{\alpha}(\phi) =V+\frac{\alpha}{2}V\left({1-2V}\right) +{\cal O}\left({\alpha^2}\right).
\label{smallalpha}
\ee

For $\alpha$ very large, we get
\be
V_{\alpha}(\phi) =\sqrt{\frac{V}{2}}+\frac{1}{2\alpha}\left({\sqrt{2V}-1}\right) +{\cal O}\left(\frac{1}{\alpha^2}\right).
\label{bigalpha}
\ee
The next step is to take this limit to analyse the modifications introduced in the previous scenarios, $V(\phi)$.

\subsection{Generalization of $V_A$: $V^A_{\alpha}(\phi)$}

For the case $V(\phi)=V_A(\phi)$ given in \eqref{Vn}, we have
\be
V^A_{\alpha}(\phi)=\frac{1}{2\alpha}\left(\sqrt{1+2\alpha \left(1+\frac12\alpha\right)\left(1- \phi^{2n}\right)^2}-1\right).
\label{v1mod}
\ee
The modified mass is $m^2_{\alpha}=4n^2\left(1+\alpha/2\right)$, which increases as either $n$ or $\alpha$ gets larger. The combined effect of these parameters alters the shape of $V^A_{\alpha}(\phi)$. For a fixed value of $\alpha$, increasing $n$ makes the potential flatter around the origin ($\phi=0$) and increasingly sharp at the minima ($\phi=\pm 1$). In the limit of $n$ very large, the influence of $\alpha$ becomes negligible. Indeed, for $|\phi|<1$ one has $\phi^{2n}\to 0$, and the potential approaches the constant value $V^A_{\alpha}(\phi)\to 1/2$. At the same time, the potential vanishes exactly at $\phi=\pm 1$, leading to a step-like form with a flat interior plateau and abrupt drops to the minima. Otherwise, for a fixed $n$, raising $\alpha$ amplifies the square-root deformation in a different way, also enhancing a non-analytic behaviour at the minima. 

\begin{figure}%
\centering
\includegraphics[scale=0.36]{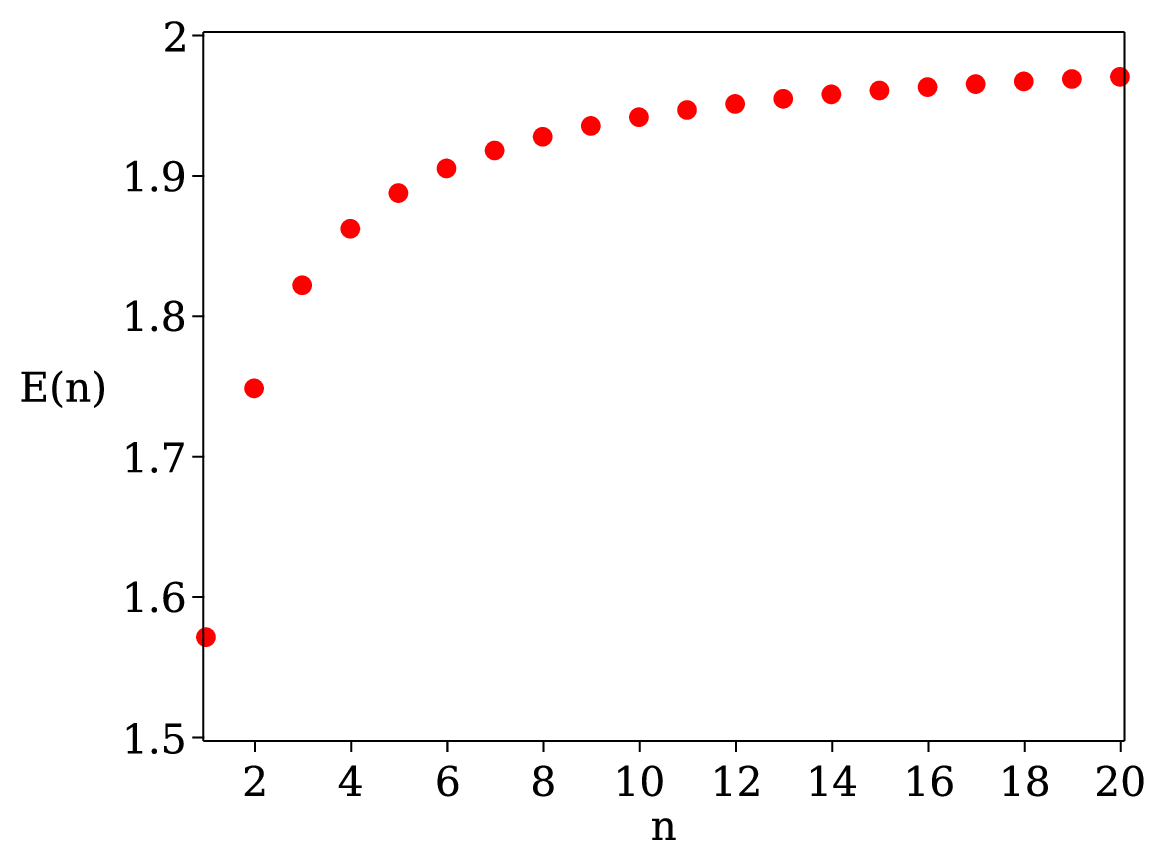}
\caption{Total energy $E_{\infty}^{A}$ in \eqref{EnMod} as a function of the parameter $n$,  for $n=1$ we have $E = {\pi}/{2}$, and the limit $n \to \infty$ gives $E=2$.}
\label{fig4a}
\end{figure}

To better understand this parameter dependence, let us examine the extreme regimes: $\alpha$ very small and $\alpha$ very large. In the limit $\alpha\rightarrow 0$, the modified potential recovers the standard one in Eq.~\eqref{Vn}. 
In contrast, taking the limit $\alpha\rightarrow \infty$, the expression for  $V^A_{\alpha}(\phi)$ assumes a modulus shape
\be
V_{\infty}^{A}(\phi)=\lim_{\alpha\to \infty} V_{\alpha}^{A}(\phi)=\frac12|1-\phi^{2n}|
\label{moduloVn}
\ee
This way, the $\alpha$–deformation provides a continuous interpolation between the standard polynomial potential $V_A(\phi)$ and its modulus counterpart. Using $W_\phi=\sqrt{|1-\phi^{2n}|}$, it is possible to find an expression for the function $W$
\be
W(\phi) =\phi \cdot { _2F_1}\left(-\frac12,\frac1{2n};1+\frac1{2n};\phi^{2n}\right).
\label{Wmod1}
\ee
This function can also be written in terms of the incomplete Beta function $W(\phi)=\frac{1}{2n}B_{\phi^{2n}}(\frac1{2n},\frac32)$. The total energy for this situation becomes
\ben
E_{\infty}^{A}=2\,W(1)=\frac{\sqrt{\pi}\,\Gamma\left(1+\frac{1}{2n}\right)}{\Gamma\left(\frac32+\frac{1}{2n}\right)},
\label{EnMod}
\een
as shown in Fig.~\ref{fig4a} ($E = {\pi}/{2}$ for $n=1$, and $E=2$ in the limit $n \to \infty$). The first-order equation becomes
\be
\frac{d\phi}{dx}=\sqrt{|1-\phi^{2n}|}.
\label{p1m}
\ee
The solution provides a set of compact kinks, given by
\be
\label{solVnX}
\phi(x) \cdot {_2F_1}\left(\frac12,\frac1{2n};1+\frac1{2n};\phi(x)^{2n}\right)= x, 
\ee
inside the interval $x\in[-L_A, L_{A}]$, where 
\be
L_{A}= \frac{\sqrt{\pi}\,\Gamma\left(1+\frac{1}{2n}\right)}{\Gamma\left(\frac12+\frac{1}{2n}\right)}.
\label{barx1}
\ee
The defect \eqref{solVnX} exists in a compact region between $-L_A$ and $L_A$ for each value of $n$,  and the minima $v_{\pm}=\pm1$ are reached at these finite points of $x$. For $n=1$ we get $L_A=\pi/2$, and  when $n\rightarrow \infty$ we have $L_A=1$. The energy density, which is given by $\rho=\phi'^2$, is exactly null outside the compact region. Besides that, there is a relation between the total energy $E_{\infty}^{A}$ and $L_{A}$, that is $E_{\infty}^{A}=2L_{A}/(1+\frac1n)$.

In this perspective, the solution in Eq.~\eqref{solVnX} represents a family of compactons obtained from a standard theory, whose spatial boundaries are controlled by $n$. Specifically, the kink solutions supported by $V_A$ are smoothly transformed into compactons associated with the modulus potential $V^A_{\infty}$, which compels the field to saturate to its vacuum values ($v_{\pm} = \pm 1$) at the finite coordinates $\pm L_A$. This extends the mechanism discussed in \cite{marques2014}, where the transition was proposed only for the $\phi^4$ case ($n=1$). Furthermore, in contrast to Ref.~\cite{lima2024}, where compact configurations emerge through nonlinear kinetic terms, here the compactification is driven entirely by the deformation of the potential itself.

\begin{figure}%
\centering
\includegraphics[scale=0.25]{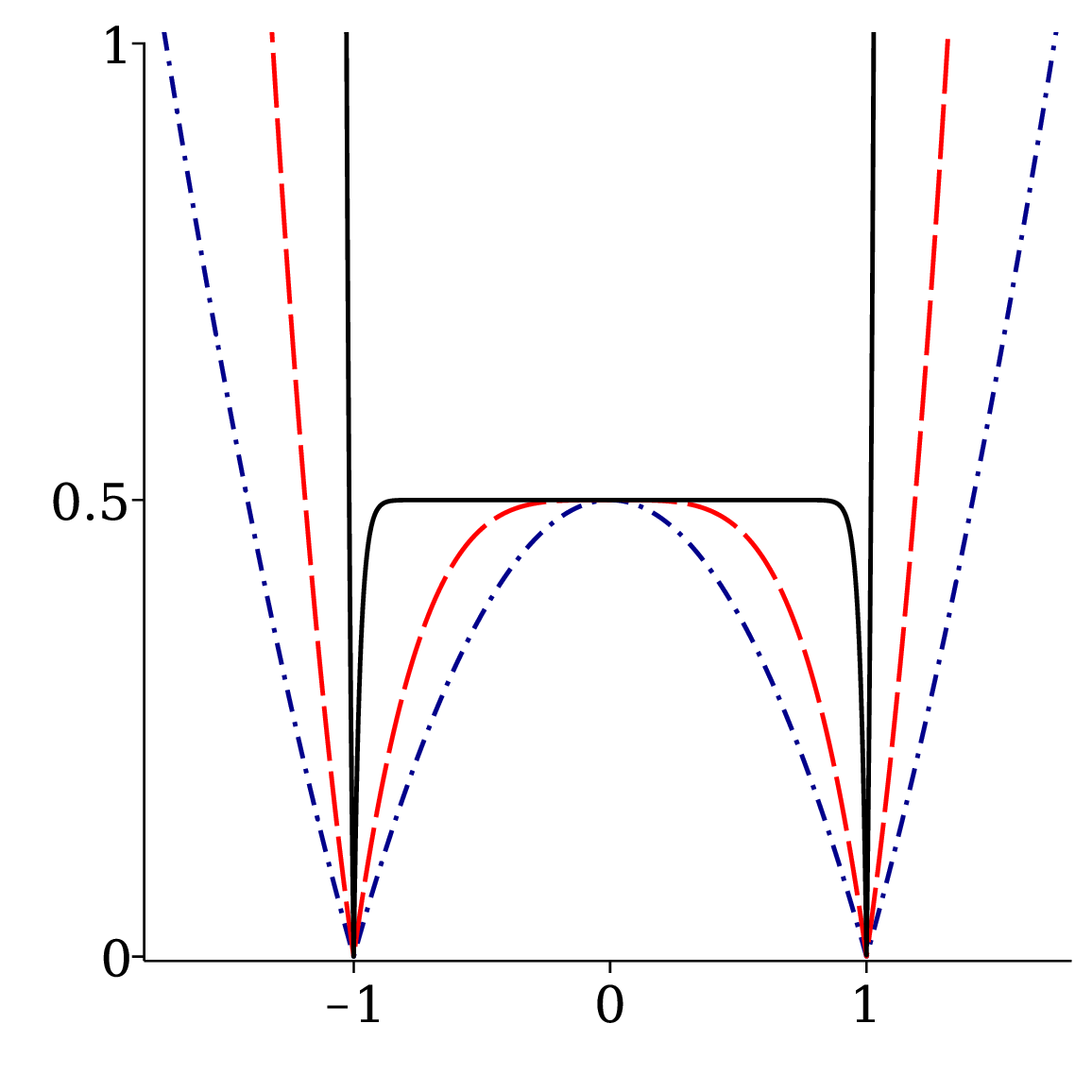}
\includegraphics[scale=0.24]{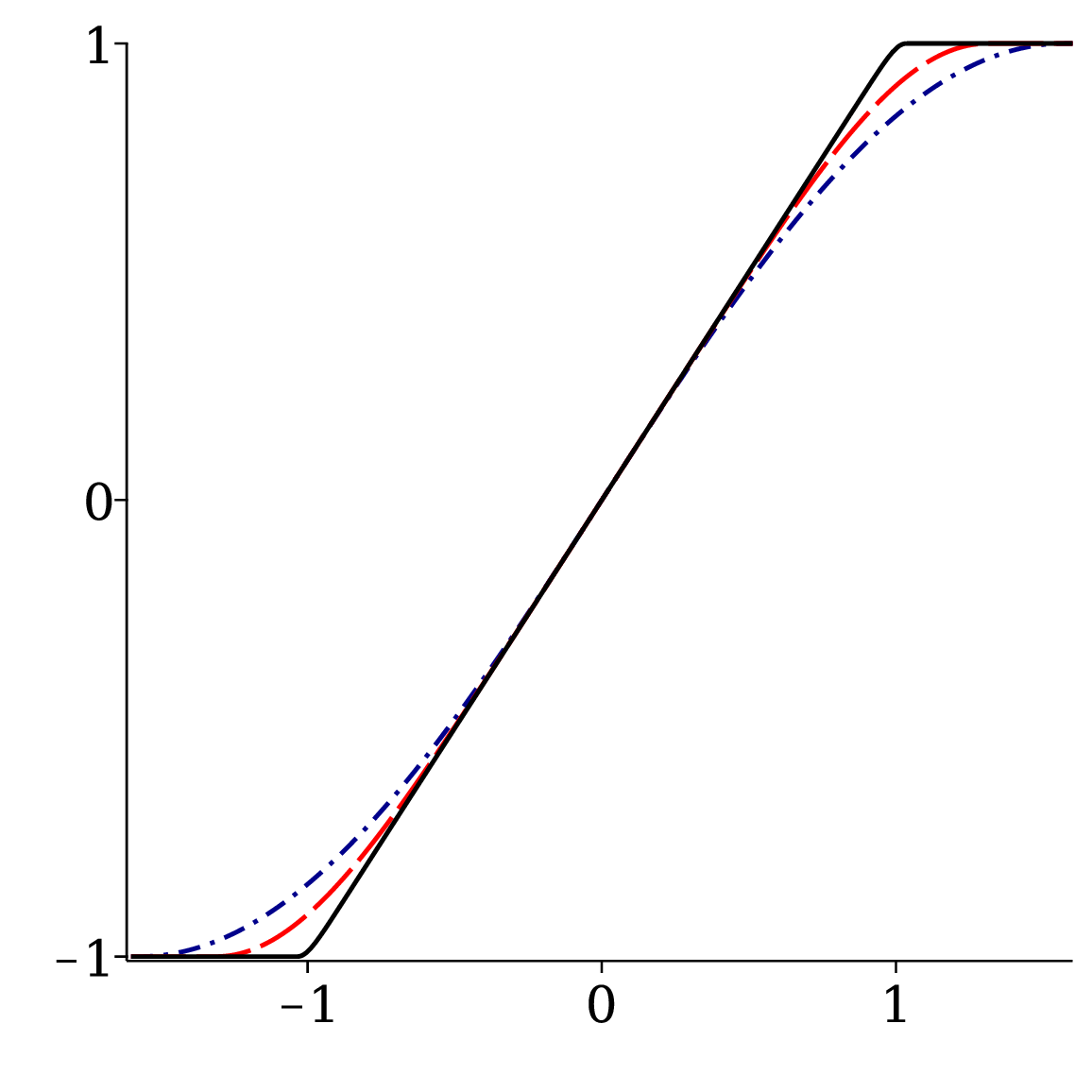}
\caption{The potential $V_{\infty}^{A}(\phi)$ (left) and the compacton solutions $\phi(x)$ (right), for $n=1,2,20$, represented by dash-dotted (blue), dashed (red), and solid (black) lines, respectively.}
\label{fig4b}
\end{figure}
\begin{figure}%
\centering
\includegraphics[scale=0.25]{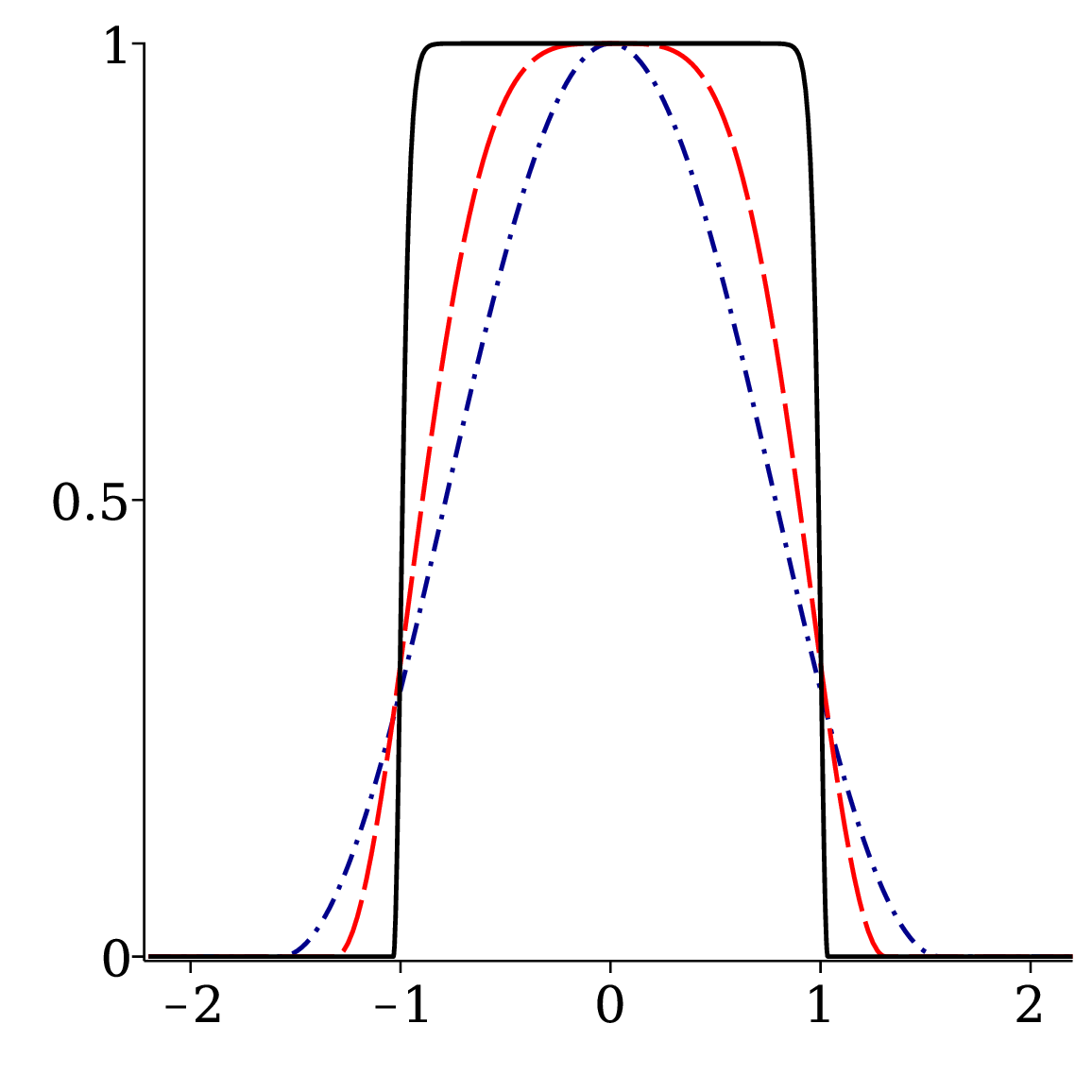}
\includegraphics[scale=0.24]{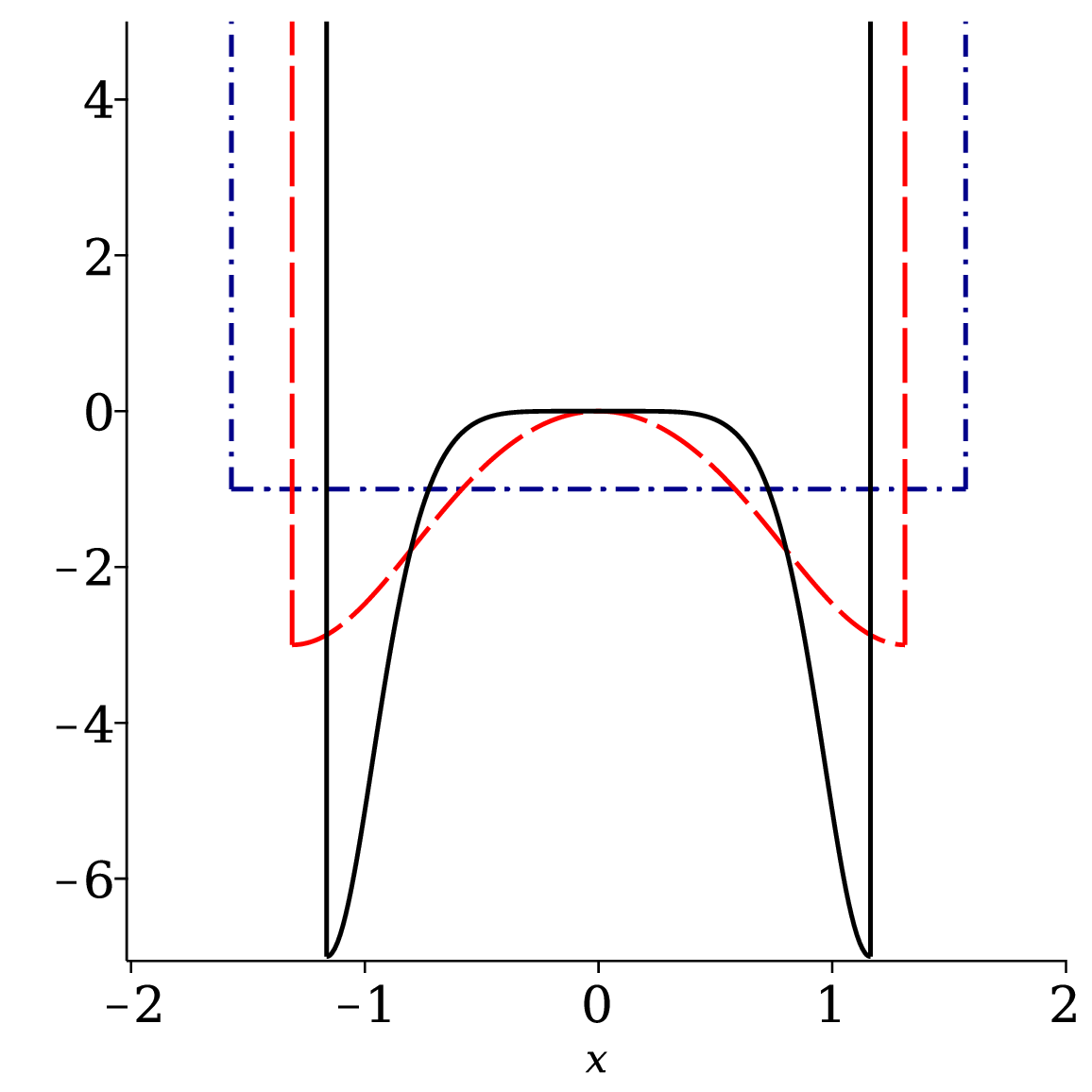}
\caption{The energy density $\rho(x)$ (left) and the stability potential $U(x)/n$ (right) for model $V_{\infty}^{A}(\phi)$. In the left panel, the same line styles and parameter values as used in Fig.~\ref{fig4b}. In the right panel, the cases $n=1,2,4$ are shown using the same line styles.}
\label{fig4c}
\end{figure}

For example, in the case $n=1$, we can write
\be
W(\phi)= \frac{\phi}{2}  \sqrt{\left|1 - \phi^2\right|}+ \frac12{\arcsin(\phi)}.
\ee
That gives the total energy $E=\pi/2$. Through
\be
{_2F_1}\left(\frac12,\frac1{2};\frac3{2};\phi^{2}\right)=\frac{\arcsin(\phi)}{\phi} 
\ee
for $x\in[-\pi/2, \pi/2]$, the solution $\phi(x)$ takes the form
\begin{eqnarray}\label{sinx}
\phi(x) = \left\{
\begin{array}{ll}
-1, & \mbox{for} \quad \, x < -\pi/2,\\ 
 \sin(x), & \mbox{for} \quad |x| \leq \pi/2,\\
1, \qquad & \mbox{for}\quad \, x > \pi/2\,.
\end{array} \right.
\end{eqnarray}

The corresponding energy density becomes
\begin{eqnarray}\label{cosx}
\rho(x) = \left\{
\begin{array}{ll}
0, & \mbox{for} \quad \, x < -\pi/2,\\ 
 \cos^2(x), & \mbox{for} \quad |x| \leq \pi/2,\\
0, \qquad & \mbox{for}\quad \, x > \pi/2\,.
\end{array} \right.
\end{eqnarray}

On the other hand, when $n$ is very large, the potential $V_{\infty}^{A}(\phi)=\tfrac12|1-\phi^{2n}|$ tends to behave in the same way of $V_A(\phi)=\tfrac12(1-\phi^{2n})^2$. In both situations, the solution is the compacton $\phi(x)=x$ for $x\in[-1,1]$, $\phi(x)=-1$ for $x<-1$, and $\phi(x)=1$ for $x>1$, with total energy $E=2$. 

Figures~\ref{fig4b} and \ref{fig4c} illustrate main features of the model $V_{\infty}^{A}(\phi)$. In Fig.~\ref{fig4b}, the potential and the corresponding compacton profiles are shown, where the field reaches the minima at finite points $\pm L_A$. Figure~\ref{fig4c} displays the associated energy densities and stability potentials. The energy density is strictly localized inside the compact region, vanishing outside it. The stability potential exhibits infinite walls at the edges, confining the fluctuation modes, where $U(x)=+\infty$ at $|x|=L_A$. 
For $n=1$, it reduces to the constant value $U(x)=-1$ inside the open interval $x\in(-\pi/2,\pi/2)$, while at the boundaries $|x|=\pi/2$ it diverges to $+\infty$. For larger $n$, it becomes progressively deeper as illustrated for $n=2$ and $n=4$.

\subsection{Generalization of $V_B$: $V^B_{\alpha}(\phi)$}

The second generalization is obtained by taking $V(\phi)=V_B(\phi)$ given in \eqref{V2}, we have
\be
V^B_{\alpha}(\phi)=\frac{1}{2\alpha}\left(\sqrt{1+2\alpha\left(1+\frac12\alpha\right)\left(1- \phi^{2}\right)^{2n}}-1\right).
\label{vBmod}
\ee
In this case, the mass term vanishes for all $n>1$, while  it is
$
m^2_{\alpha} = 4+2\alpha
$ for $n=1$.
The first nonvanishing derivative of $V^B_{\alpha}(\phi)$ at the minima occurs at order $2n$, similar to its appearance in $V_B(\phi)$. It can be expressed as
\[
\left.\frac{d^{2n} V^B_{\alpha}}{d\phi^{2n}}\right|_{\phi=\pm 1}
= \left(1+\frac{\alpha}{2}\right)\cdot a_{2n}
\]
where $a_{2n}$ is given by
\be
a_{2n}= \frac{1}{2}\sum_{j=n}^{2n} \binom{2n}{j} (-1)^j \frac{(2j)!}{(2j-2n)!}
\ee
which is the same coefficient gotten from \eqref{nonvan}. Thus, only the case $n=1$ supports massive fluctuations, whereas for $n>1$ the vacua are massless.

Taking the limit $\alpha\rightarrow \infty$, the expression for  $V^B_{\alpha}(\phi)$ becomes
\be
V_{\infty}^{B}(\phi)=\frac12|1-\phi^{2}|^n
\label{modulolong}
\ee
The corresponding first-order equation is
\be
\frac{d\phi}{dx}=\left|1-\phi^2\right|^{n/2}.
\label{VBp1}
\ee
The solution is given by
\be
\label{solVbinf}
\phi(x) \cdot {_2F_1}\left(\frac12,\frac{n}{2}; \frac{3}{2} ;\phi(x)^2 \right)= x
\ee
To determine whether the solution exhibits compact behaviour, we examine whether it reaches the vacuum at a finite coordinate. This occurs if the quantity
\[
L_B\;={_2F_1}\left(\frac12,\frac{n}{2}; \frac{3}{2}; 1 \right)
=\frac12\,B\!\left(\frac12,\,1-\frac{n}{2}\right)
\]
converges. Using the convergence criterion for the Beta integral, this holds only for $n<2$. 
Since here $n\in\mathbb{Z}^{+}$, we conclude that the solution is compacton only for $n=1$.
Indeed, for $n=1$ one finds $\phi(x)=\sin x$ defined in $x\in[-\pi/2,\pi/2]$, with $L_B=\pi/2$.
 For $n=2$, near $\phi\to 1$ we have 
\[
1-\phi(x)\sim e^{-2x}\quad(x\to\infty),
\]
yielding an exponential (short–range) tail and no compact support. For $n>2$, the asymptotic behaviour changes to a power-law,
\[
1-\phi(x)\sim \frac{1}{2\big((n-2)\,x\big)^{2/(n-2)}} \qquad (x \to \infty)
\]
so the kink develops a long–range tail. We note that, for even values of $n$, the modulus-type potential $V_{\infty}^{B}(\phi)$ coincides with $V_{B}(\phi)=\tfrac12\,(1-\phi^{2})^{2m}$ (where $n=2m$ with $m=1,2,3,...$). Accordingly, the right tail of the energy density $\rho(x)$ goes as $\rho\sim e^{-4x}$ for $n=2$, and as $\rho\sim x^{-\frac{2n}{\,n-2\,}}$ for $n>2$.

In typical scalar field configurations, taking the modulus limit ($\alpha \to \infty$) is expected to naturally convert standard kinks into solutions with compact support. However, this is not the case for the long-range model $V_B$. While the special case $n = 2$ gives a transition where the modulus potential yields the standard short-range $\phi^4$ kink; for all higher values ($n > 2$) the structure near the minima remains flat, leading to power-law tails. Our analysis proves that for long-range systems, the $\alpha$-deformation is incapable of overcoming the extreme flatness of the massless vacua. As a consequence, the kinks preserve a highly interactive character and do not develop compact support.

\begin{figure}%
\centering
\includegraphics[scale=0.25]{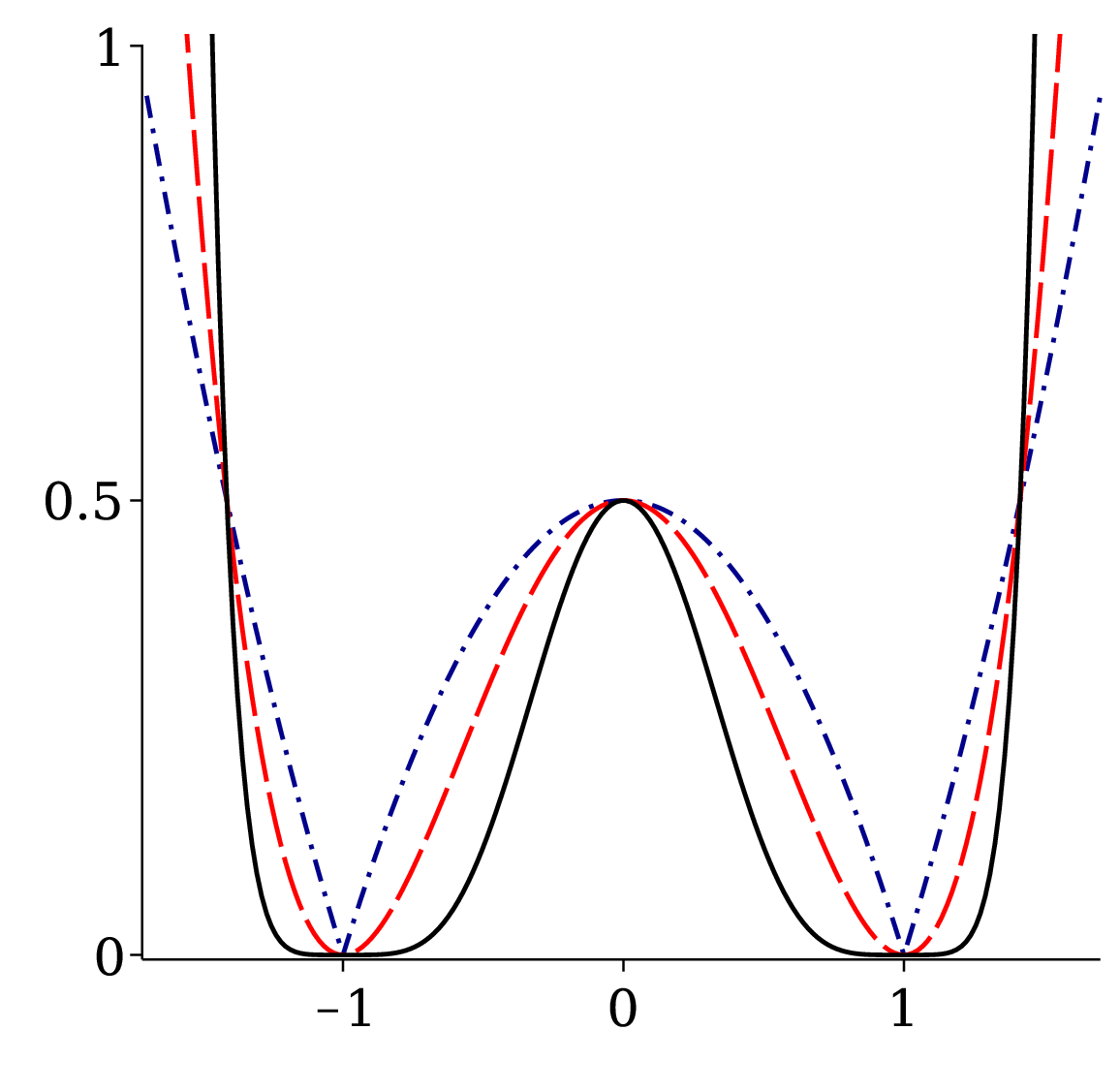}
\includegraphics[scale=0.24]{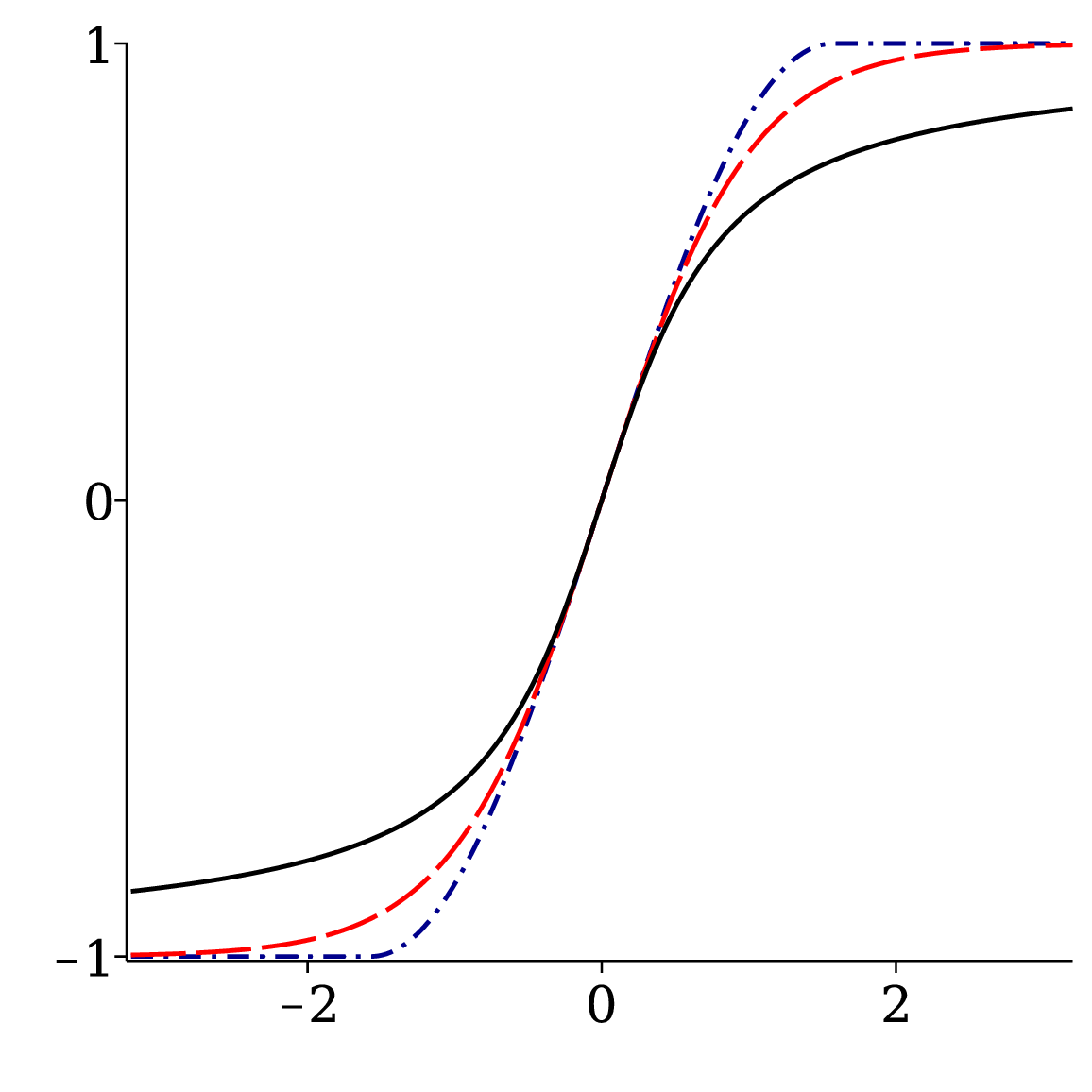}
\caption{The potential $V_{\infty}^{B}(\phi)$ (left) and the solutions $\phi(x)$ (right), for $n=1,2,5$, represented by dash-dotted (blue), dashed (red) and solid (black) lines, respectively.}
\label{fig5a}
\end{figure}
\begin{figure}%
\centering
\includegraphics[scale=0.25]{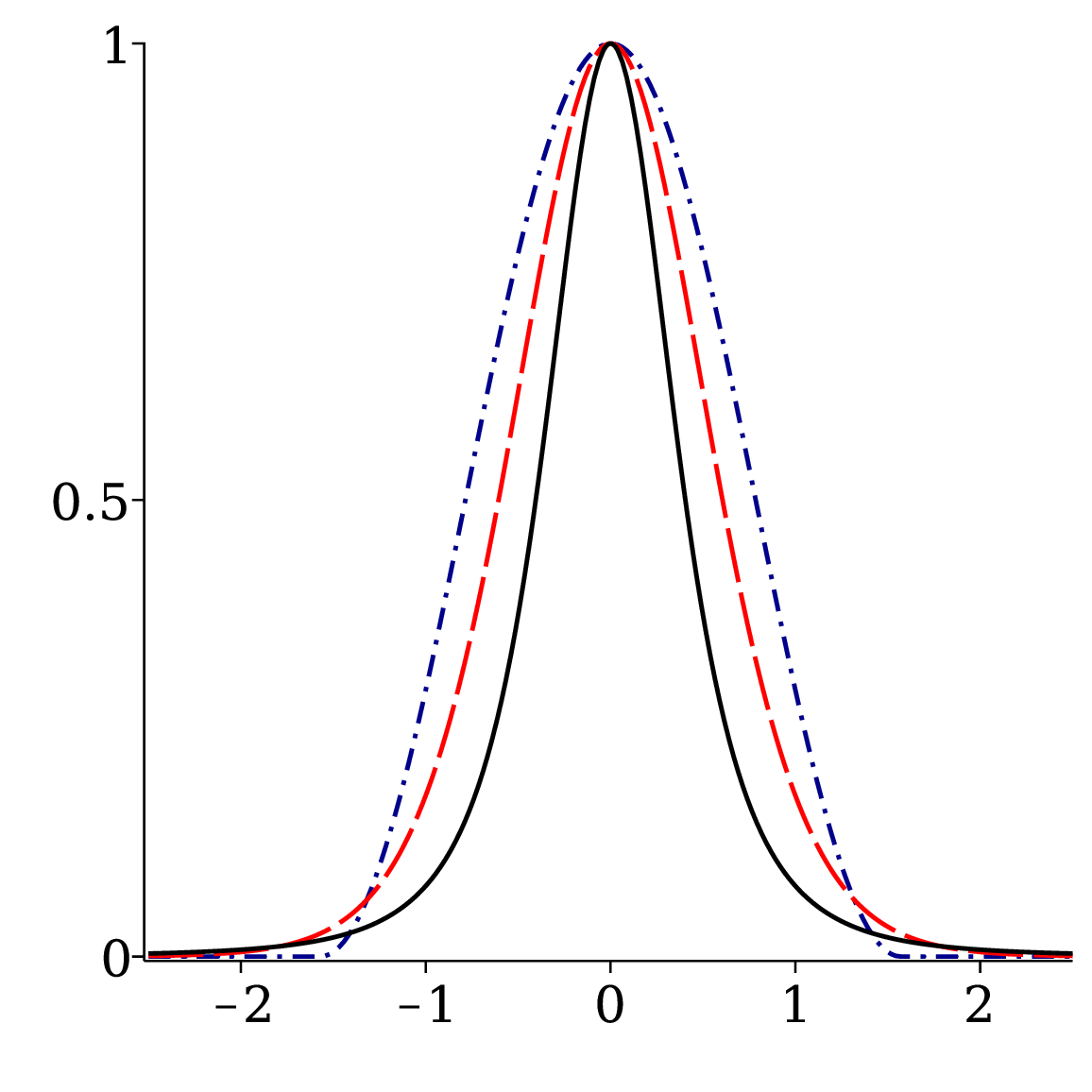}
\includegraphics[scale=0.24]{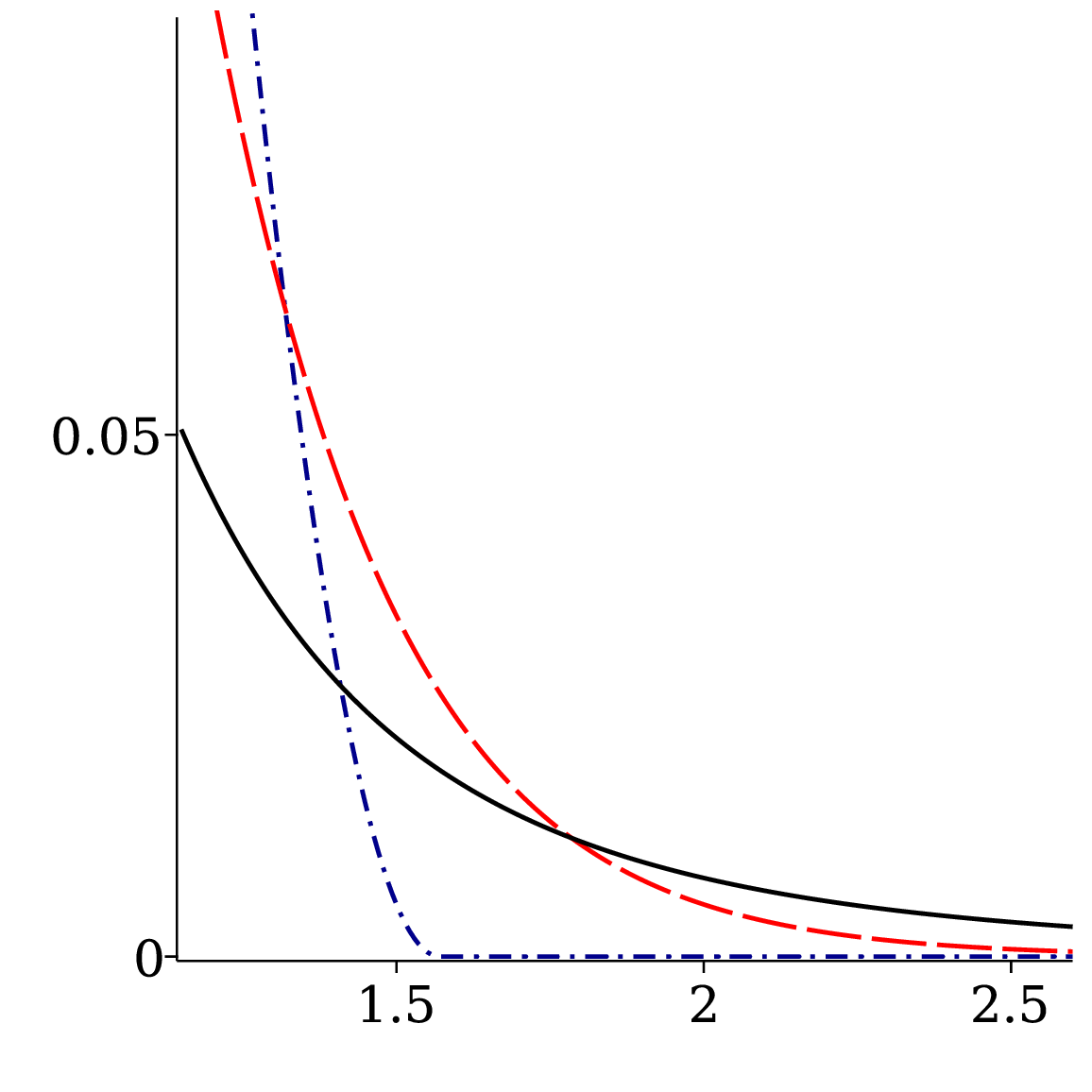}
\caption{The energy density $\rho(x)$ for the same values and line styles used in Fig.~\ref{fig5a}. The left panel shows its general behaviour, and the right panel depicts its behaviour far from the core.}
\label{fig5b}
\end{figure}

\begin{figure}%
\centering
\includegraphics[scale=0.36]{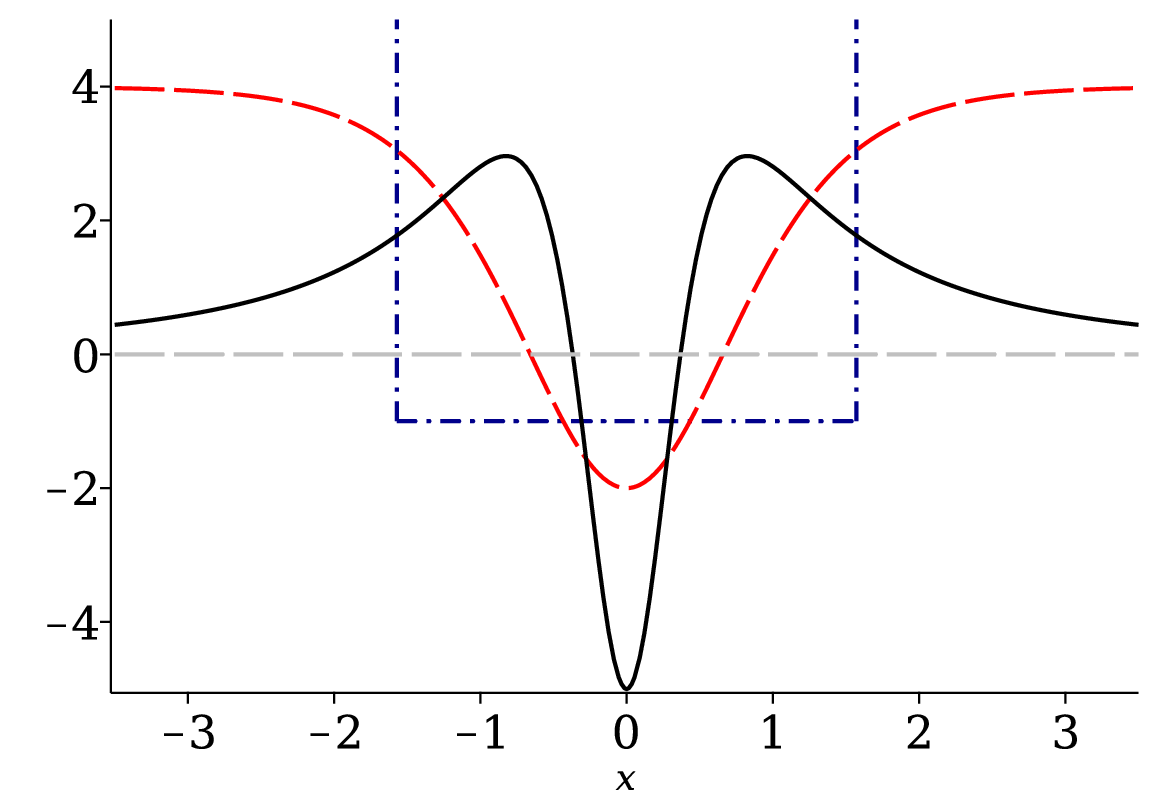}
\caption{The stability potential $U(x)$ for the same values of parameter and line styles used in Fig.~\ref{fig5a}.}
\label{fig5c}
\end{figure}

Using $W_\phi=|1-\phi^2|^{n/2}$, one finds
\be
W(\phi)= \phi \cdot {_2F_1}\!\left(\frac12,-\frac{n}{2};\frac32;\phi^2\right)
\label{WgeneB}
\ee
The total energy of the localized configuration is 
\be
E_{\infty}^{B}=\frac{\sqrt{\pi}\,\Gamma\!\left(\frac{n}{2}+1\right)}{\Gamma\!\left(\frac{n}{2}+\frac32\right)},
\ee
which has a decreasing behaviour as $n$ gets larger, starting from $E=\pi/2$ when $n=1$.

Figures~\ref{fig5a}--\ref{fig5c} illustrate the main features of the model 
$V_{\infty}^{B}(\phi)=\tfrac12|1-\phi^{2}|^{n}$. 
Figures~\ref{fig5a} and \ref{fig5b} show the potential, the corresponding solutions, and the associated energy densities for representative values of $n$. 
For $n=1$, the solution is a compacton of sine type, confined to a finite interval. 
For $n=2$, the potential reduces to the standard $\phi^{4}$ form, yielding a kink with an exponential tail. 
For higher $n$, the solution acquires a long–range tail.
The energy densities reflect this behaviour, peaking near the defect core and engendering tails that go exponentially for $n=2$ and polynomially for $n>2$. Figure~\ref{fig5c} shows the corresponding stability potential $U(x)$, which forms an infinite well for $n=1$, reducing to a modified Pöschl–Teller for $n=2$, and going as $1/x^{2}$ for larger $n$, indicating a transition from strictly localized to long–range fluctuation spectra.

\subsection{Generalization of $V_C$: $V^{C}_{\alpha}(\phi)$}

\begin{figure}%
\centering
\includegraphics[scale=0.25]{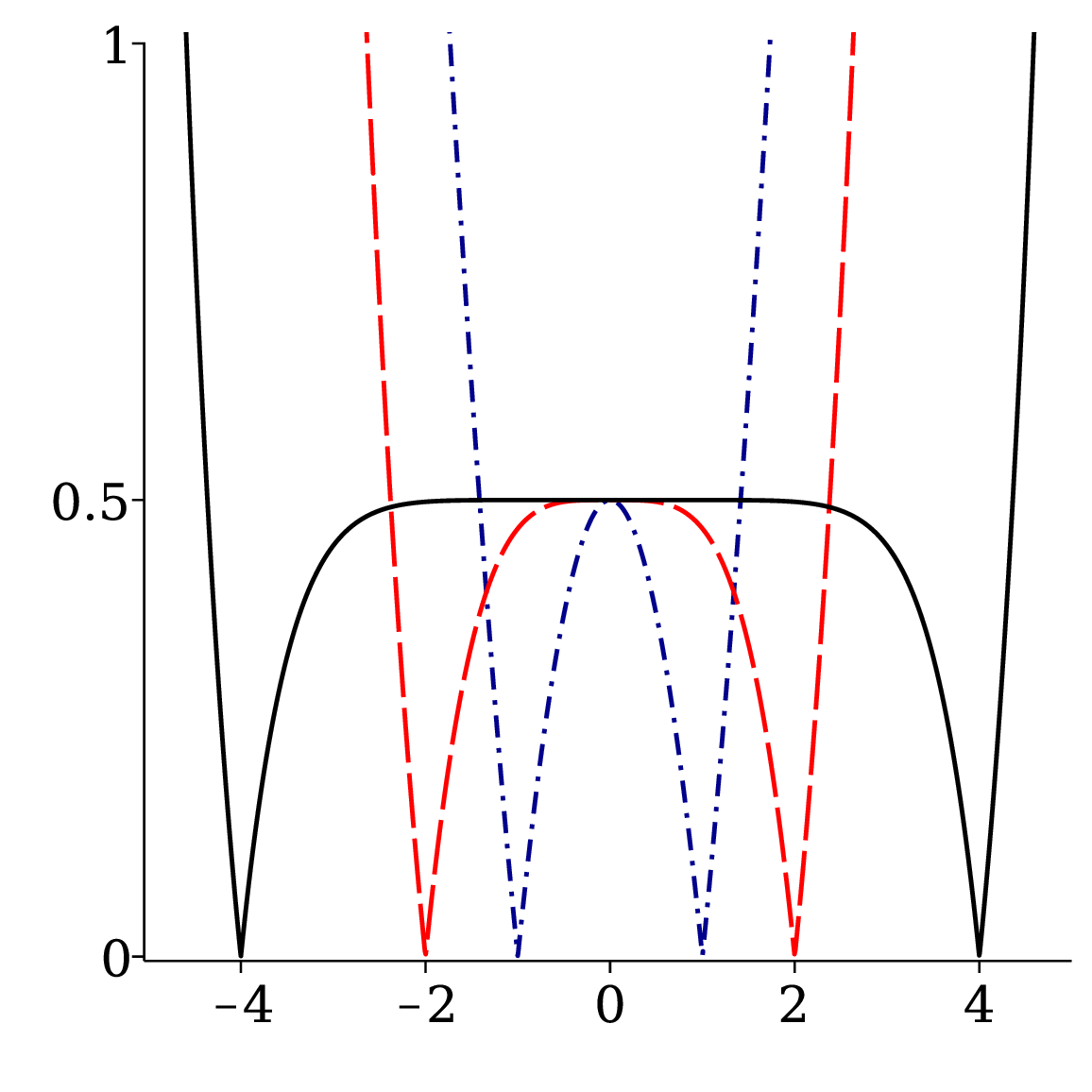}
\includegraphics[scale=0.24]{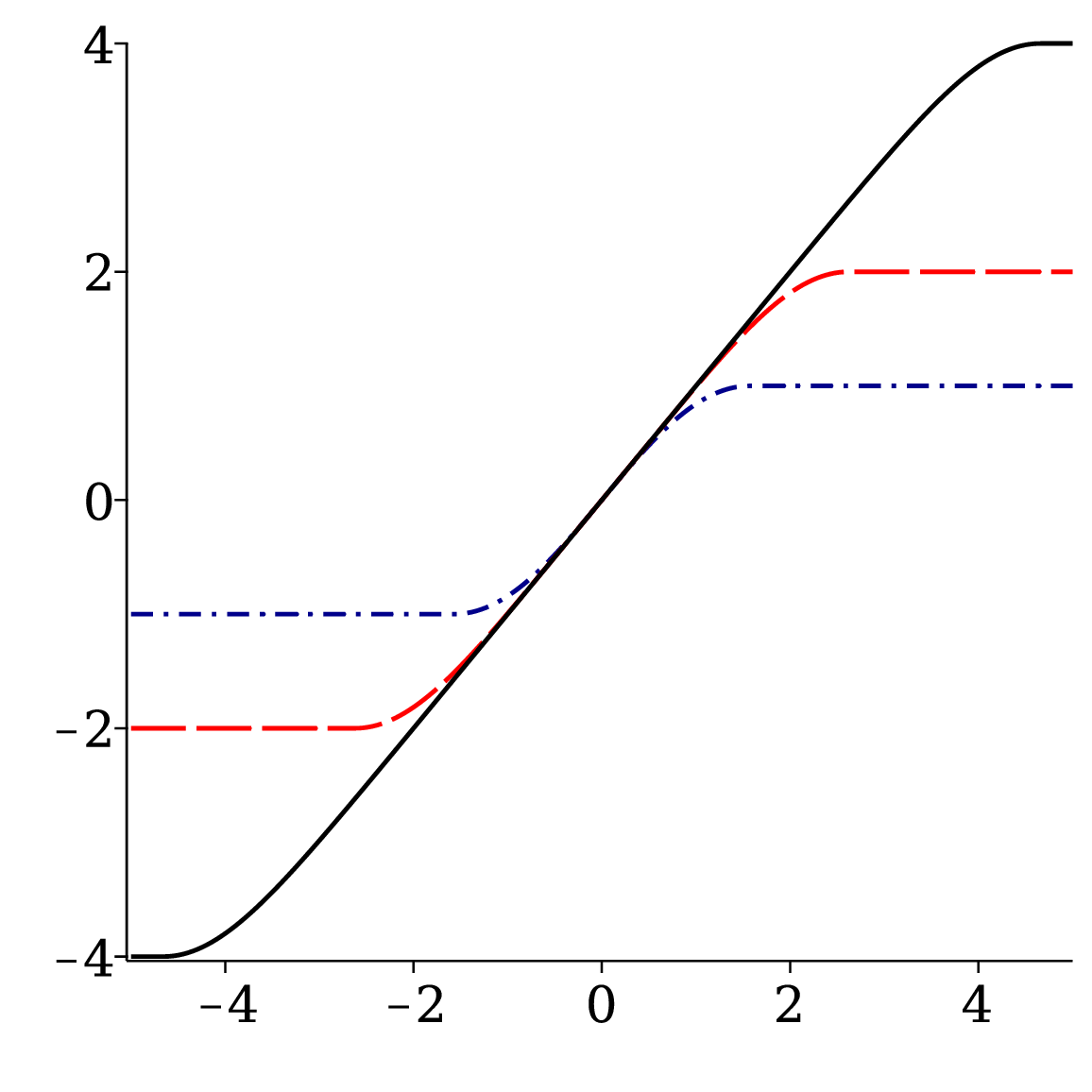}
\caption{The potential $V_{\infty}^{C}(\phi)$ in Eq.~\eqref{moduloStar} (left) and the compact solutions $\phi(x)$ (right), for $n=1,2,4$, represented by dash-dotted (blue), dashed (red) and solid (black) lines, respectively.}
\label{fig4aa}
\end{figure}

\begin{figure}%
\centering
\includegraphics[scale=0.25]{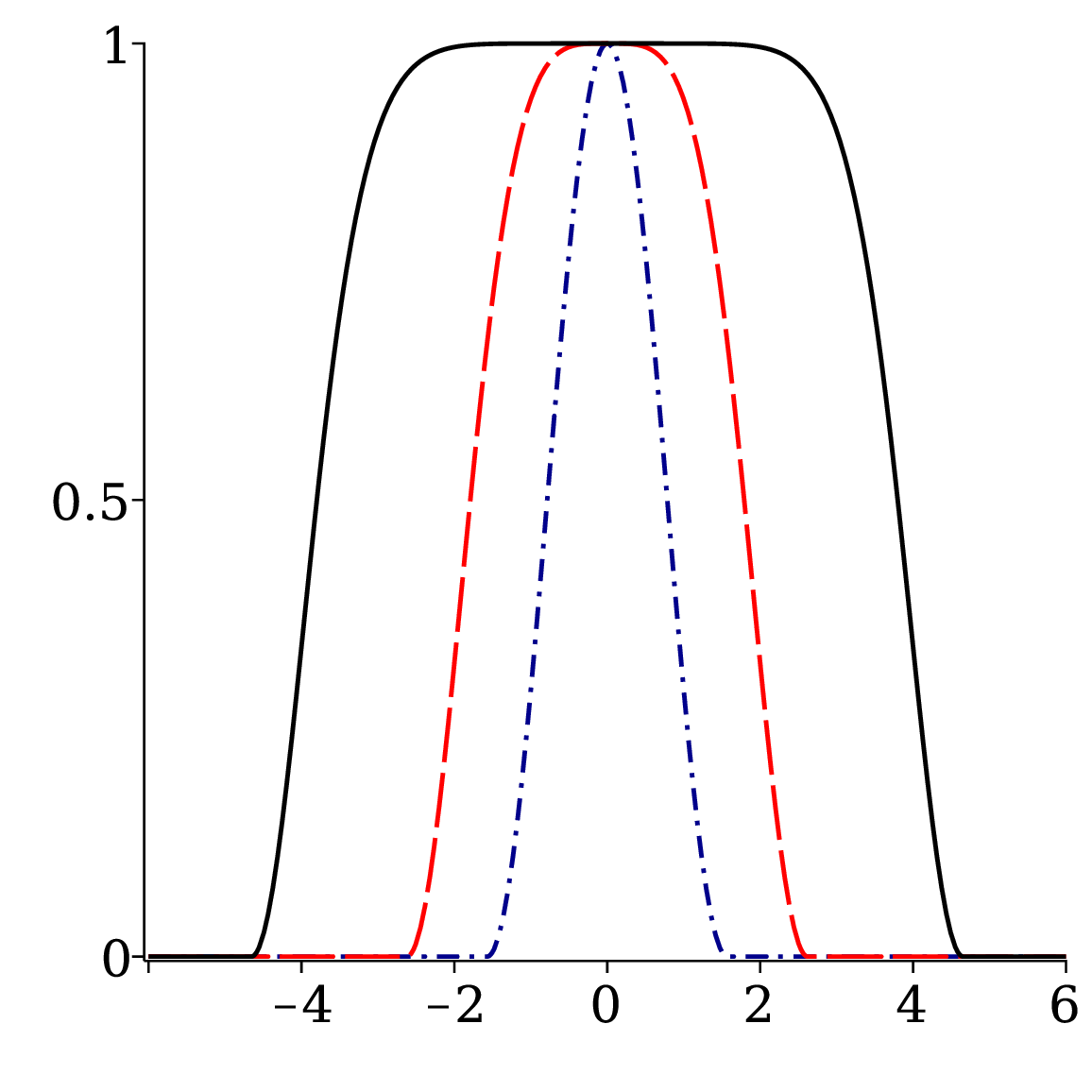}
\includegraphics[scale=0.24]{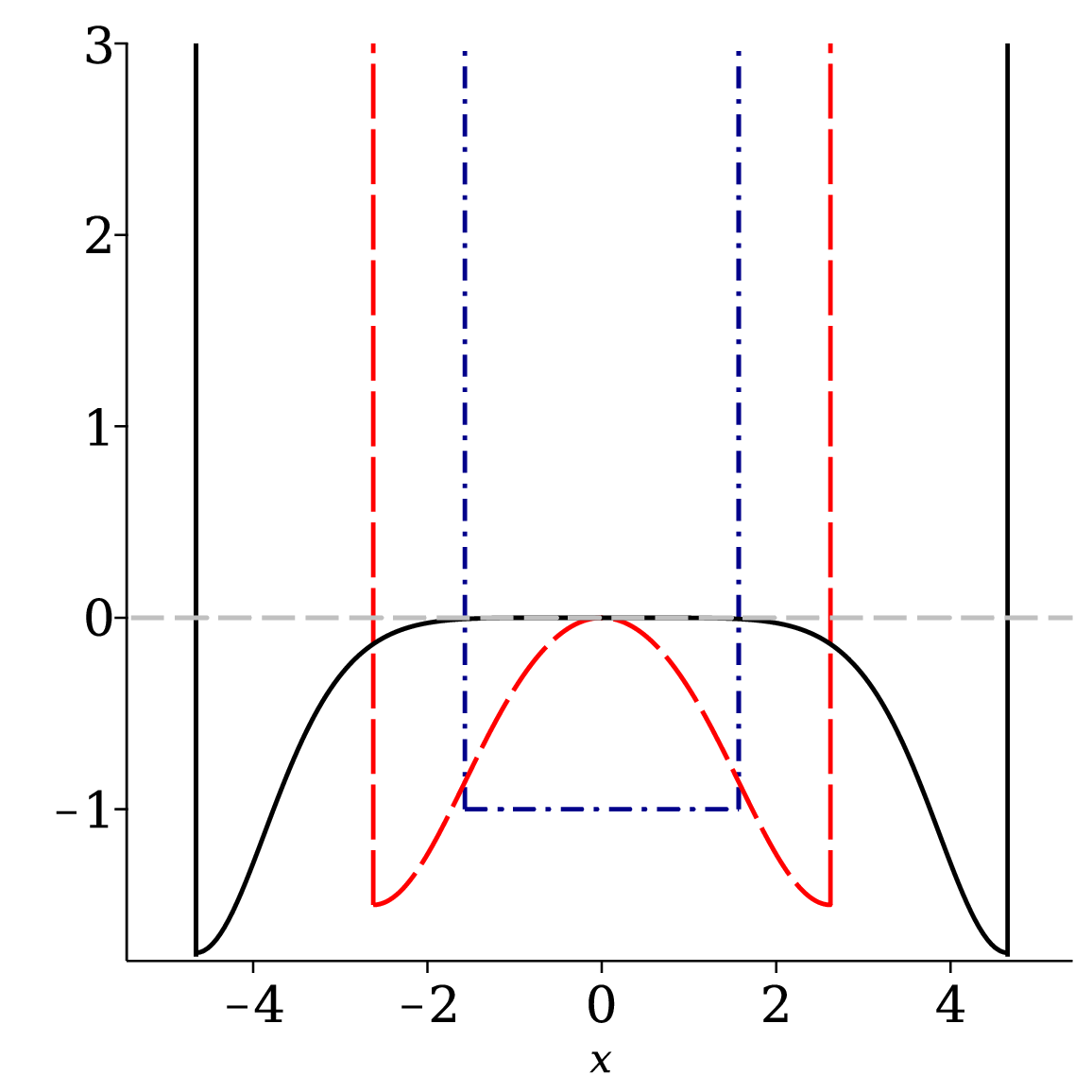}
\caption{The energy density $\rho(x)$ (left) and the stability potential $U(x)$ (right) for the model $V_{\infty}^{C}(\phi)$, using the same line styles and parameters used in Fig.~\ref{fig4aa}. }
\label{fig4bb}
\end{figure}

For the case $V(\phi)=V_C(\phi)$, the expression for $V_{\alpha}(\phi)$ reads
{\small
\be
V_{\alpha}^{C}(\phi)=\frac{1}{2\alpha}\left(\sqrt{1+2\alpha \left(1+\tfrac{\alpha}2\right)\left(1- \PC{\tfrac{\phi}{n}}^{2n}\right)^2}-1\right).
\label{Vmod2}
\ee}
This construction embeds $V_C(\phi)$ into a broader deformation framework, with the original form being recovered in the limit $\alpha\to 0$.  The argument, $\phi/n$, ensures that the modified mass is independent of $n$, yielding $m^2_{\alpha}=4+2\alpha$, while the positions of the minima remain fixed at $v_\pm=\pm n$.

The limit $\alpha\rightarrow \infty$ gives
\be
V_{\infty}^{C}(\phi)=\frac12\left|1-\PC{\frac{\phi}n}^{2n}\right|,
\label{moduloStar}
\ee
which assumes a vacuumless form as $n$ increases, in close analogy with the behaviour discussed earlier for the model $V_C{(\phi)}$ in \eqref{Starobinskyg}. This potential is shown in Fig.~\ref{fig4aa}. The corresponding function $W$ is
\be
W=\phi \cdot {_2F_1}\left(-\frac12,\frac1{2n};1+\frac1{2n};\left(\frac{\phi}n\right)^{2n}\right)
\label{Wmod2}
\ee
from which the total energy follows as
\be
E_{\infty}^{C}=2\,W(n)=\frac{n\sqrt{\pi}\,\Gamma\!\left(1+\tfrac{1}{2n}\right)}{\Gamma\!\left(\tfrac32+\tfrac{1}{2n}\right)}.
\ee
As expected, $E_{\infty}^{C}=n\,E_{\infty}^{A}$, where the linear growth with $n$ occurs due to the separation of the minima.

The first-order equation takes the form
\be
\frac{d\phi}{dx}=\left|1-\PC{\frac{\phi}{n}}^{2n}\right|^{1/2}
\ee
and the corresponding solutions exhibit compact support for every value of $n$, being determined implicitly by
\be
\phi \cdot { _2F_1}\left(\frac12,\frac1{2n};1+\frac1{2n};\left(\frac{\phi}{n}\right)^{2n}\right)= x,
\label{solstaromod}
\ee
valid in the interval $x\in[-L_{C},L_{C}]$, while $\phi(x)=-n$ for $x< -L_{C}$ and $\phi(x)=n$ for $x> L_{C}$. The vacua $(\phi_{min}=-n,n)$ are reached at the compacton edges $x=\pm L_{C}$, where
\be
L_{C}=\frac{n \sqrt{\pi}\,\Gamma\!\left(1+\tfrac{1}{2n}\right)}{\Gamma\!\left(\tfrac12+\tfrac{1}{2n}\right)}=n\,L_{A}
\ee
For $n=1$ one finds $L_{C}=\pi/2$,  while for larger $n$ the compact interval increases linearly with $n$. In the limit $n\to\infty$, the vacua move off to infinity, and the solution approaches a linear form, $\phi(x)=x$, 
corresponding to a vacuumless configuration.  Thus, this model provides a smooth interpolation between a compacton profile and a vacuumless regime.

 Figures~\ref{fig4aa} and \ref{fig4bb} illustrate the main features of the model $V_{\infty}^{C}(\phi)$. 
In Fig.~\ref{fig4aa}, the potential and corresponding compact solutions are displayed for representative values of $n$. 
As $n$ increases, the vacua move farther apart, and the compact interval confining the solutions broadens accordingly.
This effect is reflected in the energy density profiles, shown in the left panel of Fig.~\ref{fig4bb}, which remain strictly localized inside the compact region, getting wider as $n$ grows. 
The right panel of Fig.~\ref{fig4bb} shows the associated stability potential $U(x)$, which presents infinite walls. 

In Table~\ref{tab:flat_space_models}, we provide an overview of the models investigated in the two-dimensional flat spacetime, namely Models A, B, and C, along with their corresponding modulus deformations. This comparative analysis maps out the relations for their respective scalar potentials $V(\phi)$, BPS functions $W(\phi)$, total energy $E$, and analytical implicit solutions $\phi(x)$.

\begin{table}[H]
\centering
\caption{Comparative analysis of the models in two-dimensional flat spacetime.}
\label{tab:flat_space_models}
\bgroup
\renewcommand{\arraystretch}{2.4} 
\resizebox{\textwidth}{!}{%
\begin{tabular}{lllll}
\toprule
\textbf{Model} & \textbf{Potential} $V(\phi)$ & \textbf{Function} $W(\phi)$ & \textbf{Energy} $E$ & \textbf{Solution} $\phi(x)$ \\ 
\midrule

\textbf{Model A} & 
$\displaystyle \frac{1}{2}\left(1-\phi^{2n}\right)^2$ & 
$\displaystyle \phi - \frac{\phi^{2n+1}}{2n+1}$ & 
$\displaystyle \frac{4n}{2n+1}$ & 
$\displaystyle \phi \cdot {_2F_1}\left(1,\frac{1}{2n};1+\frac{1}{2n};\phi^{2n}\right) = x$ \\
& & & & Support: $x \in (-\infty, \infty)$ for finite $n$ \\
\addlinespace[0.4cm]

\textbf{Model B} & 
$\displaystyle \frac{1}{2} \left(1 - \phi^2 \right)^{2n}$ & 
$\displaystyle \phi \cdot {_2F_1} \left( \frac{1}{2}, -n; \frac{3}{2}; \phi^2 \right)$ & 
$\displaystyle \frac{\sqrt{\pi} \, \Gamma(n + 1)}{\Gamma\left(n + \frac{3}{2} \right)}$ & 
$\displaystyle \phi \cdot {_2F_1} \left( \frac{1}{2}, n; \frac{3}{2}; \phi^2 \right) = x$ \\
& & & & Support: $x \in (-\infty, \infty)$ \\
\addlinespace[0.4cm]

\textbf{Model C} & 
$\displaystyle \frac{1}{2}\left(1-\left(\frac{\phi}{n}\right)^{2n}\right)^2$ & 
$\displaystyle \phi - \frac{n}{2n+1} \left( \frac{\phi}{n} \right)^{2n+1}$ & 
$\displaystyle \frac{4n^2}{2n+1}$ & 
$\displaystyle \phi \cdot {_2F_1} \left( 1, \frac{1}{2n}; 1 + \frac{1}{2n}; \left( \frac{\phi}{n} \right)^{2n} \right) = x$ \\
& & & & Support: $x \in (-\infty, \infty)$ \\
\midrule

\textbf{Model $V_{\infty}^{A}$} & 
$\displaystyle \frac{1}{2}\left|1-\phi^{2n}\right|$ & 
$\displaystyle \phi \cdot { _2F_1}\left(-\frac{1}{2},\frac{1}{2n};1+\frac{1}{2n};\phi^{2n}\right)$ & 
$\displaystyle \frac{\sqrt{\pi}\,\Gamma\left(1+\frac{1}{2n}\right)}{\Gamma\left(\frac{3}{2}+\frac{1}{2n}\right)}$ & 
$\displaystyle \phi \cdot {_2F_1}\left(\frac{1}{2},\frac{1}{2n};1+\frac{1}{2n};\phi^{2n}\right)= x$ \\
& & & & Support: $x \in [-L_A, L_A]$, \\
& & & &  $L_{A}= {\sqrt{\pi}\,\Gamma\left(1+\frac{1}{2n}\right)}/{\Gamma\left(\frac{1}{2}+\frac{1}{2n}\right)}$ \\
\addlinespace[0.4cm]

\textbf{Model $V_{\infty}^{B}$} & 
$\displaystyle \frac{1}{2}\left|1-\phi^{2}\right|^n$ & 
$\displaystyle \phi \cdot {_2F_1}\left(\frac{1}{2},-\frac{n}{2};\frac{3}{2};\phi^2\right)$ & 
$\displaystyle \frac{\sqrt{\pi}\,\Gamma\left(\frac{n}{2}+1\right)}{\Gamma\left(\frac{n}{2}+\frac{3}{2}\right)}$ & 
$\displaystyle \phi \cdot {_2F_1}\left(\frac{1}{2},\frac{n}{2}; \frac{3}{2} ;\phi^2 \right)= x$ \\
& & & & Support: $x \in (-\infty, \infty)$ for $n \ge 2$ \\
\addlinespace[0.4cm]

\textbf{Model $V_{\infty}^{C}$} & 
$\displaystyle \frac{1}{2}\left|1-\left(\frac{\phi}{n}\right)^{2n}\right|$ & 
$\displaystyle \phi \cdot {_2F_1}\left(-\frac{1}{2},\frac{1}{2n};1+\frac{1}{2n};\left(\frac{\phi}{n}\right)^{2n}\right)$ & 
$\displaystyle \frac{n\sqrt{\pi}\,\Gamma\left(1+\frac{1}{2n}\right)}{\Gamma\left(\frac{3}{2}+\frac{1}{2n}\right)}$ & 
$\displaystyle \phi \cdot { _2F_1}\left(\frac{1}{2},\frac{1}{2n};1+\frac{1}{2n};\left(\frac{\phi}{n}\right)^{2n}\right)= x$ \\
& & & & Support: $x \in [-L_C, L_C]$, \\
& & & & $L_{C}={n \sqrt{\pi}\,\Gamma\left(1+\frac{1}{2n}\right)}/{\Gamma\left(\frac{1}{2}+\frac{1}{2n}\right)}$ \\
\bottomrule
\end{tabular}%
}
\egroup
\end{table}

\section{Braneworld Scenarios} \label{sec-5}
Let us now consider the scalar field in AdS\(_5\) (five-dimensional Anti–de Sitter space) warped geometry with a single extra dimension of infinite extent. We follow \cite{Br2,bra,Br9}, in which the metric takes the form
\begin{equation}
ds^2 = e^{2A(y)} \eta_{\mu\nu} dx^\mu dx^\nu - dy^2,
\end{equation}
with \(A = A(y)\) being the warp function, \(\eta_{\mu\nu}\) describing the four-dimensional (\(\mu, \nu = 0, 1, 2, 3\)) Minkowski spacetime, and \(y\) standing for the extra dimension. In this case, the total action, consisting of the Einstein–Hilbert term plus the source scalar sector, is
\begin{equation}
S = \int d^4x \, dy \, \sqrt{g} \left( -\frac14R + {\cal L}(\phi, \partial_\mu \phi) \right),
\end{equation}
where $g$ is the determinant of the five-dimensional metric tensor using $(+,-,-,-,-)$ signature., \(R\) is the scalar curvature,  \({\cal L}(\phi, \partial_\mu \phi)\) describes the scalar field, and we are using \(4\pi G_5 = 1\).

We suppose that the scalar field only depends on the extra dimension, \(\phi = \phi(y)\). In this case, the scalar field equation of motion has the form
\begin{equation}
\frac{d^2\phi}{dy^2} + 4 \frac{dA}{dy} \frac{d\phi}{dy} = \frac{dV}{d\phi}.
\end{equation}
In this context, the Einstein’s equations are reduced to
\begin{equation}
\frac{d^2 A}{dy^2} = -\frac{2}{3} \left(\frac{d\phi}{dy}\right)^2,
\end{equation}
\begin{equation}
\left(\frac{dA}{dy}\right)^2 = \frac{1}{6} \left(\frac{d\phi}{dy}\right)^2 - \frac{1}{3} V(\phi).
\end{equation}
We now take
\begin{equation}
\frac{dA}{dy} = -\frac23 W(\phi) \qquad \mbox{and} \qquad 
\frac{d\phi}{dy} = W_\phi
\label{foebrane}
\end{equation}
These first-order equations solve the equations of motion if the potential is written as
\begin{equation}
V(\phi) = \frac{1}{2} W_\phi^2 - \frac{4}{3} W^2(\phi).
\end{equation}
We can also find
\be
\frac{dA}{d\phi}=-\frac23\frac{W}{W_\phi},
\label{Aphi}
\ee
and the energy density is of the form
\be
\rho = e^{2A}\PC{W_\phi^2 - \frac43 W^2}=\frac{d}{dy}\!\left(e^{2A}W\right)
\ee
Consequently, the total energy is given by 
\begin{equation}
E=\int_{-\infty}^{+\infty} dy\, \rho(y)
=\left.e^{2A}W\right|_{-\infty}^{+\infty}
\end{equation}
which vanishes due to the behaviour of the warp factor $e^{2A}$ that asymptotically goes to zero for physically acceptable solutions.

Since the $\alpha$–deformation introduced in Sec.~\ref{sec-4} provides a unified framework, which contains the models in Sec.~\ref{sec-3} as the limiting case $\alpha\ll 1$, we shall adopt this description. Accordingly, the braneworld analysis will be organized in two regimes, $\alpha\ll 1$ (associated with the undeformed models) and $\alpha\gg 1$ (which yields the potentials of modulus-type), allowing us to treat all these scenarios within a single formulation.

\subsection{Case $\alpha \ll 1$}

\begin{figure}%
\centering
\includegraphics[scale=0.25]{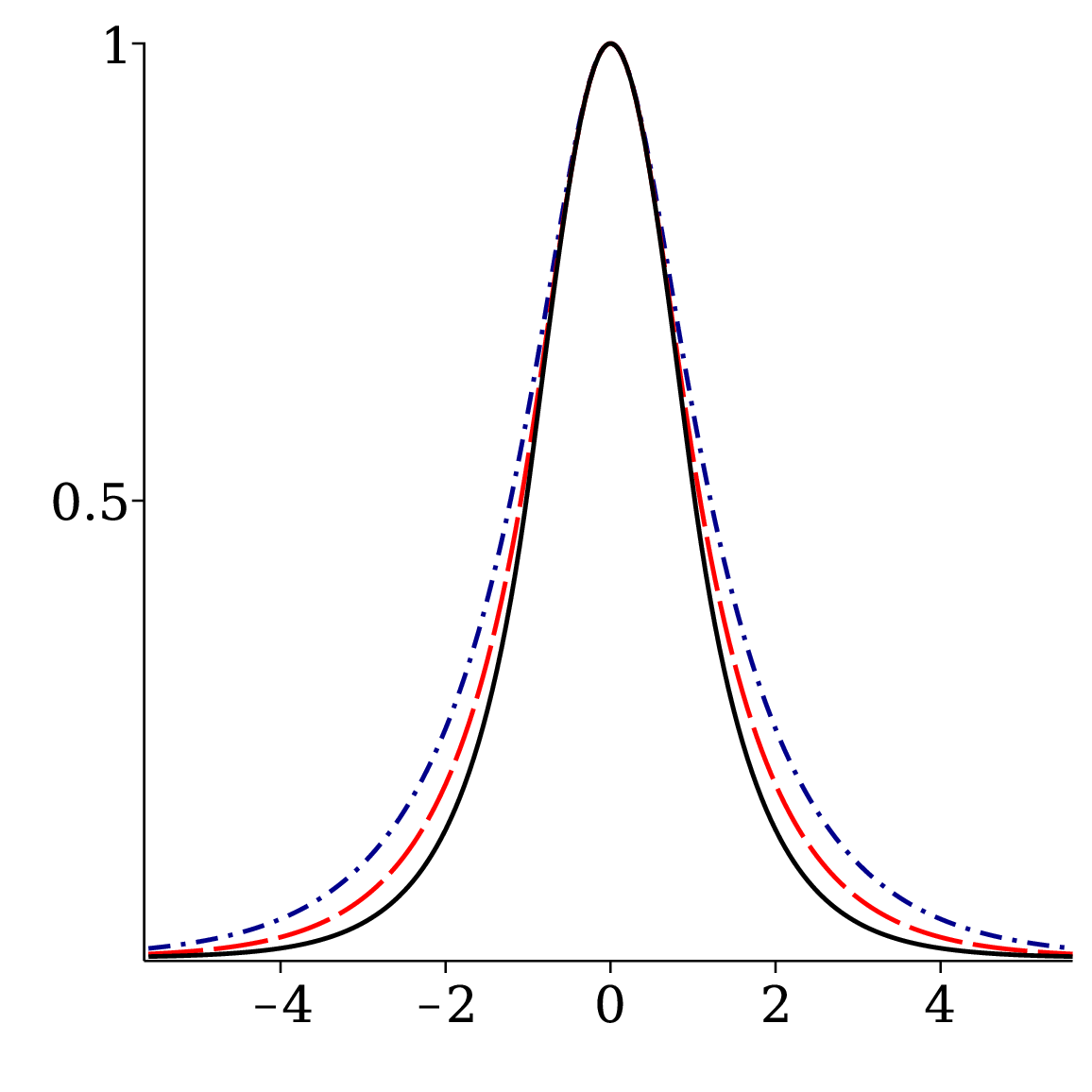}
\includegraphics[scale=0.24]{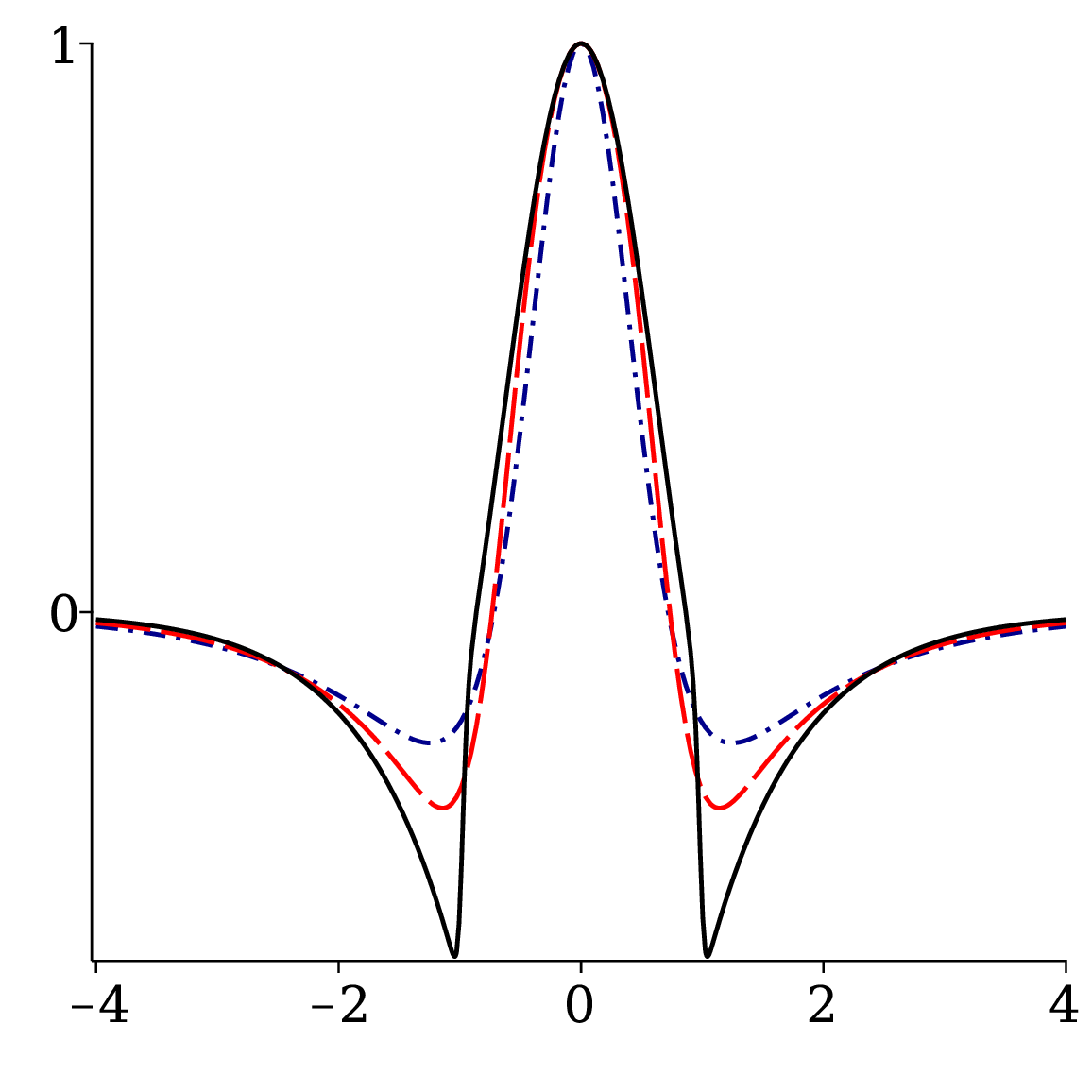}
\caption{The warp factor $e^{2A}$ (left) and the energy density $\rho$ (right) as a function of $y$ for model \eqref{Vbrane1}, adopting $n=1,2,20$, represented by dash-dotted (blue), dashed (red), and solid (black) lines, respectively (the same parameters and styles used in Fig.~\ref{fig1a}).}
\label{fig5}
\end{figure}

Here, we insert the undeformed scenarios into the five-dimensional framework. First, using the superpotential $W$ associated with $V_A$, then the one corresponding to $V_B$, and finally that of $V_C$, where each case is analysed separately. We start considering $W_{\phi}=1 - \phi^{2n}$, which gives $W$ as in~\eqref{W1}: 
\be
W(\phi) = \phi - \frac{\phi^{2n+1}}{2n+1},
\label{Wphin}
\ee
so $V(\phi)$ becomes
\begin{equation}
V(\phi) = \frac12\left( 1 - \phi^{2n} \right)^2 - \frac43\phi^2\PC{1-\frac{\phi^{2n}}{2n+1}}^2
\label{Vbrane1}
\end{equation}
The scalar field solution follows the same profile as in Eq.~\eqref{solVnstandard}. The first–order differential equation for the warp function,
\be
\frac{dA}{dy} = -\frac{2\phi}{3}\!\left(1 - \frac{\phi^{2n}}{2n+1}\right),
\ee
cannot be solved analytically for arbitrary $n$ because the explicit dependence $\phi(y)$ is not known.
Nonetheless, by employing the relation $A_\phi= -\tfrac{2}{3}\tfrac{W}{W_\phi}$, one can express the warp function analytically in terms of the scalar field as
\be
A(\phi)=-\frac{\phi^2}{3(2n+1)}-\frac{2n\phi^2}{3(2n+1)}{ _2F_1}\left(1,\tfrac1{n};1+\tfrac1{n};\phi^{2n}\right)
\ee
Particularly, $A(y)=-\frac{1}{9} \tanh^2(y) - \frac{4}{9} \ln(\cosh(y))$ when $n=1$. Because the kink transmutes into a compact form for very large values of \( n \), we note that the warp factor behaves as a thin brane when the extra dimension goes outside the compact interval \([-1, 1]\). In fact, when the kink becomes compacton, the warp function has the form
\ben
A(y) =
\begin{cases}
-\frac{1}{3} y^2, & |y| \leq 1; \\
-\frac{2}{3} |y| + \frac{1}{3}, & |y| > 1.
\end{cases}
\label{wfactor1}
\een
In the same limit, the warp function can be used to write the energy density analytically as
\ben
\rho(y) = 
\begin{cases}
\PC{1-\frac43 y^2}e^{-\frac{2}{3} y^2}, & |y| \leq 1; \\
-\frac43 e^{-\frac{4}{3} |y| + \frac{2}{3}}, & |y| > 1.
\end{cases}
\label{enerbrane1} 
\een
As we can note, the function $A(y)$ behaves quadratically near the brane core,  producing a thick–brane structure for $|y|\leq 1$, but changes to a linear profile for $|y|>1$, characteristic of thin–brane models~\cite{marques2014}. 
The energy density develops a finite discontinuity at the compacton edges in the limit $n\to\infty$, signaling the transition between the thick and thin brane phases.   Figure~\ref{fig5} shows the warp factor and the corresponding energy density.  As $n$ increases, the scalar field transition allows the model to develop the hybrid structure.

\begin{figure}%
\centering
\includegraphics[scale=0.25]{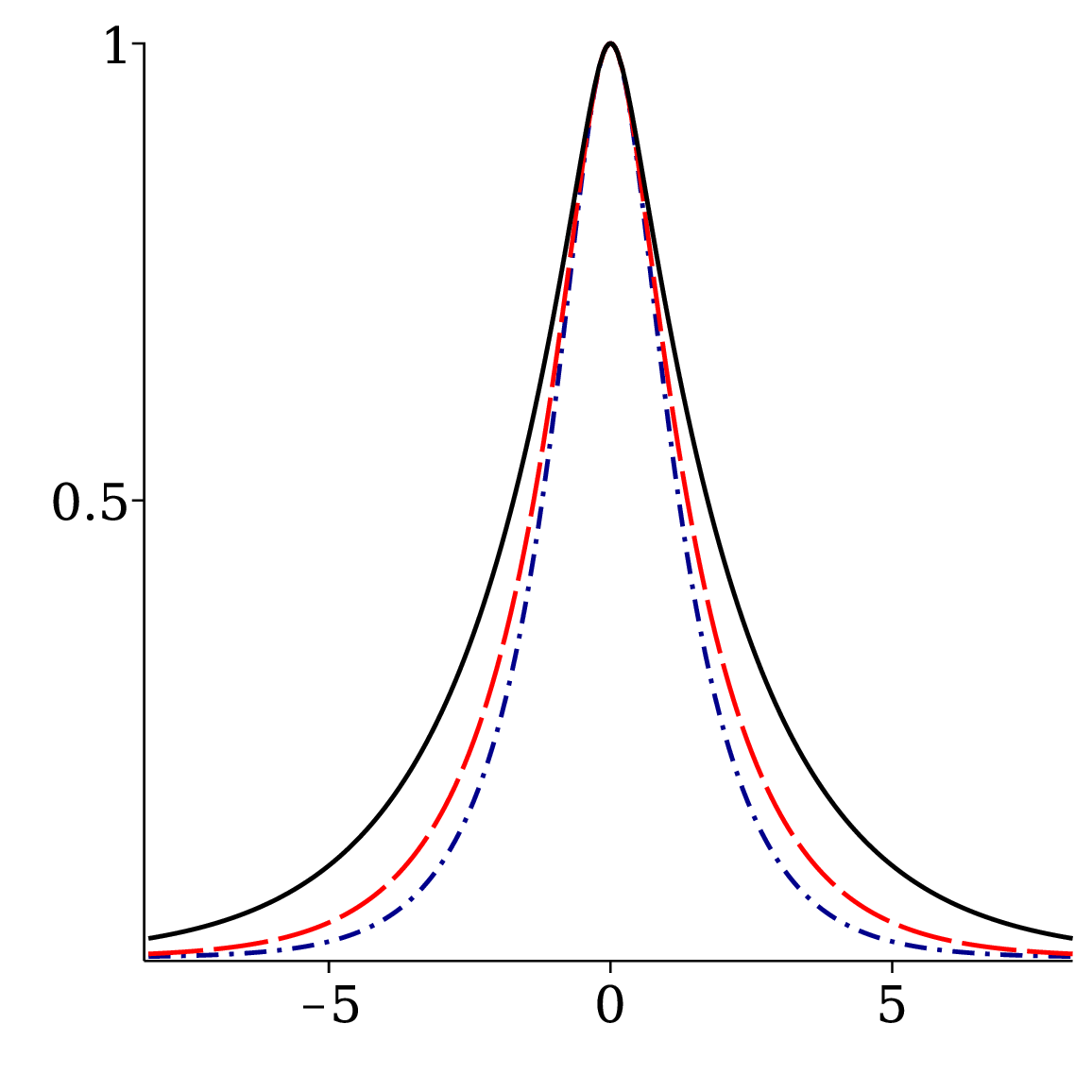}
\includegraphics[scale=0.29]{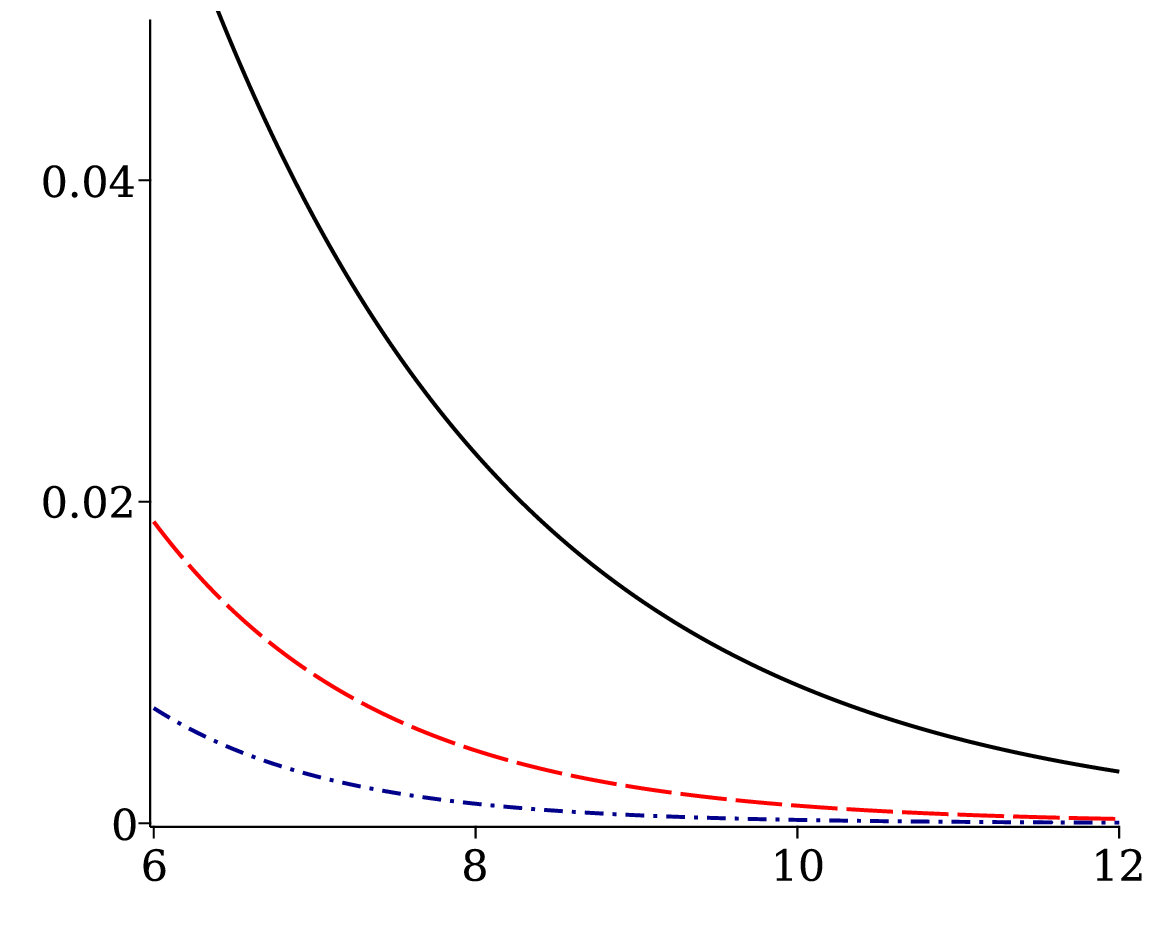}
\includegraphics[scale=0.25]{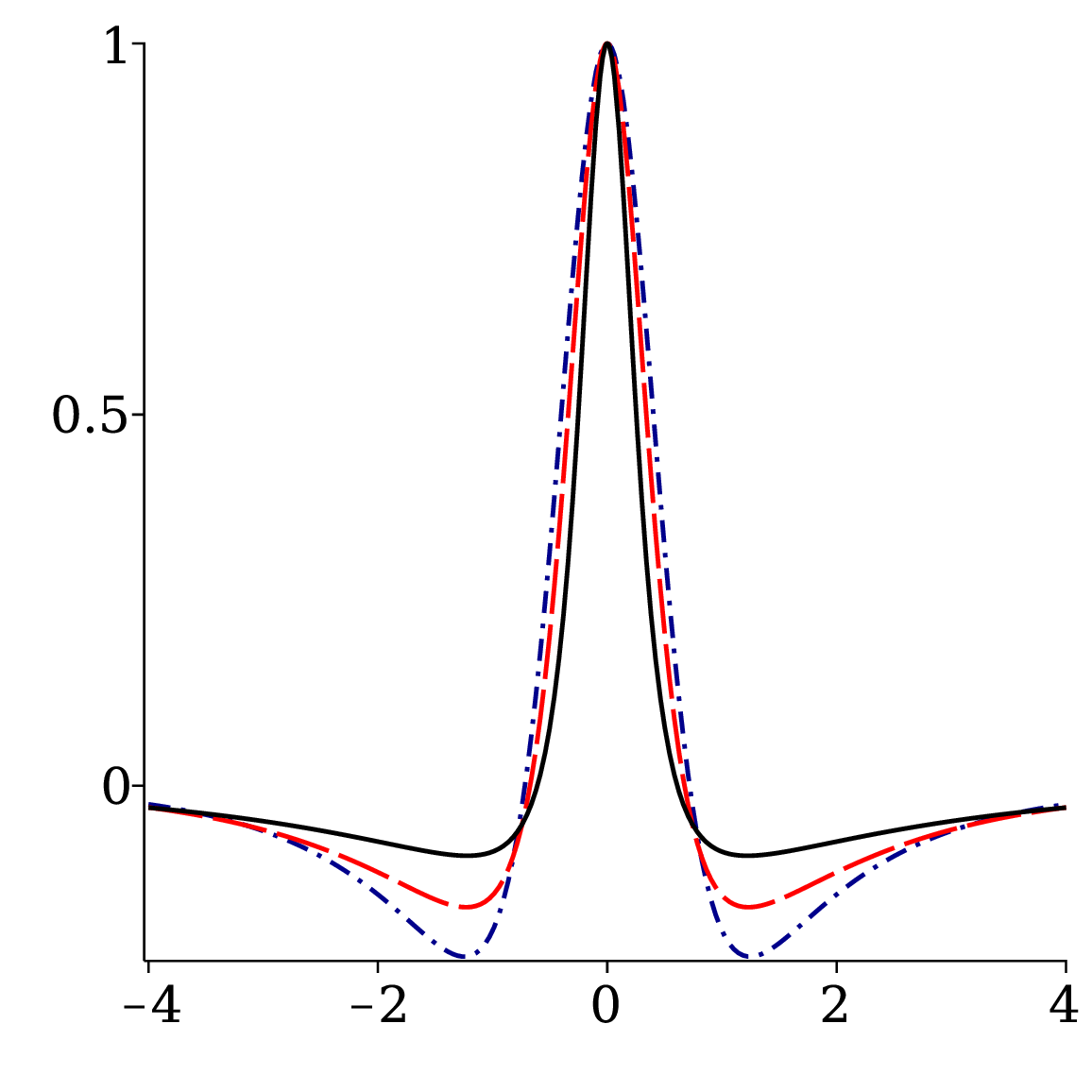}
\includegraphics[scale=0.29]{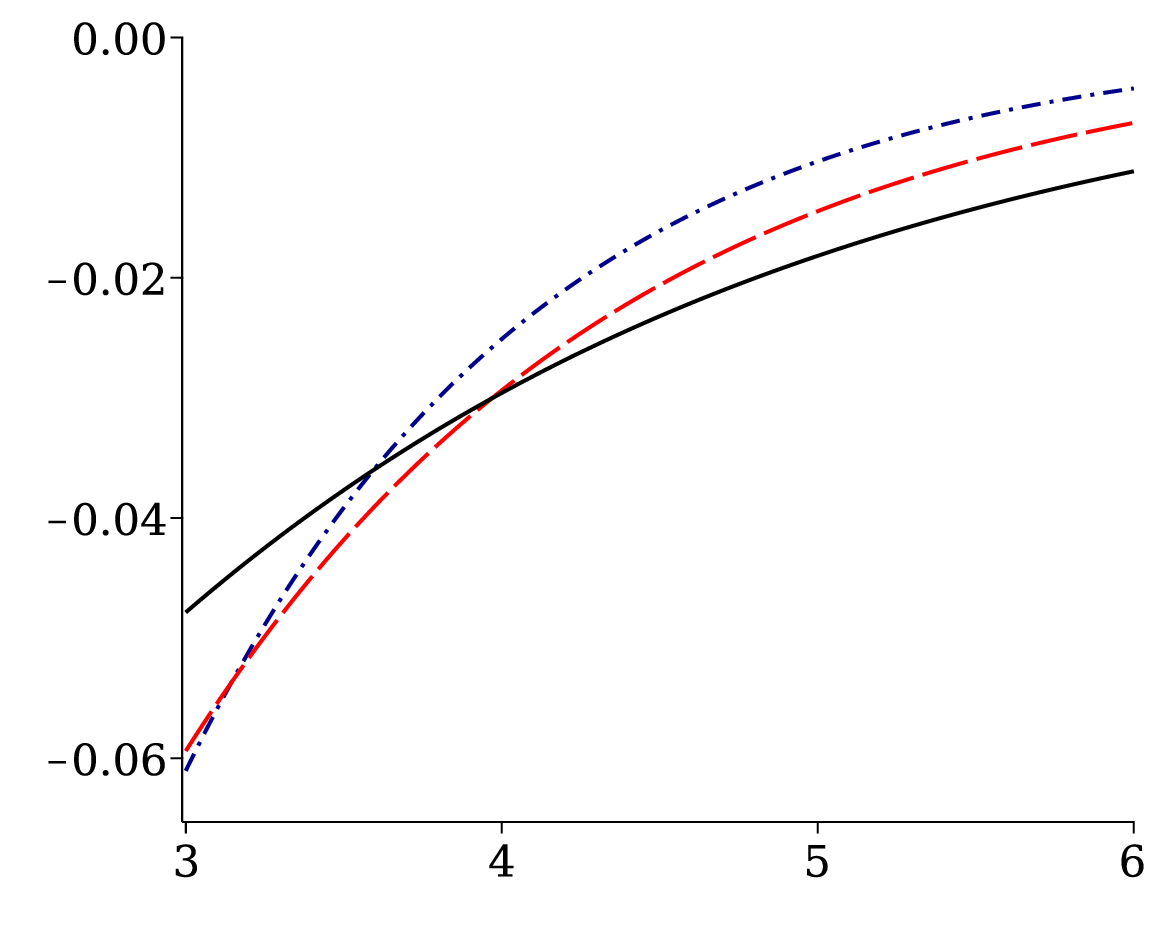}
\caption{The warp factor $e^{2A}$ (top panels) and the energy density $\rho$ (bottom panels) for model \eqref{longrangebrane}. We use $n=1,2,5$, represented by dash-dotted (blue), dashed (red) and solid (black) lines, respectively (the same values of $n$ used in Fig.~\ref{fig5a}). The left panels show the general behaviour, and the right panels the behaviour far from the centre.}
\label{fig6aa}
\end{figure}

The second model within the warped geometry considers $W_{\phi}=(1-\phi^2)^n$, whose integral gives 
\be
W(\phi) = \phi \cdot {_2F_1} \left( \frac{1}{2}, -n; \frac{3}{2}; \phi^2 \right)
\label{longrangebrane}
\ee
The kink profile obeys the same first-order equation as in \eqref{foe2}, where, for $n>1$, it has a long-range (power-law) tail in $y$. Since obtaining $A(\phi)$ by direct integration of \eqref{Aphi} is not possible using $W(\phi)$ in the hypergeometric notation, we applied the series representation of ${}_2F_1$ to get
\be
W(\phi) = \sum_{k=0}^{n}(-1)^k \binom{n}{k} \frac{\phi^{\,2k+1}}{2k+1}
\label{Wseries}
\ee
Then, $A(\phi)$ yields the analytic expression
\ben
A(\phi)\;=\;-\frac{1}{3}\sum_{k=0}^{n}(-1)^k\binom{n}{k}\;
\frac{\phi^{\,2k+2}}{(2k+1)(k+1)} \;
_2F_{1}\!\big(k+1,\,n;\,k+2;\,\phi^{2}\big),
\label{Aphi-hypergeom-sum}
\een
valid for integer $n\ge 1$. Figure~\ref{fig6aa} shows the warp factor and the corresponding energy density for this long–range model. 
The warp factor broadens as $n$ increases, reflecting the extended profile of the scalar field. 
Here, the gravitational field falls off slowly along the extra dimension.  
The energy density gets close to the origin near the centre, surrounded by slowly vanishing tails, 
consistent with the power-law asymptotic behaviour of the kinks. 
Although gravity remains localized, the long–range nature of the scalar field causes a softer and broader localization of the brane.

\begin{figure}%
\centering
\includegraphics[scale=0.25]{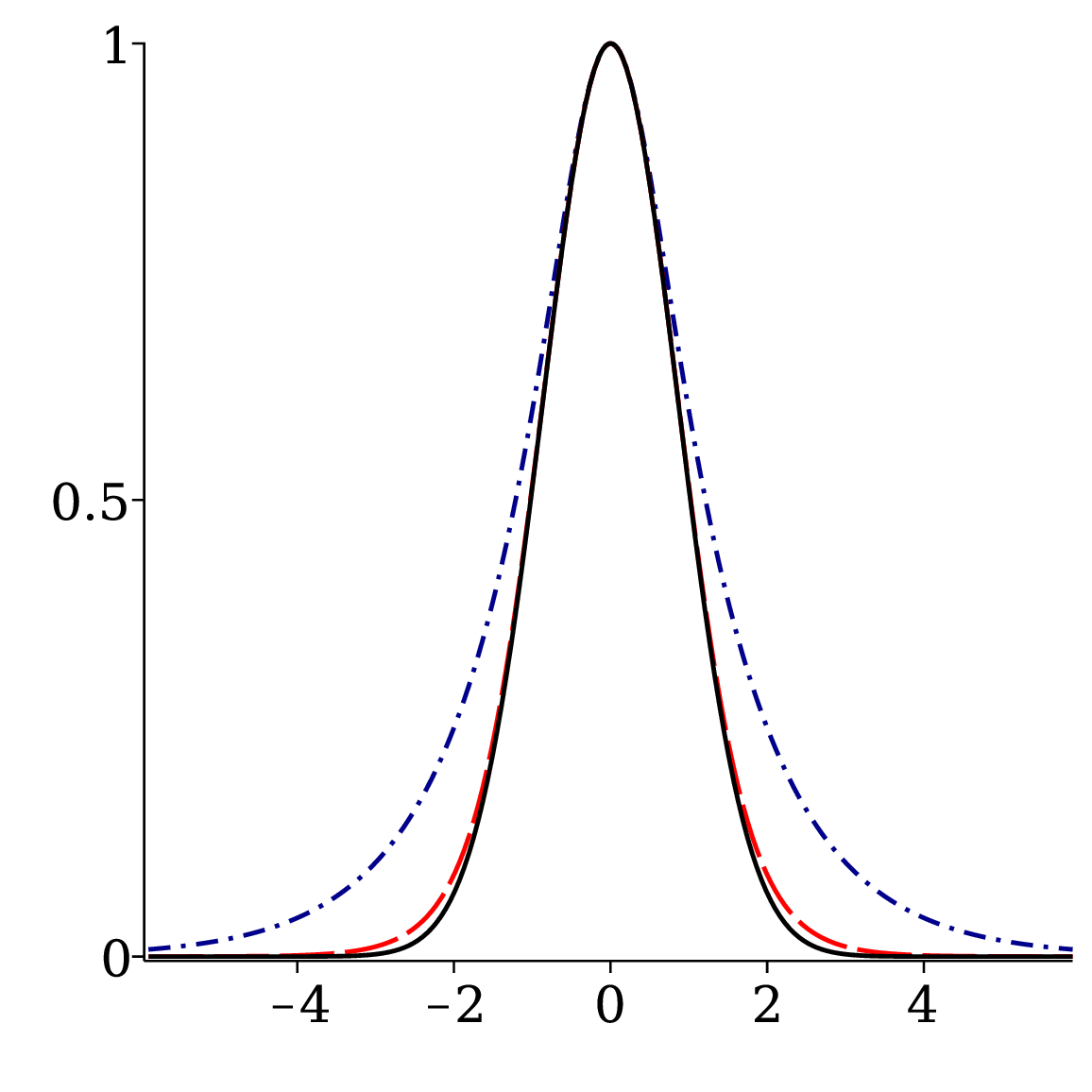}
\includegraphics[scale=0.24]{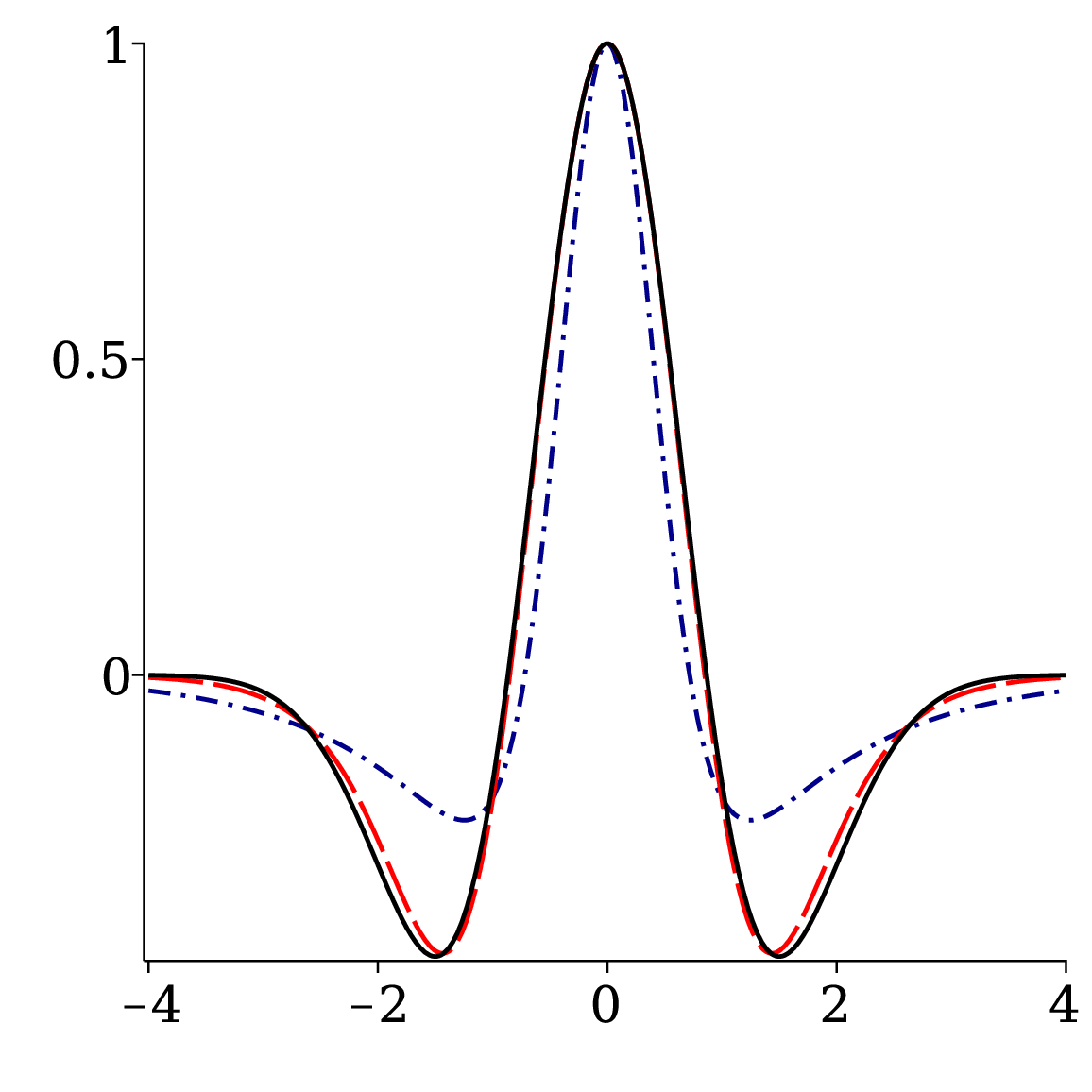}
\caption{The warp factor $e^{2A}$  (left) and the energy density $\rho$ (right) for model \eqref{Vbrane2}. We choose $n=1,2,4$, represented by dash-dotted (blue), dashed (red) and solid (black) lines, respectively (the same values of $n$ used in Fig.~\ref{fig3a}).}
\label{fig6}
\end{figure}

The next model is built with $W$ as in Eq.~\eqref{W2},
\be
W(\phi) = \phi - \frac{n}{2n+1} \left( \frac{\phi}{n} \right)^{2n+1}
\label{Vbrane2}
\ee
so that $W_{\phi} = 1 - \left({\phi}/{n}\right)^{2n}$. The warp function obtained from \eqref{Aphi} is smooth for all $y$, so the geometry exhibits thick–brane behaviour throughout the whole space, where
\be
\small
A(\phi)=-\frac{\phi^2}{3(2n+1)}-\frac{2n\phi^2}{3(2n+1)}{ _2F_1}\left(1,\tfrac1{n};1+\tfrac1{n};\PC{\tfrac{\phi}n}^{2n}\right)
\ee
In the limit of very large $n$, the warp function and the energy density are 
\be
A(y)= -\frac{1}{3}y^{2} \qquad \text{and} \qquad \rho(y) =  \PC{1-\frac43 y^2}e^{-\frac{2}{3} y^2}
\label{warp2}
\ee
This drives the warp factor to the Gaussian profile $e^{-2y^{2}/3}$ over the extra dimension, as seen in Fig.~\ref{fig6}. The gravity becomes more localized around the brane as $n$ gets larger, saturating in the vacuumless limit. This contrasts with the hybrid case, where for the same limit, the warp factor is Gaussian near the origin but becomes exponential outside the compact region. Here, we also note a contrast between how the vacuumless limit behaves in flat and warped spaces. In the two-dimensional flat space scenario, the vacuumless limit ($n \to \infty$) causes the total energy $E_C$ of the defect to diverge because the minima are removed to infinity. However, when this same field profile is embedded into the five-dimensional warped geometry, the gravitational sector is unaffected by this divergence, generating a localized thick brane configuration.

\subsection{Case $\alpha\gg 1$}

\begin{figure}%
\centering
\includegraphics[scale=0.25]{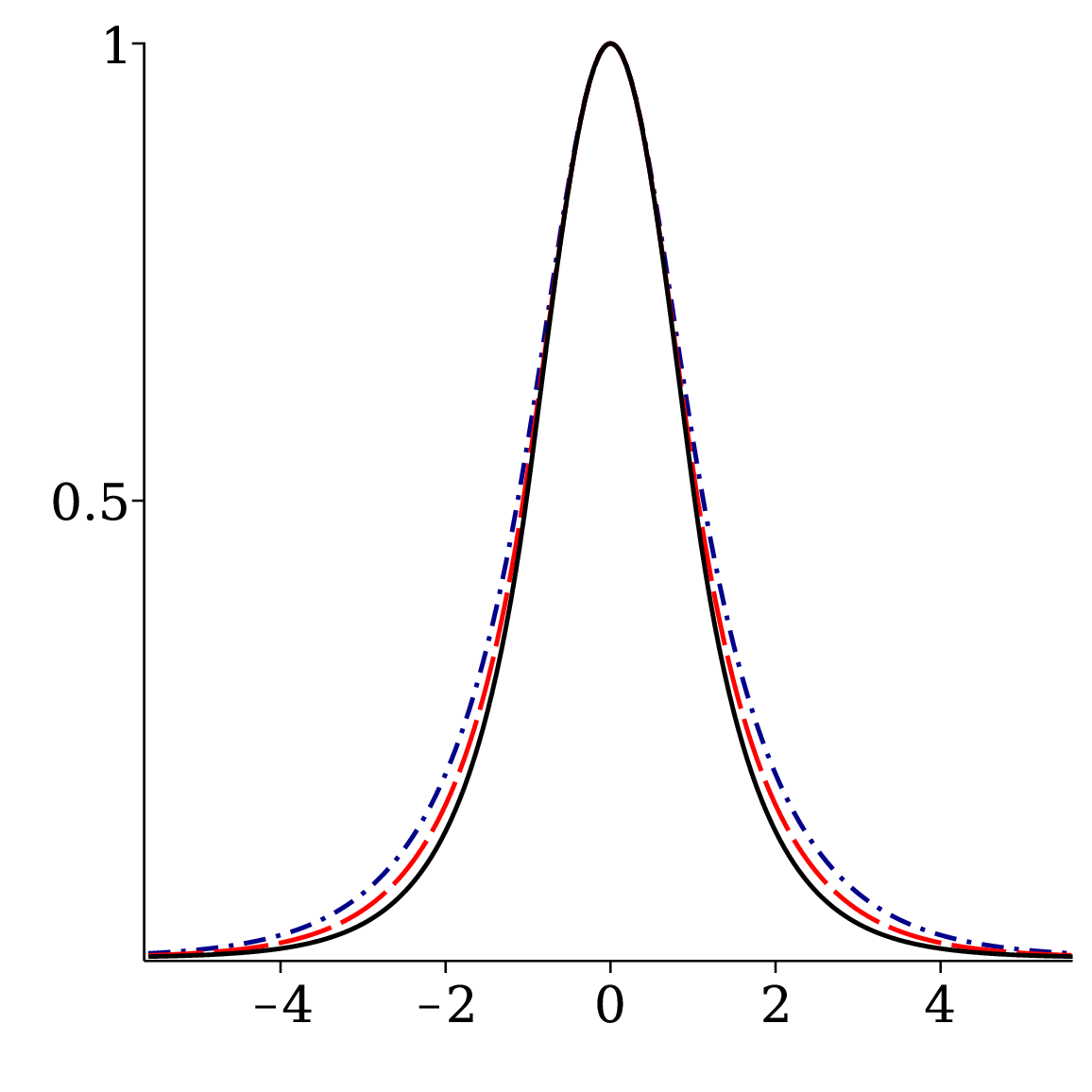}
\includegraphics[scale=0.24]{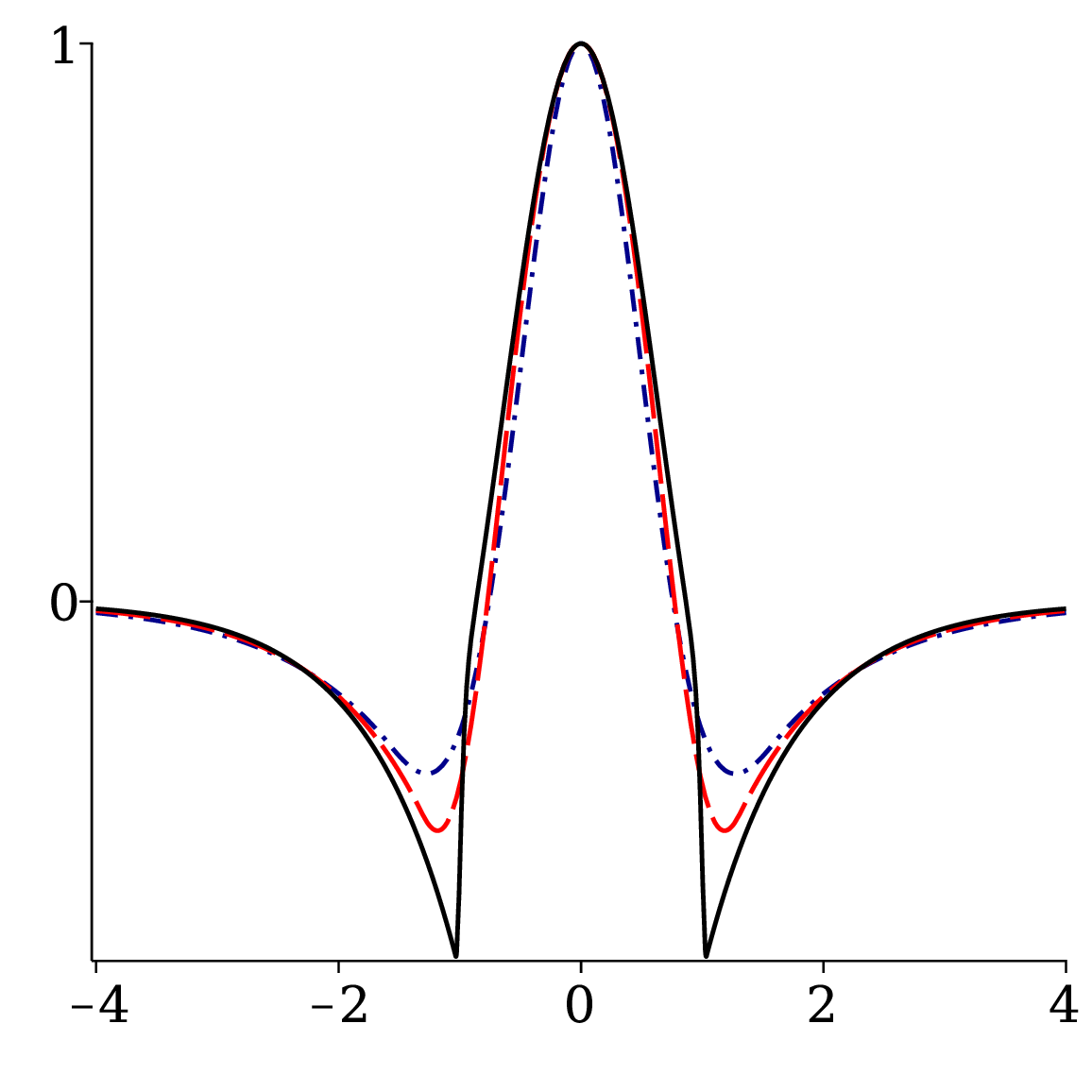}
\caption{The warp factor $e^{2A}$ (left) and the energy density $\rho$ (right) as a function of $y$ for model \eqref{Wmod1brane}, adopting $n=1,2,20$, represented by dash-dotted (blue), dashed (red), and solid (black) lines, respectively (the same parameters and styles used in Fig.~\ref{fig1a}).}
\label{fig7}
\end{figure}

In this situation,  we shall focus on the potentials of modulus type $V_{\infty}^{A}$ and  $V_{\infty}^{C}$. The brane associated with $V_{\infty}^{B}$ is not considered in detail here, since its contributions to the warped geometry are very similar to those already obtained from the other models.
We first consider $W_{\phi} =\sqrt{|1 - \phi^{2n}|}$, which gives
\be
W(\phi) =\phi \cdot { _2F_1}\left(-\frac12,\frac1{2n};1+\frac1{2n};\phi^{2n}\right)
\label{Wmod1brane}
\ee
The solution for $\phi$ has a compacton profile along the extra dimension, similar to \eqref{solVnX}.
In the situation $|y| \leq L_A$, we get
\[
\phi(y) \cdot {_2F_1}\left(\frac12,\frac1{2n};1+\frac1{2n};\phi(y)^{2n}\right)= y, 
\]
with $\phi\in[-1,1]$ and
\[
L_A= \frac{\sqrt{\pi}\,\Gamma\left(1+\frac{1}{2n}\right)}{\Gamma\left(\frac12+\frac{1}{2n}\right)}
\]
The hypergeometric series gives
\be
W(\phi) =-\frac{1}{2\sqrt{\pi}}\sum_{k=0}^{\infty}\Gamma\left(-\tfrac12+k\right)\frac{\phi^{2nk+1}}{(2nk+1)k!}
\ee
Then, one finds the expression for $A(\phi)$
\ben
A(\phi)=\frac{\phi^2}{6\sqrt{\pi}}\sum_{k=0}^{\infty}\frac{\Gamma\left(-\frac12+k\right)\phi^{2nk}}{(2nk+1)(nk+1)k!} \; 
{_2F_1}\left(\frac12,k+\frac1{n};k+1+\frac1{n};{\phi}^{2n}\right),
\een
valid for $|y| \leq L_A$.  When the field reaches $\phi=\pm1$, one can write the boundary condition $\bar{A}=A(\phi=\pm1)$ in terms of the generalized hypergeometric function  $_3F_2$
\ben
\bar{A}= - \frac{\sqrt{\pi}\,\Gamma\left(1+\frac{1}{n}\right)}{3\,\Gamma\left(\frac12+\frac{1}{n}\right)}
\;{_3F_2}\left(-\frac12,\frac1{n},\frac1{2n};\frac12+\frac1{n},1+\frac1{2n}; 1\right) 
\label{Abar}
\een
where $\bar{A}=-\frac{\pi^2}{24}-\frac16$ for $n=1$ and $\bar{A}=-1/3$ for $n\to\infty$. 
In contrast, for the case $|y| > L_A$, the scalar field saturates the vacua, $\phi(y)=\mathrm{sgn}(y)=\pm1$. Then, one has
\be
W(\pm 1) = \pm \frac{\sqrt{\pi}\,\Gamma\left(1+\frac{1}{2n}\right)}{2\,\Gamma\left(\frac32+\frac{1}{2n}\right)} = \pm \frac{nL_A}{n+1}
\ee
Accordingly, the first-order relation for $A(y)$ gives
\ben
\frac{dA}{dy}=
\begin{cases}
-\dfrac{2nL_A}{3(n+1)}, & y>L_A,\\[6pt]
\;\;\dfrac{2nL_A}{3(n+1)}, & y<-L_A,
\end{cases}
\een
whose integration yields the warp function
\be
A_{out}(y)=-\frac{2n\,L_A}{3(n+1)}
\big(|y|-L_A\big)+\bar{A},
\ee
where $A_{\text{out}}(y)$ represents the warp function outside the compact region $|y|>L_A$, and $\bar{A}=A(\phi=\pm1)=A_{out}(\pm L_A)$. This linear behaviour corresponds to a thin brane regime, which outside the compact domain goes as $e^{2A(y)}\!\sim\! e^{-\tfrac{4nL_A}{3(n+1)}|y|}$. The energy density becomes $\rho=-\frac43W^2e^{2A}$, then
\ben
\rho_{out}(y)=-\frac{4n^2 L_A^2}{3(n+1)^2}\,\exp\PC{-\frac{4nL_A}{3(n+1)}\PC{|y|-L_A} + 2\bar{A}}   
\een
These results are shown in Fig.~\ref{fig7} for some values of $n$. 
In the present case, we obtain a whole family of hybrid branes, since 
for each finite $n$ the compact solution induces a hybrid brane configuration. This contrasts with Ref. \cite{marques2014}, where the hybrid behaviour emerges only in the limit $n\to\infty$.

Explicitly, for the case $n=1$, we find
\ben
A(y) =
\begin{cases}
-\dfrac{y^{2}}{6} - \dfrac{\sin^{2}(y)}{6}, & |y| \leq \dfrac{\pi}{2}, \\[6pt]
-\dfrac{\pi}{6}|y| + \dfrac{\pi^{2}}{24} - \dfrac{1}{6}, & |y| > \dfrac{\pi}{2}.
\end{cases}
\een
The energy density inside the compact interval $y\in[-\pi/2,\pi/2]$ is 
\ben
\rho_{in}(y) = \Big(\cos^{2}y - \tfrac{1}{3}\big(y + \sin (y) \,\cos (y)\big)^{2}\Big)
\,e^{-\frac{1}{3}\big(y^{2} + \sin^{2}(y)\big)}  
\een
while outside this region it takes the form
\ben
\rho_{out}(y) =
-\dfrac{\pi^{2}}{12}\,
\exp\Big(-\frac{\pi}{3}|y| + \frac{\pi^{2}}{12} - \frac{1}{3}\Big), \quad
 |y| > \dfrac{\pi}{2}.  
\een
In this situation, there is no discontinuity in the energy density, since $\rho_{in}(y=\pi/2)=\rho_{out}(y=\pi/2)=-\frac{\pi^2}{12}\,\exp(-\frac{\pi^2}{12}-\frac13)$. However, in the limit $n\to \infty$, the warp function and energy density develop the same expressions~\eqref{wfactor1} and \eqref{enerbrane1}, and a finite discontinuity appears in $\rho(x)$, as shown in Fig.~\ref{fig7}. Consequently,  although the superpotentials defined in Eqs.~\eqref{Wphin}~and~\eqref{Wmod1brane} are distinct for finite $n$; for sufficiently large $n$, the two models share the same scalar field solution, energy density, and warp factor profiles.

We could also consider the model in Eq.~\eqref{WgeneB}, which follows from $W_{\phi} = |1 - \phi^{2}|^{n/2}$. However, its contribution to the warped geometry is essentially already analysed. In particular, $n=1$ corresponds to the hybrid behaviour obtained from $W_{\phi}=\sqrt{|1-\phi^{2}|}$. The case $n>1$ leads to thick–brane configurations, where $n=2$ is modulated by the short–range (exponential) $\phi^{4}$ kink, and $n>2$ by kinks with power-law (long–range) tails. For this reason, we do not repeat the analysis.

\begin{figure}%
\centering
\includegraphics[scale=0.25]{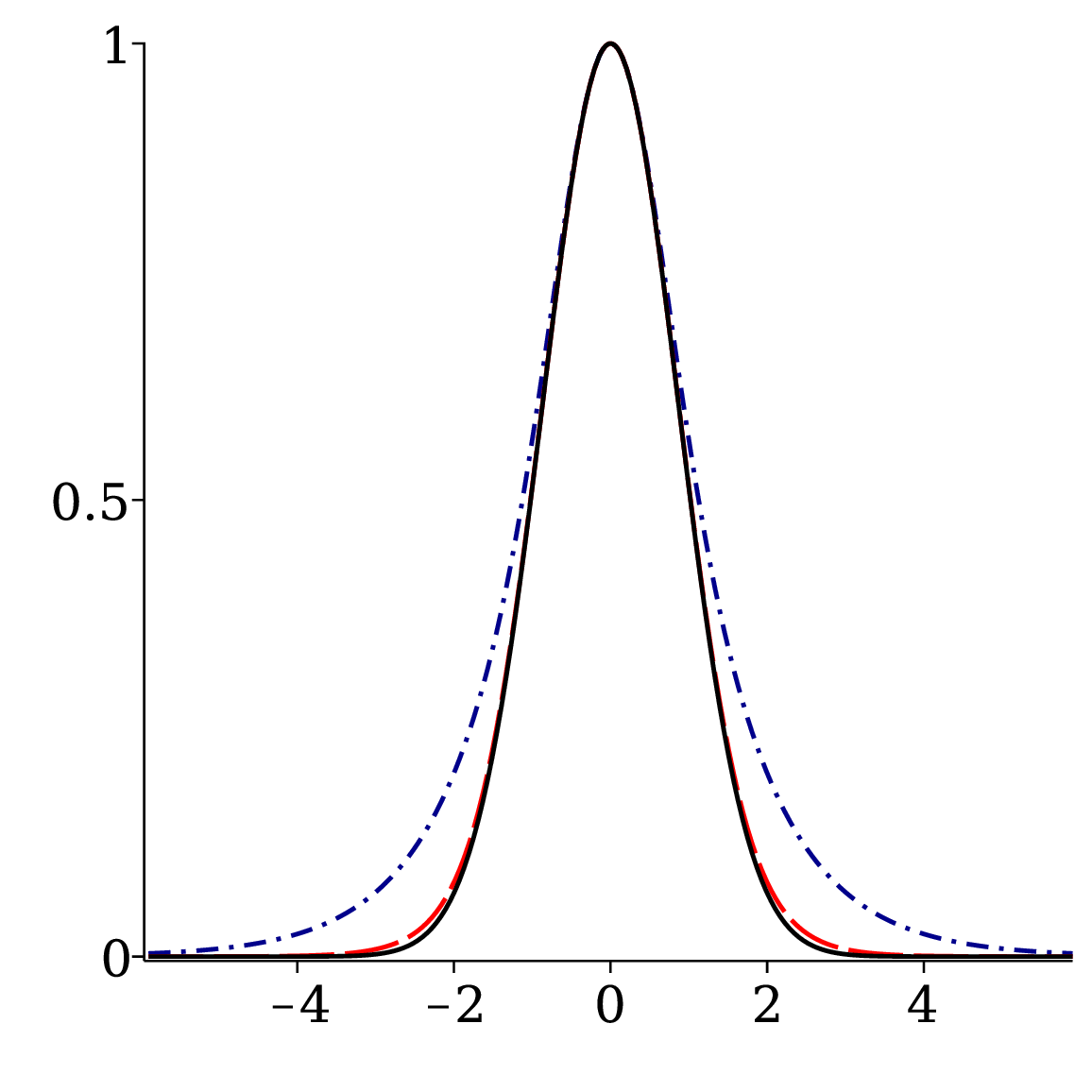}
\includegraphics[scale=0.24]{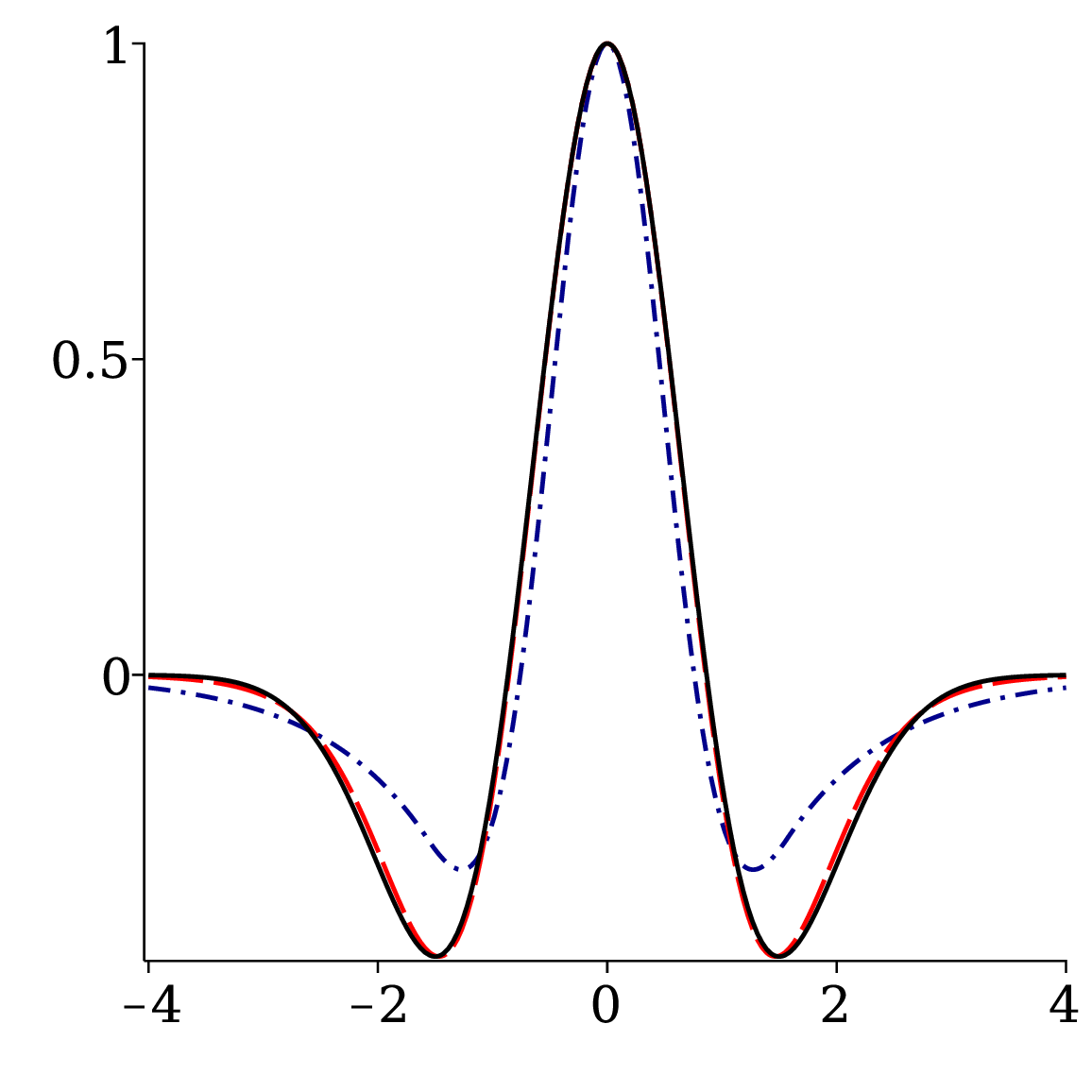}
\caption{The warp factor $e^{2A}$ (left) and the energy density $\rho$ (right) as a function of $y$ for model \eqref{Wmod2}, with $n=1,2,4$, represented by dash-dotted (blue), dashed (red) and solid (black) lines, respectively (the same parameters and styles used in Fig.~\ref{fig3a}).
}
\label{fig8}
\end{figure}

The last model considers $W$ given in Eq.~\eqref{Wmod2}, that is
\be
W=\phi \cdot {_2F_1}\left(-\frac12,\frac1{2n};1+\frac1{2n};\left(\frac{\phi}n\right)^{2n}\right),
\label{Wbranemods}
\ee
so $W_{\phi}=\sqrt{\left|1-({\phi}/{n})^{2n}\right|}$. The scalar field has the same compact profile as that gotten from Eq.~\eqref{solstaromod}. Accordingly, the field reaches the vacua at the finite points $\phi(\pm L_C)=\pm n$, where $L_C=nL_A$. The expression for $A(\phi)$ is obtained by rewriting $W$ in hypergeometric series, then
\ben
A(\phi)=\frac{\phi^2}{6\sqrt{\pi}}\sum_{k=0}^{\infty}\frac{\Gamma\left(-\frac12+k\right)}{(2nk+1)(nk+1)k!}\PC{\frac{\phi}n}^{2nk} \times \nonumber \\ 
{_2F_1}\left(\frac12,k+\frac1{n};k+1+\frac1{n};\PC{\frac{\phi}n}^{2n}\right),
\een
for $\phi\in[-n,n]$. When the field reaches $\phi=\pm n$, we find $A(\phi=\pm n)=n^2 \bar{A}$, where $\bar{A}$ is already given in \eqref{Abar}. 
For the case $|y| > L_C$, we get $\phi(y)=n \,\mathrm{sgn}(y)$, so the analytical expression for the warp function outside the compact support is
\ben
A_{out}(y)=-\frac{2nL_C}{3(n+1)}\PC{|y|-L_C} + n^2\bar{A}, 
\een
and the energy density is
\ben
\rho_{out}(y)\!=-\frac{4n^2 L_C^2}{3(n+1)^2}\,\exp\!\PC{\!-\frac{4nL_C}{3(n+1)}\!\PC{|y|\!-\!L_C}\! +\! 2n^2\bar{A}\!}  
\een
This model gives rise to another class of branes exhibiting hybrid behaviour. 
However, as $n \to \infty$, the compact region $L_C$ diverges, making the warp factor approach the Gaussian profile $e^{-2y^{2}/3}$. Therefore, the geometry that initially presents a hybrid regime becomes thick for very large $n$. The two models defined by Eqs.~\eqref{Vbrane2} and \eqref{Wbranemods}, which are distinct for finite $n$, converge to the same scalar field solution, energy density, and warp factor profiles when $n$ gets large. The warp factor and the associated energy density for this case are displayed in Fig.~\ref{fig8}.

Therefore, the $\alpha$-deformation allows to generate two distinct families of hybrid branes at finite and discrete values of $n$ (such as $n=1, 2, 4$), features that have not yet been addressed in the literature. In Ref~\cite{marques2014}, a hybrid brane profile described by a single scalar field was proposed in the limit where the parameter $n$ goes to infinity. The present work expands this perspective: in the case of \eqref{Wmod1brane}, the brane exhibits a hybrid profile for any finite value of $n$; whereas in the case of \eqref{Wbranemods}, the brane begins as hybrid for $n=1$ but continuously transitions to a thick configuration in the limit $n \to \infty$.

\section{Outlook}\label{sec-6}

In this work, we investigated scalar field models that can interpolate among kinks, compactons, long-range, or vacuumless configurations, depending on how an integer parameter $n$ is included. Starting from three representative models ($V_A$, $V_B$, and $V_C$), we discuss their main properties and demonstrate how the solutions are altered as $n$ increases. We then introduced a deformation controlled by a real parameter $\alpha$, which modifies the shape of the potential while preserving its minima. The limit $\alpha \to \infty$ yields new families of models whose first-order equations can still be solved. 

Next, we embedded the two-dimensional models into five-dimensional warped geometries, allowing us to investigate a rich set of braneworld scenarios. Through this approach, new families of hybrid branes are obtained, and their distinct physical features are explicitly described. The results show that the compact, long-range, and vacuumless regimes proposed in a two-dimensional flat geometry can also play an interesting role in the case of a five-dimensional warped background. This suggests that we can explore other possibilities, in particular, the case of branes in two-dimensions, considering generalized Jackiw-Teitelboim gravity, coupling the dilaton field with scalar matter fields \cite{JT1,JT2,JT3,JT4,JT5}. It is also currently of interest to study branes in four-dimensions, including modified Einstein-Gauss-Bonnet gravity, as recently investigated in Ref. \cite{B4d}.
We can also think of studying braneworlds generated by several scalar fields, focusing on how the fields can be used to modify the internal structure of the brane, in a way similar to the case previously described in Ref. \cite{Lob}. These and other possibilities are presently under consideration, and we hope to report on them in the near future. 

\section*{Acknowledgments}

The authors would like to thank the Conselho Nacional de Desenvolvimento Científico e Tecnológico, CNPq, (Grants 303469/2019-6 and 402830/2023-7) and PRPGI/IFBA for the financial support.



\begin{thebibliography}{99}

\bb{B1} R. Rajaraman, \textit{Solitons and Instantons} (North-Holland, 1982).
\bb{B2} A. Vilenkin and E. P. S. Shellard, \textit{Cosmic Strings and Other Topological Defects} (Cambridge University Press, 1994).
\bb{B3} N. Manton and P. Sutcliffe, \textit{Topological Solitons} (Cambridge University Press, 2004).
\bb{B4}T. Vachaspati, Kinks and Domain Walls (Cambridge
University Press, 2007).
\bb{B5}Y. M. Shnir, Topological and Non-Topological Solitons
in Scalar Field Theories (Cambridge University Press,
2018).
\bb{BB1} D. Walgraef, \textit{Spatio-Temporal Pattern Formation} (Springer, 1997).
\bb{BB2} C. J. Pethick and H. Smith, \textit{Bose–Einstein Condensation in Dilute Gases} (Cambridge University Press, 2002).
\bb{BB3} G. F. Nataf \textit{et al.}, Nat. Rev. Phys. 2, 634 (2020).
\bb{BB4} Y. Tokura and N. Kanazawa, Chem. Rev. 121, 2857 (2021).

\bb{rosenau} P. Rosenau and J. M. Hyman, Phys. Rev. Lett. 70, 564 (1993).
\bb{C1}H. Arodz, Acta Phys. Polon. B 33, 1241 (2002).
\bb{C2}H. Arodz, P. Klimas, and T. Tyranowski, Acta Phys.
Polon. B 36, 3861 (2005).
\bibitem{compactkinks} D. Bazeia, E. da Hora, R. Menezes, H. P. de Oliveira, and C. dos Santos, Phys. Rev. D 81, 125016 (2010).
\bb{marques2014} D. Bazeia, L. Losano, M. A. Marques, and R. Menezes, Phys. Lett. B 736, 515 (2014).
\bb{lima2017} D. Bazeia, E. E. M. Lima, and L. Losano, Eur. Phys. J. C 77, 127 (2017).

\bb{Br1}L. Randall and R. Sundrum, Phys. Rev. Lett. 83, 4690 (1999).
\bb{Br2}O. DeWolfe, D. Z. Freedman, S. S. Gubser, and A. Karch, Phys. Rev. D 62, 046008 (2000).
\bb{Br3}G. Dvali, G. Gabadadze, and M. Porrati, Phys. Lett. B 485, 208 (2000).
\bb{bra}C. Csaki, J. Erlich, T. J. Hollowood, Y. Shirman
Nucl. Phys. B 581, 309 (2000).
\bb{Br4}F. A. Brito, M. Cvetic, and S.-C. Yoon, Phys. Rev. D 64, 064021 (2001).
\bb{Br5}M. Cvetic and N. D. Lambert, Phys. Lett. B 540, 301 (2002).
\bb{Br6}A. Campos, Phys. Rev. Lett. 88, 141602 (2002).
\bb{Br7}D. Bazeia, F. A. Brito, and J. R. Nascimento, Phys. Rev. D 68, 085007 (2003).
\bb{Br8}D. Bazeia, C. Furtado, and A. R. Gomes, JCAP 0402, 002 (2004).
\bb{Br9}D. Bazeia and A. R. Gomes, JHEP 05, 012 (2004).


\bb{ChoVacuumless1999}I. Cho and A. Vilenkin, Phys. Rev. D 59, 021701(R) (1999).
\bb{ChoV}I. Cho and A. Vilenkin, Phys. Rev. D 59, 063510 (1999).

\bb{dbazeia1999} D. Bazeia, Phys. Rev. D 60, 067705 (1999).

\bb{Gomes2012} A. R. Gomes, R. Menezes, and J. C. R. E. Oliveira, Phys. Rev. D 86, 025008 (2012).
\bb{BMM2018} D. Bazeia, R. Menezes, and D. C. Moreira, J. Phys. Commun. 2, 055019 (2018).
\bb{Blinov2022} P. A. Blinov, T. V. Gani, A. A. Malnev, V. A. Gani, and V. B. Sherstyukov, Chaos Solitons Fractals 165, 112805 (2022).



\bb{bogomol} E. B. Bogomolny, Sov. J. Nucl. Phys. 24, 449 (1976).
\bb{Bazeia} D. Bazeia, \textit{Defect Structures in Field Theory}, arXiv:hep-th/0507188.


\bb{Kevrekidis2019} P. G. Kevrekidis and J. Cuevas-Maraver (eds.), \textit{A Dynamical Perspective on the $\phi^4$ Model}, Springer (2019).
\bb{superlongrangeKinks} I. Andrade, M. A. Marques, and R. Menezes, Chaos, Solitons \& Fractals 192, 116040 (2025).
\bb{twin2012} D. Bazeia, A. S. Lobão Jr., and R. Menezes, Phys. Rev. D 86, 125021 (2012).

\bb{teller} G. Pöschl and E. Teller, Z. Phys. 83, 143 (1933).
\bb{flugge} S. Flügge, \textit{Practical Quantum Mechanics} (Springer, 1994).
\bb{mathrefs} M. Abramowitz and I. A. Stegun, \textit{Handbook of Mathematical Functions} (Dover, 1972); I. S. Gradshteyn and I. M. Ryzhik, \textit{Table of Integrals, Series, and Products}, 7th ed. (Academic, 2007).




\bb{epjp-2021} S. S. da Costa \textit{et al.}, Eur. Phys. J. Plus 136, 84 (2021).

\bb{starobinsky} A. A. Starobinsky, Phys. Lett. 91B, 99 (1980).

\bb{lima2022} E. E. M. Lima and F. A. Brito, Ann. Phys. 439 (2022).
\bb{lima2024} D. Bazeia and E. E. M. Lima, Eur. Phys. J. C 84, 668 (2024).

\bb{JCAP2018} M. A. Santos \textit{et al.}, JCAP 03, 023 (2018).
\bb{ketov2020} S. V. Ketov, J. Phys. A 53, 084001 (2020).
\bb{ketov2025} S. V. Ketov, \textit{On Legacy of Starobinsky Inflation}, arXiv:2501.06451 [gr-qc] (2025).

\bb{JT1}Y. Zhong, JHEP 04, 118 (2021).
\bb{JT2} Y. Zhong, F. Y. Li and X. D. Liu, Phys. Lett. B
822, 136716 (2021).
\bb{JT3}Y. Zhong, Phys. Lett. B 827, 136947 (2022).
\bb{JT5}I. Andrade, D. Bazeia, A. S. Lobão Jr, and R. Menezes, Chin. Phys. C 46, 125102 (2022).
\bb{JT4}J. Feng and Y. Zhong, EPL 137, 49001 (2022).
\bb{B4d}D. Bazeia, R. Menezes, A. Yu. Petrov, and P. J. Porfirio, Int. J. Mod. Phys. D 34, 2550039 (2024).
\bb{Lob}D. Bazeia, A. S. Lobão Jr, Eur. Phys. J. C 82, 579L (2022).
\end{thebibliography}
\end{document}